\documentclass[aps,pre,preprint,groupedaddress,amssymb]{revtex4-2}
\usepackage{graphicx} 
\usepackage{dcolumn}
\usepackage{bm}
\usepackage{amsmath}
\usepackage{epsfig}
\usepackage{color,hyperref}

\begin{document}
\title{Ergodic properties of occupation times in heterogeneous media}

\author{V. M\'endez}
\email{vicenc.mendez@uab.cat}
\author{R. Flaquer-Galm\'es}

\affiliation{Grup de F\'{\i}sica Estad\'{\i}stica, Departament de F\'{\i}sica. Facultat de Ci\`{e}ncies, Universitat Aut\`{o}noma de Barcelona, 08193 Barcelona, Spain}

\begin{abstract}
We investigate the ergodic properties of Brownian motion in heterogeneous media through the statistics of occupation times. Using the Feynman–Kac formalism, we derive analytical expressions for the distributions, moments, and ergodicity-breaking parameters of occupation times in two models with spatially varying diffusion coefficient: a piecewise-constant profile and a power-law profile. In the piecewise model, the half occupation time and the occupation time within an interval follow asymmetric arcsine and half-Gaussian distributions, respectively, indicating non-ergodic behavior. For the power-law case, the corresponding distributions are the Lamperti and Mittag–Leffler. In both models, we identify a transition from non-ergodic to ergodic dynamics as the exponent vary. Numerical simulations fully corroborate the analytical results, demonstrating the effectiveness of the Feynman–Kac approach for quantifying ergodicity in heterogeneous diffusion processes.
\end{abstract}

\maketitle

\section{Introduction}
Heterogeneous diffusion processes have received growing attention in recent years, motivated by the need to understand transport in complex systems where microscopic properties vary in space or time. Such systems arise naturally in physics, chemistry, and biology, encompassing transport in porous materials, colloidal suspensions, and intracellular environments. The nonuniform structure of these media often induces strong spatial dependencies in local transport coefficients, leading to diffusion dynamics that deviate significantly from classical Brownian motion.
Studies of diffusion in heterogeneous media cover a wide range of scenarios, including particles confined between nearly parallel plates \cite{LaBa01}, diffusion under temperature gradients \cite{Ya12}, nanoporous solids \cite{Ma21}, and confined hard-sphere fluids \cite{Mi08}. In parallel, experimental developments—particularly single-particle tracking techniques \cite{ChMe20,Che13,Ma15,We17,No17,Che17,Kr19}—have enabled the direct observation of individual trajectories, uncovering pronounced spatial variations of the diffusion coefficient. For instance, both experimental and numerical investigations have shown that the diffusion coefficient of proteins within the cell cytoplasm depends systematically on position \cite{Ku11,En11}. These results highlight the importance of accounting for space-dependent diffusion coefficient in realistic descriptions of transport.

Models with space-dependent diffusion coefficients have been used successfully in several contexts, such as Richardson diffusion in turbulence \cite{Ri26}, mesoscopic transport in disordered porous media \cite{De12}, and diffusion on random fractal structures \cite{Lo09}. The inclusion of spatial heterogeneity leads to rich and often anomalous dynamical behavior. Recent theoretical analyses have reported phenomena such as non-Gaussian propagators, anomalous scaling of the mean-squared displacement (MSD), and weak ergodicity breaking between time-averaged and ensemble-averaged MSDs \cite{Che13,Che14,WaDeCh19,ChMe20}. These observations have established heterogeneous diffusion as a minimal and versatile framework for studying deviations from standard ergodic diffusion.

However, most previous studies have characterized ergodicity breaking through ensemble- and time-averaged MSDs, while less attention has been devoted to the complementary perspective based on occupation times—the fraction of time a particle spends in a given spatial region. Occupation-time statistics provide an alternative probe of dynamical sampling and ergodic properties, particularly in systems where the diffusion coefficient varies spatially. They offer a direct measure of how the particle explores distinct regions of heterogeneous space and thus encode information inaccessible from MSD-based analysis alone.

In this work, we investigate the statistical properties and ergodic behavior of a Brownian particle diffusing in heterogeneous media, focusing on the distribution and moments of occupation times. Using the Feynman–Kac formalism, we derive exact and asymptotic results for two classes of functionals: the half occupation time (the time spent by the particle in the positive half-space) and the occupation time within a finite interval. We consider two representative spatial dependencies of the diffusion coefficient: (i) a piecewise-constant dependence, where the diffusion coefficient differs between the positive and negative halves of space, and (ii) a power-law dependence. 
From the Feynman–Kac equation, we derive the ergodicity-breaking parameter in terms of the first two moments of the occupation time, and examine its dependence on the form of the spatial heterogeneity. The results elucidate how variations in the diffusion coefficient give rise to nontrivial statistics of occupation times, thereby contributing to a more complete theoretical understanding of anomalous transport and ergodicity breaking in heterogeneous media.
While in this work we focus on these two explicit cases, the formalism is general, and can be applied to any other explicit expression for the diffusion coefficient.

Our results are organized as follows. In Sec. II, we show the Feynman-Kac equation for the characteristic function of a functional of a Brownian particle with a space-dependent diffusion coefficient. We also introduce the two explicit expressions for the diffusion coefficient analyzed in this paper. In Sec. III we illustrate how to analyze the ergodic properties from the two first moments of a functional. In Secs. IV and V we obtain the results for the occupation times in the half space and in an interval, respectively. 
%In Sec. IV we apply the infinite ergodic theory to our study. 
We conclude in Sec. VIII. 

\section{Generating function of a Brownian functional}
Consider the stochastic functional
\begin{equation}
    Z(t|x_0)=\int_{0}^{t}U[x(\tau)]d\tau
    \label{Z}
\end{equation}
where $U[x(\tau)]$ is a positive function of the stochastic trajectory $\{x(\tau); 0\leq \tau\leq t\}$ of a particle which is initially at $x_0$, this is, $x(t=0)=x_0$. Let $P(Z,t|x_0)$ be the probability density function (PDF) of the functional $Z(t)$ and $Q(p,t|x_0)$ its characteristic function or generating function, this is, 
\begin{eqnarray}
    Q(p,t|x_0)=\int_{0}^{\infty}e^{-pZ}P(Z,t|x_0)dZ=\left\langle e^{-pZ(t)}\right\rangle 
    =\left\langle e^{-p\int_{0}^{t}U[x(\tau)]d\tau}\right\rangle 
    \label{mgf0}
\end{eqnarray}
which is nothing but the Laplace transform with respect to $Z$. Expanding the exponential in power series
\begin{equation}
    Q(p,t|x_0)=\left\langle \sum_{n=0}^{\infty}\frac{(-p)^{n}}{n!}Z(t|x_0)^{n}\right\rangle =\sum_{n=0}^{\infty}\frac{(-p)^{n}}{n!}\left\langle Z(t|x_0)^{n}\right\rangle .
    \label{mgf}
\end{equation}
We also introduce another Laplace transform conjugate to time as
\begin{eqnarray}
  \tilde{Q}(p,s|x_0)=\mathcal{L}_{t\to s}[Q(p,t|x_0)]
  =\int_0 ^\infty dt~ e^{-st}Q(p,t|x_0).
  \label{Qps}
\end{eqnarray}
The moments of $Z (t|x_0)$ can be obtained from the derivatives of $\tilde{Q}(p,s|x_0)$  in a systematic manner. To do this, we first define the Laplace transform of the moments 
\begin{align}
    \left\langle Z(s|x_{0})^{n}\right\rangle &\equiv\mathcal{L}_{t\to s}[\left\langle Z(t|x_{0})^{n}\right\rangle ] \nonumber \\
   & = \int_0^\infty dt ~e^{-st} \int_{0}^{\infty}Z(t|x_0)^{n}P(Z,t|x_{0})dZ.
\end{align}
From \eqref{mgf0} it is straightforward to show that the moments are represented in terms of the generating function in the following way
\begin{align}
    \left\langle Z(s|x_0)^{n}\right\rangle =(-1)^{n} \frac{\partial^{n}\tilde{Q}(p,s|x_0)}{\partial p^{n}}\bigg|_{p=0}. 
    \label{moments}
\end{align}
Now we need to find the expression for the generating function $\tilde{Q}(p,s|x_0)$ which depends of the details of the underlying random walk. Let $x(t)$ be the position of a Brownian particle moving in a one-dimensional heterogeneous media. It is a stochastic process evolving according the overdamped Langevin equation with a space-dependent diffusion coefficient, namely, 
\begin{eqnarray}
        \frac{dx(t)}{dt}=\sqrt{2D(x)}\xi(t)
    \label{le2}
\end{eqnarray}
where $\xi(t)$ is a Gaussian noise with zero mean and autocorrelation $\left\langle \xi(t)\xi(t')\right\rangle =\delta(t-t')$. In the following we consider the Langevin equation in the Itô interpretation \cite{We79}. 

Such spatially dependent diffusion coefficients model many processes, where a partial list includes
random walks in an inhomogeneous medium \cite{LaBa01,PeJi16,ReSh16} ,
chemical reactions \cite{Ga04}, diffusion (in momentum space) in laser cooling processes \cite{BaBo01}, dissipative particle dynamics \cite{FaGr16}, vortex-antivortex annihilations \cite{Br00}, biophysics \cite{PiHe16,BeMa17,Ku11} e.g. measurements of proteins' diffusivity in mammalian cells \cite{Ku11}, and modeling of $1/f$ noise \cite{KaAl09}.

When the particle's position evolves according to \eqref{le2}, the generating function  $\tilde{Q}(p,s|x_0)$ of the functional \eqref{Z}  obeys the so called backward Feynman-Kac equation \cite{De18}
\begin{eqnarray}
\frac{\partial Q(p,t|x_{0})}{\partial t}=D(x_{0})\frac{\partial^{2}Q(p,t|x_{0})}{\partial x_{0}^{2}}-pU(x_{0})Q(p,t|x_{0}).
    \label{FK}
\end{eqnarray}
For each explicit expression of $U(x)$, we have to solve Eq. \eqref{FK} for the generating function using suitable boundary conditions, and then use Eq. \eqref{moments} to derive the moments. To do this we need also to specify the explicit expressions for $D(x)$. First
we assume the power law dependence \cite{Sr07,Che13,LeBa19,Radice23,Si22,WaDeCh19}
\begin{eqnarray}
    D(x)=D_0|x|^\alpha,\quad \alpha<2.
    \label{dpl}
\end{eqnarray}

The corresponding Fokker-Planck equation for $P(x,t)$ in the Itô interpretation is
$$
\frac{\partial P(x,t)}{\partial t}=D_{0}\frac{\partial^{2}}{\partial x^{2}}\left[|x|^{\alpha}P(x,t)\right]
$$
whose solution is (see Appendix C in Ref. \cite{WaDeCh19})
\begin{eqnarray}
    P(x,t)=\frac{N}{|x|^\alpha t^{\frac{1-\alpha}{2-\alpha}}}e^{-\frac{|x|^{2-\alpha}}{4D_{0}(2-\alpha)t}}
    \label{pdfdx}
\end{eqnarray}
where 
$$N=\frac{(2-\alpha)^{\frac{\alpha}{2-\alpha}}}{2D_{0}^{\frac{1-\alpha}{2-\alpha}}\Gamma\left(\frac{1-\alpha}{2-\alpha}\right)}.$$
The mean square displacement (MSD) becomes
\begin{eqnarray}
\left\langle x^{2}(t)\right\rangle =\int_{-\infty}^{\infty}x^{2}P(x,t)dx=\frac{2N}{2-\alpha}\Gamma\left(\frac{3-\alpha}{2-\alpha}\right)\left[4D_{0}(2-\alpha)\right]^{\frac{3-\alpha}{2-\alpha}}t^{\frac{2}{2-\alpha}}.
    \label{msd}
\end{eqnarray}

Thus, for $\alpha<0$ we find subdiffusion, and superdiffusion for $\alpha>0$. Brownian motion with
normal diffusion emerges for $\alpha=0$, and $\alpha=1$ produces ballistic motion.  The diffusion becomes subdiffusive if $\alpha<0$ and superdiffusive for $0<\alpha<1$. In simulations, we need to regularize the diffusion coefficient by considering 

\begin{eqnarray*}
    D(x)=
    \left\{ \begin{array}{cc}
        D_0(|x|^\alpha+\epsilon), & \alpha>0 \\
        \frac{D_0}{|x|^{-\alpha}+\epsilon}, & \alpha<0 
    \end{array}\right.
\end{eqnarray*}

with $\epsilon\ll 1$ to prevent the particles from trapping or divergences at $x = 0$ \cite{ChMe20,Me14}.

Another interesting choice for the spatial dependence of the diffusion coefficient is the piecewise constant spatial form \cite{BrLa17,Pa24,vM05}
\begin{eqnarray}
    D(x)=\left\{ \begin{array}{cc}
D_{-}, & x<0\\
D_{+}, & x>0
\end{array}\right. 
\label{pwd}
\end{eqnarray}
with $D_\pm>0$. This form of position-dependent diffusion coefficient
was previously used to find the solution of the Langevin equation in an open system, without a trapping potential \cite{Pa24}. The MSD in this case is $\left\langle x^{2}(t)\right\rangle =2\sqrt{D_+D_-}t$ so that it is diffusive \cite{Pa24}.

Below, we illustrate how to compute the generating function and the two first moments for specific functionals. Let us first show how to analyze ergodic properties in terms of the two first moment of a functional.

\section{Ergodicity Breaking}
Let us consider a stochastic trajectory $x(\tau)$  observed from $\tau=0$ up to time $\tau = t$. Consider an observable $\mathcal{O}[x(\tau)]$, a function of the trajectory $x(\tau)$. Since $x(\tau)$ is stochastic in nature, the observable $\mathcal{O}[x(\tau)]$ will also be fluctuating between the realizations. An observable of the random walk is said to be ergodic if the ensemble average equals the time average $\left\langle \mathcal{O}\right\rangle =\ensuremath{\overline{\mathcal{O}}}$ in the long time limit. This means that if $\mathcal{O}[x(\tau)]$ is ergodic then its time average $\ensuremath{\overline{\mathcal{O}}}$ is not a random variable. As a consequence, the limiting PDF of $\ensuremath{\overline{\mathcal{O}}}$ is \begin{eqnarray}
P(\overline{\mathcal{O}},t\to\infty)=\delta\left(\overline{\mathcal{O}}-\left\langle \mathcal{\overline{O}}\right\rangle \right).
     \label{lpdf}
 \end{eqnarray} 
 At this point, let us define the density $P(x,t)$ which is the probability to find the particle at the point $x$ at time $t$, i.e., it is the propagator.
 If the observable is integrable with respect to the density $P(x,t)$, then the ensemble average is
\begin{equation}
    \left\langle \mathcal{O}[x(t)]\right\rangle =\int_{-\infty}^{\infty}\mathcal{O}[x]P(x,t)dx.
\end{equation}
The time average of $\mathcal{O}[x(t)]$ is defined as
\begin{equation}
    \ensuremath{\overline{\mathcal{O}[x(t)]}=}\frac{1}{t}\int_{0}^{t}\mathcal{O}[x(\tau)]d\tau.
    \label{tav}
\end{equation}
For non-ergodic observables, since $\overline{\mathcal{O}}$ is random, its variance $\textrm{Var}(\overline{\mathcal{O}})$ is non-zero in the long time limit. Otherwise, for an ergodic observable $\textrm{Var}(\overline{\mathcal{O}})=0$ in the long time limit. Keeping this in mind, one can define the ergodicity breaking parameter EB
in the following way
\begin{eqnarray}
\textrm{EB}=\lim_{t\to\infty}\frac{\textrm{Var}(\overline{\mathcal{O}})}{\left\langle \overline{\mathcal{O}}\right\rangle ^{2}}=\lim_{t\to\infty}\frac{\left\langle \overline{\mathcal{O}}^{2}\right\rangle -\left\langle \overline{\mathcal{O}}\right\rangle ^{2}}{\left\langle \overline{\mathcal{O}}\right\rangle ^{2}}.
  \label{EB}
\end{eqnarray}
For ergodic observables, one should have $\textrm{EB}= 0$. 

In the examples below we consider the observable $\mathcal{O}[x(t)]=U[x(t)]$ so that the time average of the observable is from \eqref{tav}
\begin{eqnarray}
    \overline{\mathcal{O}[x(t)]}=\frac{1}{t}\int_{0}^{t}U[x(\tau)]d\tau=\frac{Z(t)}{t}.
    \label{o}
\end{eqnarray}
and so
\begin{eqnarray}
   \left\langle \overline{\mathcal{O}}\right\rangle =\frac{\left\langle Z(t)\right\rangle }{t},\quad\left\langle \overline{\mathcal{O}}^{2}\right\rangle =\frac{\left\langle Z(t)^{2}\right\rangle }{t^{2}}. 
   \label{oo}
\end{eqnarray}
Finally, from \eqref{EB} we can find the ergodicity breaking parameter in terms of the two first moments of the functional
\begin{eqnarray}
    \textrm{EB}=\frac{\left\langle Z(t)^{2}\right\rangle }{\left\langle Z(t)\right\rangle ^{2}}-1
    \label{EB2}
\end{eqnarray}
as $t\to \infty$.

Another quantity of interest which characterizes ergodicity is the PDF of the time averaged observable $ \overline{\mathcal{O}[x(t)]}$ around
its mean for long times, so we define the dimensionless random variable 
\begin{eqnarray}
    \eta =\lim_{t\to \infty} \frac{ \overline{\mathcal{O}}}{\left\langle \overline{\mathcal{O}}\right\rangle}
    \label{etag}
\end{eqnarray}
and from \eqref{o} and \eqref{oo}, the relative time averaged observable $U[x(t)]$ is defined by 
\begin{eqnarray}
\eta =\lim_{t\to \infty}\frac{Z(t)}{\left\langle Z(t)\right\rangle }.
    \label{eta}
\end{eqnarray}
Once $P(Z,t|0)$ is found by solving the Feynman-Kac equation \eqref{FK}, the PDF of $\eta$ follows from
\begin{eqnarray}
   P(\eta)=P(Z=\eta \left\langle Z(t)\right\rangle,t)\left\langle Z(t)\right\rangle.
   \label{peta}
\end{eqnarray}
It is interesting to note that the variance of $\eta$ is nothing but the ergodicity breaking parameter:
$$
\textrm{Var}(\eta)=\left\langle \eta^{2}\right\rangle -\left\langle \eta\right\rangle ^{2}=\left\langle \eta^{2}\right\rangle -1=\frac{\left\langle Z(t)^{2}\right\rangle }{\left\langle Z(t)\right\rangle ^{2}}-1=\textrm{EB}.
$$
Therefore, for ergodic observables (EB = 0) one has  
\begin{eqnarray}
 P(\eta)=\delta (\eta -1),
 \label{deltaeta}
\end{eqnarray}
in agreement with \eqref{lpdf}.

%\section{The First passage time}
%The time $t_f$ when a particle starting at $x_0 = 0$ first hits $x = x_t$ (with $x_t>0$) is called the
%first passage time, and is a quantity subject to many studies in physics and
%other fields \cite{Redner}. In the ecological context, the target (food) is located at $x=x_t$ and the first passage time to the target position determines the minimum time required to find the target, which is a measure of the search efficiency. The distribution of first passage times for a Brownian particle in the presence of spatial heterogeneity
%can be obtained from Feynman-Kac equation \eqref{FK} using an identity
%due to Kac \cite{Kac}.\\

%The survival probability, i.e., the probability that the target is not yet detected up to time $t$ is the probability that the first passage time $t_f$ is higher than $t$ which is equivalent to say that the particle position has been always below the threshold $x=x_t$, i.e., the particle has never crossed the target position 
%\begin{equation}
%    \textrm{Prob}\{t_{f}>t\}=\textrm{Prob}\{\max_{0\leq\tau\leq t}x(\tau)<x_t\}.
    \label{fp1}
%\end{equation}
%The Kac's identity \cite{Kac} establishes that
%\begin{equation}
%   \textrm{Prob}\{t_{f}>t\} =\lim_{p\to\infty}Q(p,t|x_{0})
%   \label{fp2}
%\end{equation}
%where $Q(p,t|x_{0})$ evolves according to the Feynman-Kac equation \eqref{FK} with $U(x_0)=\theta (x_0-x_t)$ ($\theta(\cdot)$ is the Heaviside step function) and the initial condition is $Q(p,t=0|x_{0})=1$. 
%To find $Q(p,t|x_{0})$ we perform the Laplace transform transform of Eq. \eqref{FK} with respect to time defined as
%$$
%\tilde{Q}(p,s|x_{0})=\mathcal{L}_s\left[Q(p,t|x_0)\right]=\int_0^{\infty} e^{-st}Q(p,t|x_{0})dt.
%$$
%Thus, from Eq. \eqref{FK}
%\begin{eqnarray}
%    s\tilde{Q}(p,s|x_{0})-1=D(x_{0})\frac{d^2\tilde{Q}(p,s|x_{0})}{dx_{0}^{2}}-p\theta(x_{0}-x_{t})\tilde{Q}(p,s|x_{0})
%    \label{ode}
%\end{eqnarray}
%due to the presence of discontinuous functions in the above differential equation we need to solve it in the regions $x_0<x_t$ and $x_0>x_t$ separately and impose the  matching conditions for $\tilde{Q}(p,s|x_{0})$ and its derivative. 
%Since equation \eqref{fp2} provides the cumulative probability for the first-passage time $t$, its PDF is 
%\begin{eqnarray}
%    P_f(t|x_0)=-\frac{\partial}{\partial t}\textrm{Prob}\{t_{f}>t\}
%    \label{forP}
%\end{eqnarray}
%and the mean first passage time $\left\langle t\right\rangle $ can be expressed in the form
%\begin{eqnarray}
%   \left\langle t\right\rangle =\int_{0}^{\infty}tP_f(t|x_{0})dt=-\left(\frac{\partial \Tilde{P_f}(s|x_{0})}{\partial s}\right)_{s=0}. 
   \label{mfpt0}

\section{Half occupation time}
We consider the occupation time $T^+ (t|x_0)$ of the particle above the origin (i.e., on the positive half-space) within a time window of size $t$ if it was at $x=x_{0}$ at $t=0$. It is defined as
$$
\left\langle T^{+}(t)\right\rangle =\int_{0}^{t}\theta\left[x(\tau)\right]d\tau 
$$
so that $U(x_0)=\theta (x_0)$ in this case. The characteristic function of $T^+$ has to be solved under the appropriate boundary conditions \cite{Ma05}. If the starting point $x_{0}\rightarrow+\infty$ the particle will stay on the positive side with $x(t)>0$ for all finite $t$ implying $T^{+}(t|x_{0}\rightarrow+\infty)=t$ and accordingly $P(T^+,t|x_{0}\rightarrow+\infty)=\delta(T^+-t),$ i.e., $\tilde{Q}(p,s|x_{0}\rightarrow+\infty)=1/(s+p)$. On the other hand, if the starting point $x_{0}\rightarrow-\infty$ the particle will never reach the positive side and it will stay on the negative half-space implying $T^{+}(t|x_{0}\rightarrow-\infty)=0$ and hence $P(T^+,t|x_{0}\rightarrow-\infty)=\delta(T^+)$, so that $\tilde{Q}(p,s|x_{0}\rightarrow-\infty)=1/s$.
Performing the Laplace transform of \eqref{FK} with respect to $t$, the Feynman-Kac equation for the characteristic function $\tilde{Q}(p,s|x_{0})$ reads
\begin{eqnarray}
    s\tilde{Q}(p,s|x_{0})-1=D(x_{0})\frac{d\tilde{Q}(p,s|x_{0})}{dx_{0}^{2}}-p\theta(x_{0})\tilde{Q}(p,s|x_{0})
    \label{ode1}
\end{eqnarray}
which has to be solved according to the boundary conditions
\begin{eqnarray}
    \tilde{Q}(p,s|x_{0}\rightarrow+\infty)=\frac{1}{s+p},\quad\tilde{Q}(p,s|x_{0}\rightarrow-\infty)=\frac{1}{s}.
    \label{BC1}
\end{eqnarray}
To this end we need to specify the expression of $D(x)$. Once $\tilde{Q}(p,s|x_{0})$ is known, one can find the PDF of $y=T^+/t$ in the long time limit by using the method in Ref. \cite{GoLu01}.

\subsection{Piecewise heterogeneity}
\begin{figure}[htbp]
    \centering
    \includegraphics[width=0.33\linewidth]{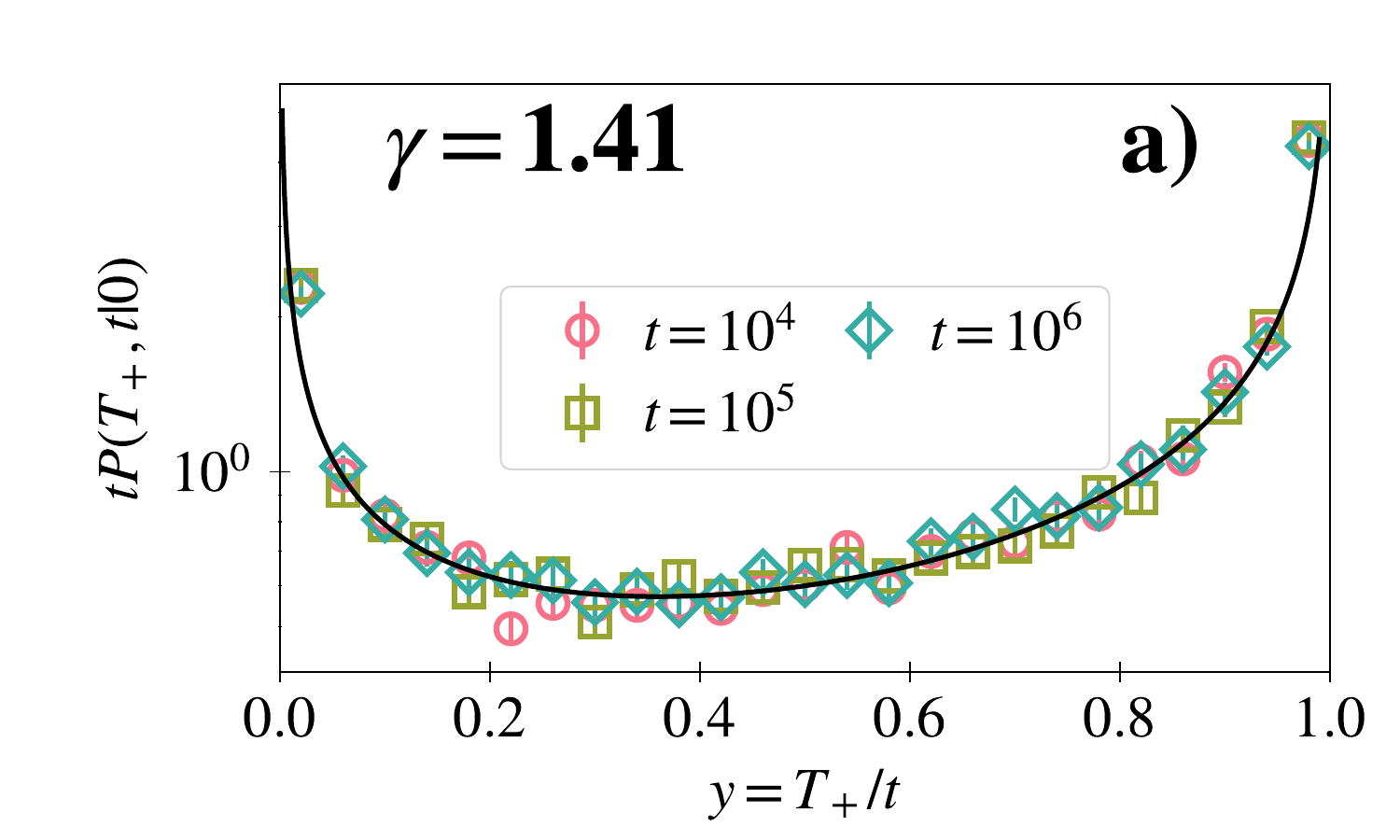}%
    \includegraphics[width=0.33\linewidth]{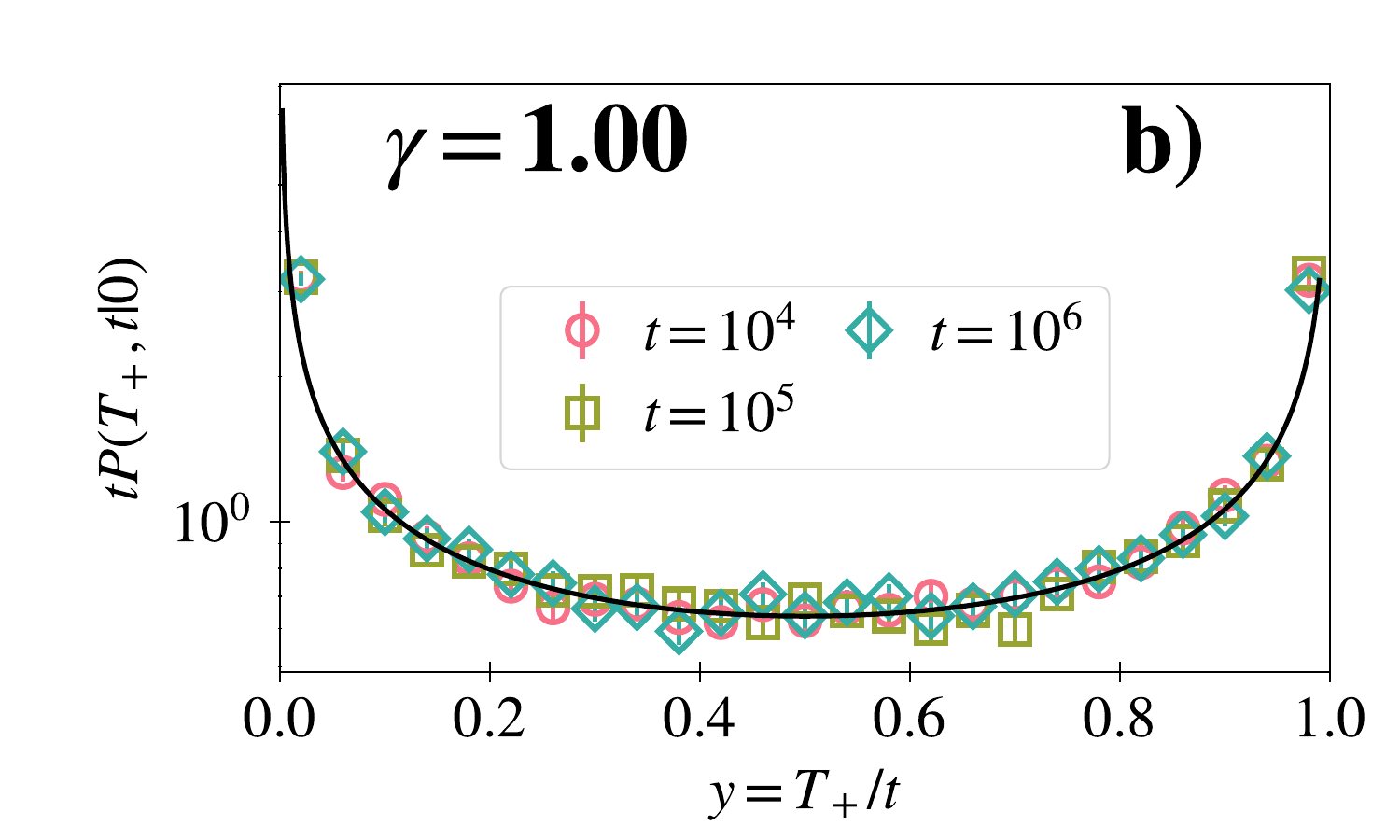}%
    \includegraphics[width=0.33\linewidth]{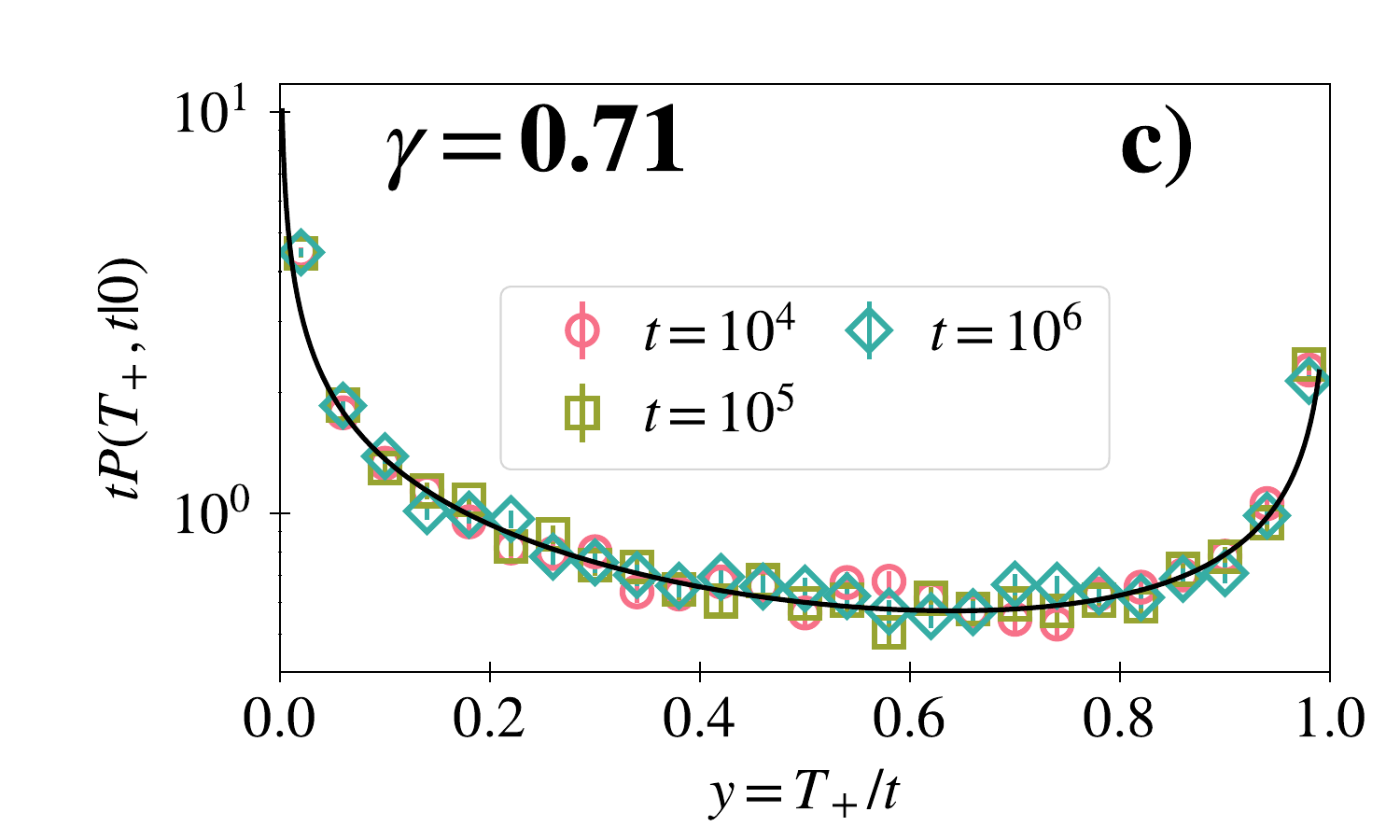}
    \caption{Limiting distribution of the half-occupation time for the piecewise heterogeneity. The different symbols are for trajectories with $t=10^4,10^5,$ and $10^6$. The black solid line corresponds to Eq. \eqref{ld1}. $D_{+}=0.5,1.0$, and $2.0$ in panels (a), (b), and (c), respectively. In all panels $D_{-}=1.0$, $x_0=0$, simulation timestep $dt=0.1$, and $N=10^4$ trajectories.}
    \label{fig:Limitting_pw_Tpl}
\end{figure}

For the piecewise diffusion coefficient \eqref{pwd} we solve \eqref{ode1} in the regions $x_0<0$ and $x_0>0$ separately taking into account \eqref{BC1}. We find
\begin{eqnarray}
    \tilde{Q}_{-}(p,s|x_{0})&=&\frac{1}{s}+A_{1}e^{x_{0}\sqrt{\frac{s}{D_{-}}}},\;x_{0}<0\nonumber\\
    \tilde{Q}_{+}(p,s|x_{0})&=&\frac{1}{s+p}+B_{2}e^{-x_{0}\sqrt{\frac{s+p}{D_{+}}}},\;x_{0}>0.
    \label{solmm}
\end{eqnarray}
The constants $A_1$ and $B_2$ can be found by requiring continuity of $\tilde{Q}$ and its derivative $\partial _{x_0}\tilde{Q}$ across $x_0=0$. This implies
\begin{eqnarray}
    \tilde{Q}_{-}(p,s|x_{0}\to0)&=&\tilde{Q}_{+}(p,s|x_{0}\to0)\nonumber\\
    \left[\frac{\partial\tilde{Q}_{-}(p,s|x_{0})}{\partial x_{0}}\right]_{x_{0}=0}&=&\left[\frac{\partial\tilde{Q}_{+}(p,s|x_{0})}{\partial x_{0}}\right]_{x_{0}=0}.
    \label{bc2}
\end{eqnarray}
Then introducing \eqref{solmm} into \eqref{bc2} and solving the resulting system of equations we obtain $A_1$ and $B_2$ and setting $x_0=0$ for simplicity into \eqref{solmm} we finally get
\begin{eqnarray}
    \tilde{Q}(p,s|0)=\frac{1}{s+p}\left(1+\frac{p}{s+\gamma\sqrt{s(s+p)}}\right)
    \label{QTmas}
\end{eqnarray}
where
$$
\gamma=\sqrt{\frac{D_{-}}{D_{+}}}
$$
can be regarded as the asymmetry parameter.
To invert the characteristic function \eqref{QTmas} back to $T^+$ and $t$ we note that it has the following scaling behavior 
$$
\tilde{Q}(p,s|0)=\frac{1}{s}g\left(\frac{p}{s}\right).
$$
Then the random variable $y=T^+/t$ possesses the distribution 
$$
P(y|0)=P\left(y=\frac{T^+}{t},t|0\right)=tP\left(T^+=yt,t|0\right)
$$
which can be computed as was done in \cite{GoLu01} by noting
\begin{eqnarray}
P(y|0)=-\frac{1}{\pi y}\lim_{\epsilon\rightarrow0}\textrm{Im}\left[ g\left(-\frac{1}{y+i\epsilon}\right)\right].
\label{ld}
\end{eqnarray}
Thus, from \eqref{QTmas} 
$$
g(\chi)=\frac{1}{1+\chi}\left(1+\frac{\chi}{1+\gamma\sqrt{1+\chi}}\right)
$$
where $\chi =p/s$. Setting $\chi ^{-1}=-y-i\epsilon$ and computing \eqref{ld} we find the \textit{asymmetric arcsine} PDF
\begin{eqnarray}
    P(T^+,t|0)= \frac{1}{t}F\left( \frac{T^+}{t}\right)
\end{eqnarray}
where
\begin{eqnarray*}
    F(y)= \frac{\gamma}{\pi\sqrt{y(1-y)}\left[y+\gamma^{2}(1-y)\right]}.
\end{eqnarray*}
Thus, from \eqref{ld},
\begin{eqnarray}
    P(y|0)=\frac{\gamma}{\pi\sqrt{y(1-y)}\left[y+\gamma^{2}(1-y)\right]},\quad 0<y<1.
    \label{ld1}
\end{eqnarray}
The same limiting distribution has been obtained for a random walk in one dimension under an asymmetric force field acting on a finite region \cite{Ba06}. Although there is no a general direct correspondence, in a way, the confining effect of the asymmetric force field is physically analogous to a heterogeneous media with constant asymmetric diffusion coefficient. In Figure \ref{fig:Limitting_pw_Tpl} we test the result \eqref{ld1} against numerical simulations. We see that for sufficiently large values of $t$,  $tP(T^+,t|0)$ converges to the limiting distribution $F(y)$. Unlike for the arcsine case, the minium of $F(y)$ is not located at $y=1/2$ as can be also observed in figure \ref{fig:Limitting_pw_Tpl}. It is located at
$$
y_{min}=\frac{5\gamma^{2}-3-\sqrt{9\gamma^{4}-14\gamma^{2}+9}}{8(\gamma^{2}-1)}.
$$
From this expression and also from Figure \ref{fig:Limitting_pw_Tpl} we see that if $\gamma>1$ then $y_{min}$ moves to the left while if $\gamma<1$ it moves to the right. For $\gamma=1$, $y_{min}=1/2$ as expected. 

The two first moments of $P(T^+,t|0)$ can be found from \eqref{moments} with \eqref{QTmas} or from \eqref{ld1}. In any case we find 
\begin{eqnarray}
    \left\langle T^{+}(t)\right\rangle =\frac{\gamma}{\gamma+1}t,\quad     \left\langle T^{+}(t)^{2}\right\rangle =\frac{\gamma(1+2\gamma)}{2(\gamma+1)^{2}}t^{2}.
    \label{t1mas}
\end{eqnarray}
Note first that $ \left\langle T^{+}(t)\right\rangle >t/2$ (the time spent at $x>0$ is in average higher than half of the total time) if $\gamma >1$, this is, if $D_->D_+$. This is expected since in this case the motion of the particle is slower in the region $x>0$ than in the region $x<0$. Note also that the coefficient $\gamma/(1+\gamma)$ in \eqref{t1mas} corresponds to the probability that the particle is at $x>0$.

If $D_-=D_+$ one has $\gamma =1$ and the two moments \eqref{t1mas}  reduce to the values corresponding to the BM in a homogeneous media. In addition, when $\gamma =1$ Eq. \eqref{ld1} turns into the \textit{arcsine law} \cite{Le40}. Now, we are in position to compute the ergodicity breaking parameter. From \eqref{EB2} and \eqref{t1mas} we finally find
\begin{eqnarray}
\textrm{EB}_+=\frac{\left\langle T^{+}(t)^{2}\right\rangle }{\left\langle T^{+}(t)\right\rangle ^{2}}-1=\frac{1}{2\gamma},
\label{EBmas2}
\end{eqnarray}
which predicts a non-ergodic behavior regardless of the values of $\gamma$.
\begin{figure}[htbp]
    \centering
    \includegraphics[width=0.33\linewidth]{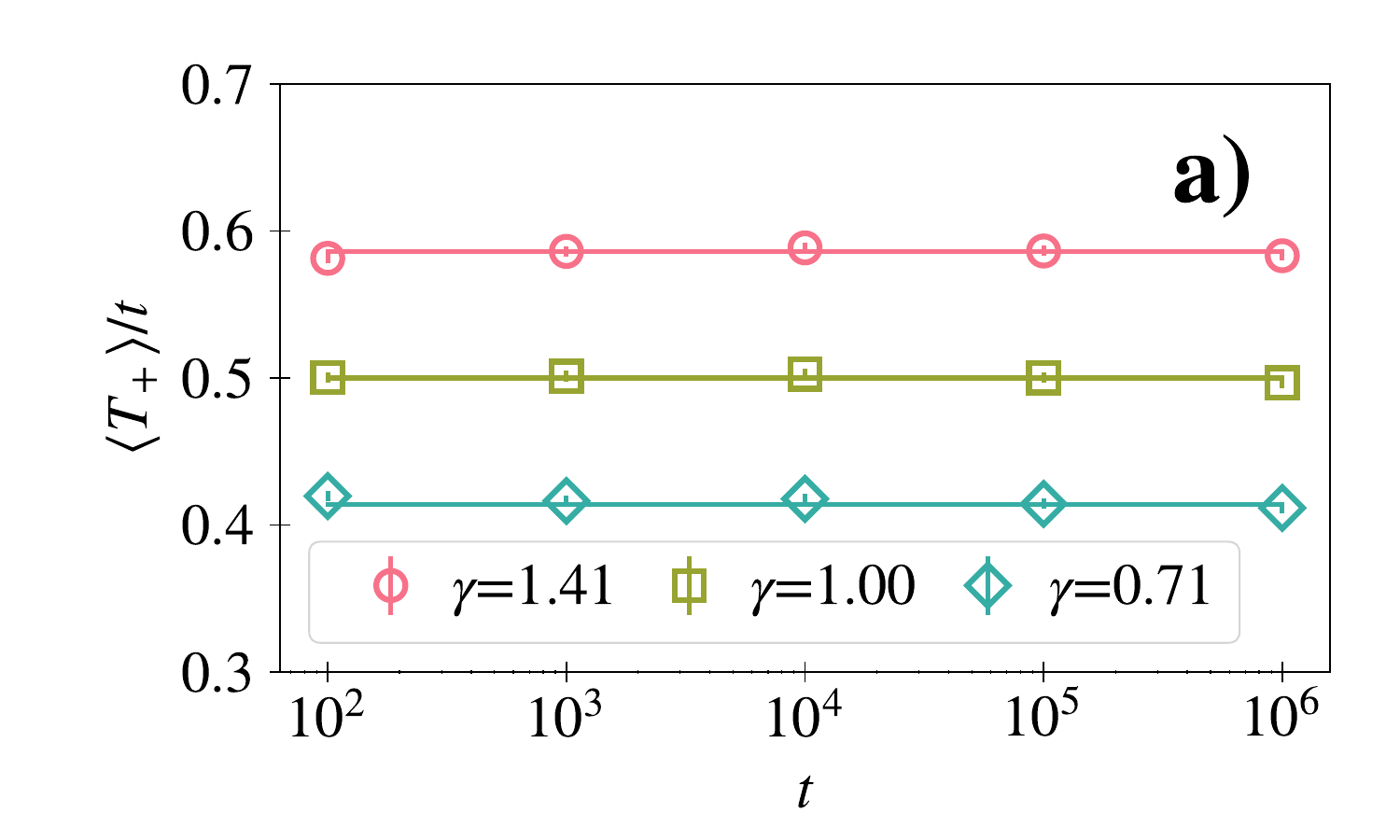}%
    \includegraphics[width=0.33\linewidth]{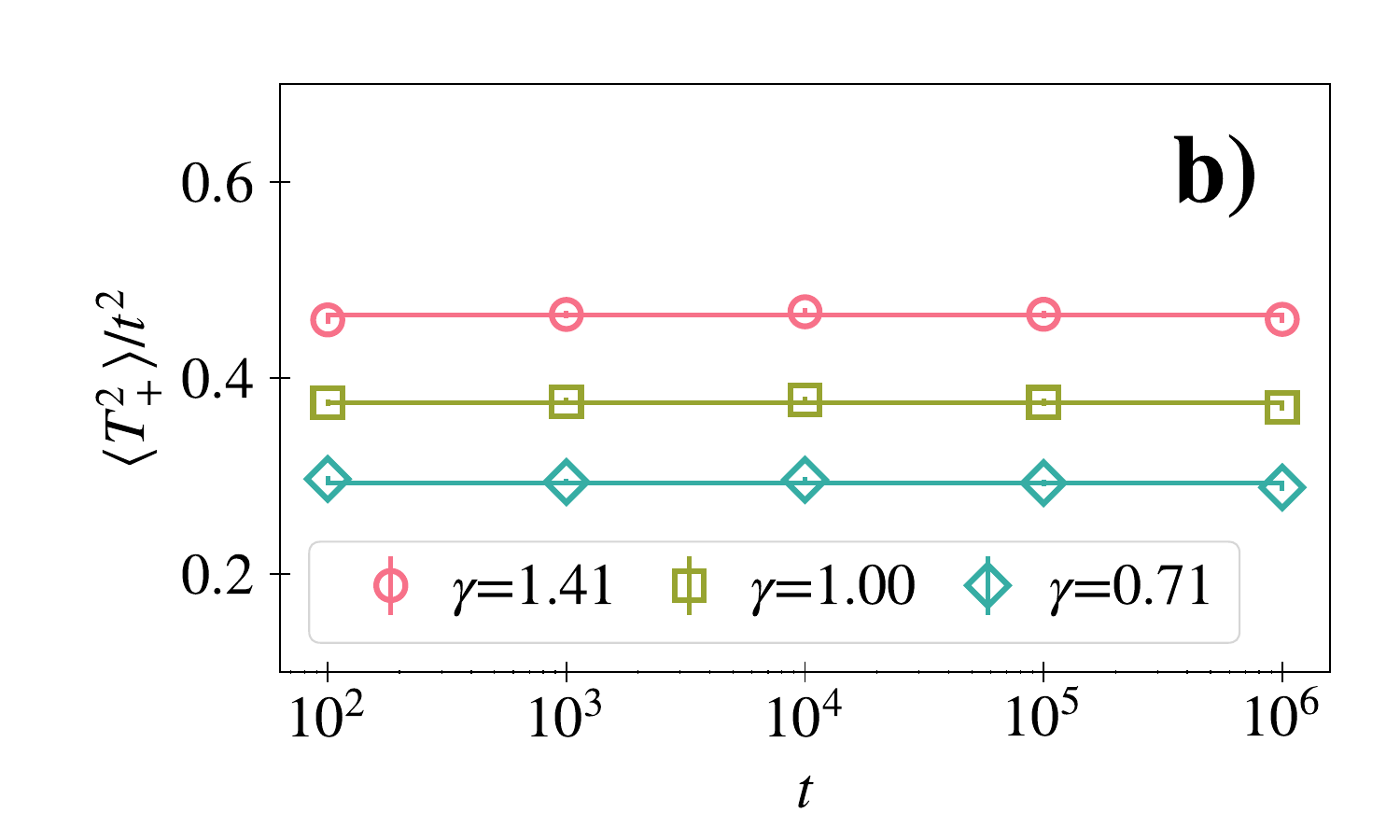}%
    \includegraphics[width=0.33\linewidth]{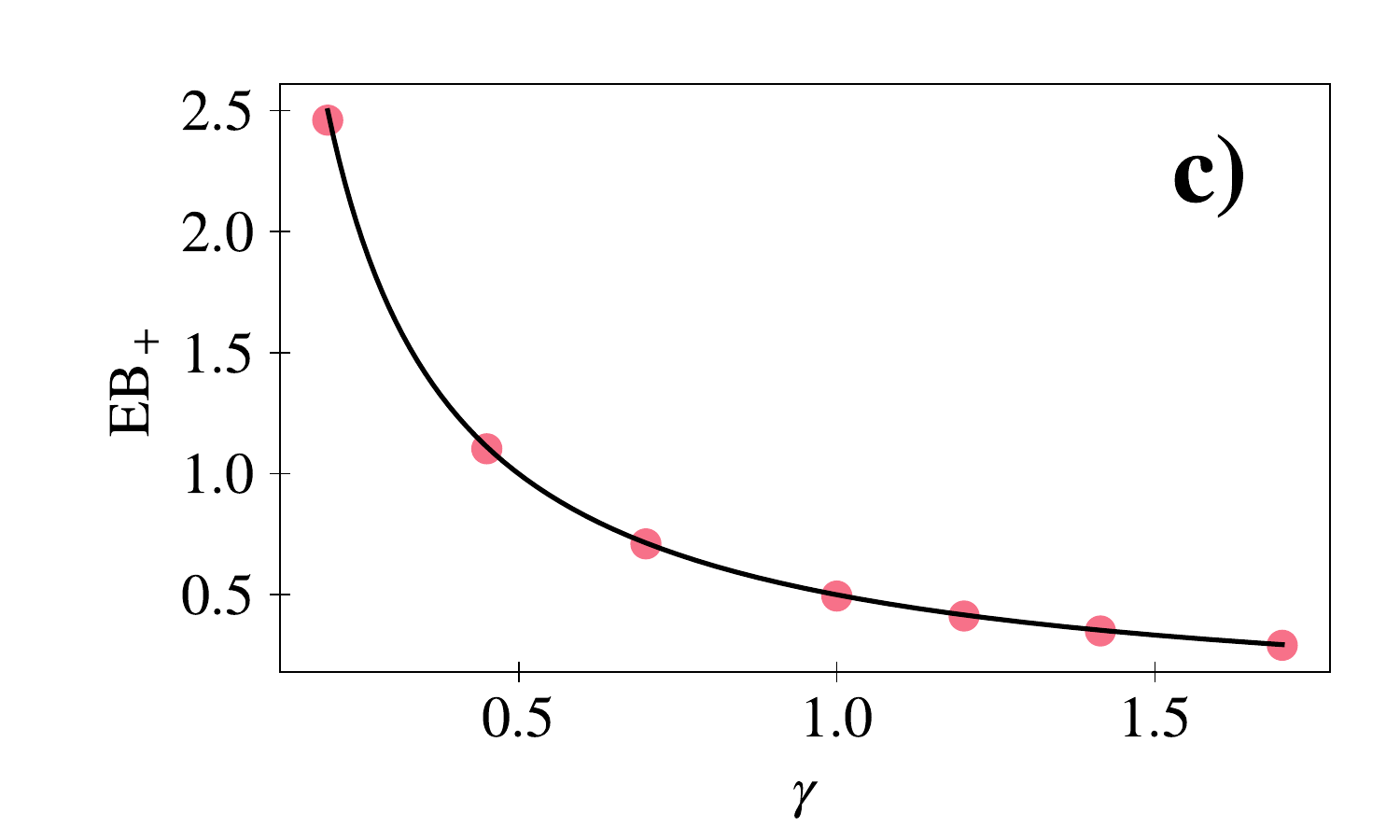}
    \caption{(a) $\langle T^{+}(t)\rangle$, (b) $\langle T^{+}(t)^2\rangle$, and (c) $\textrm{EB}_+$ of the half-occupation time for the piecewise heterogeneity. In panels (a), and (b) the solid lines are computed with Eq. \eqref{t1mas}, $D_{+}=0.5,1.0$, and $2.0$; In panel (c) the solid line is Eq. \eqref{EBmas2}. In all panels $D_{-}=1.0$, $x_0=0$, simulation timestep $dt=0.1$, and $N=10^4$ trajectories.}
    \label{fig:Pw-Tp}
\end{figure}

In Figure \ref{fig:Pw-Tp} we compare \eqref{t1mas} and \eqref{EBmas2} with numerical simulations. The greater $D_-$ is in comparison to $D_+$, the greater $\gamma$ is, therefore the probability that the particle is at $x>0$ increases, and thus the time spent there also is greater. This is captured by EB$_+$ which decreasses with $\gamma$ as expected.

\begin{figure}[htbp]
    \centering
    \includegraphics[width=0.33\linewidth]{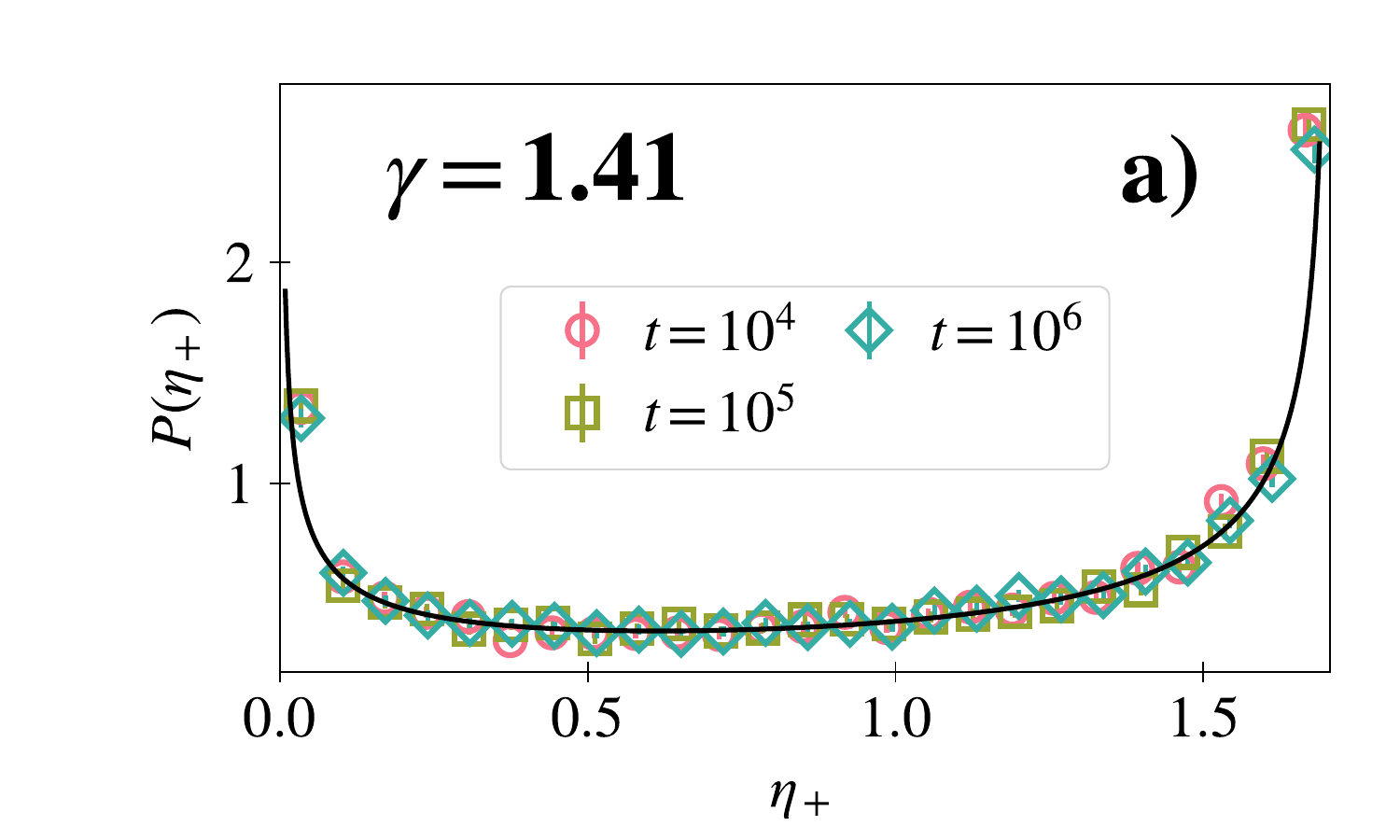}%
    \includegraphics[width=0.33\linewidth]{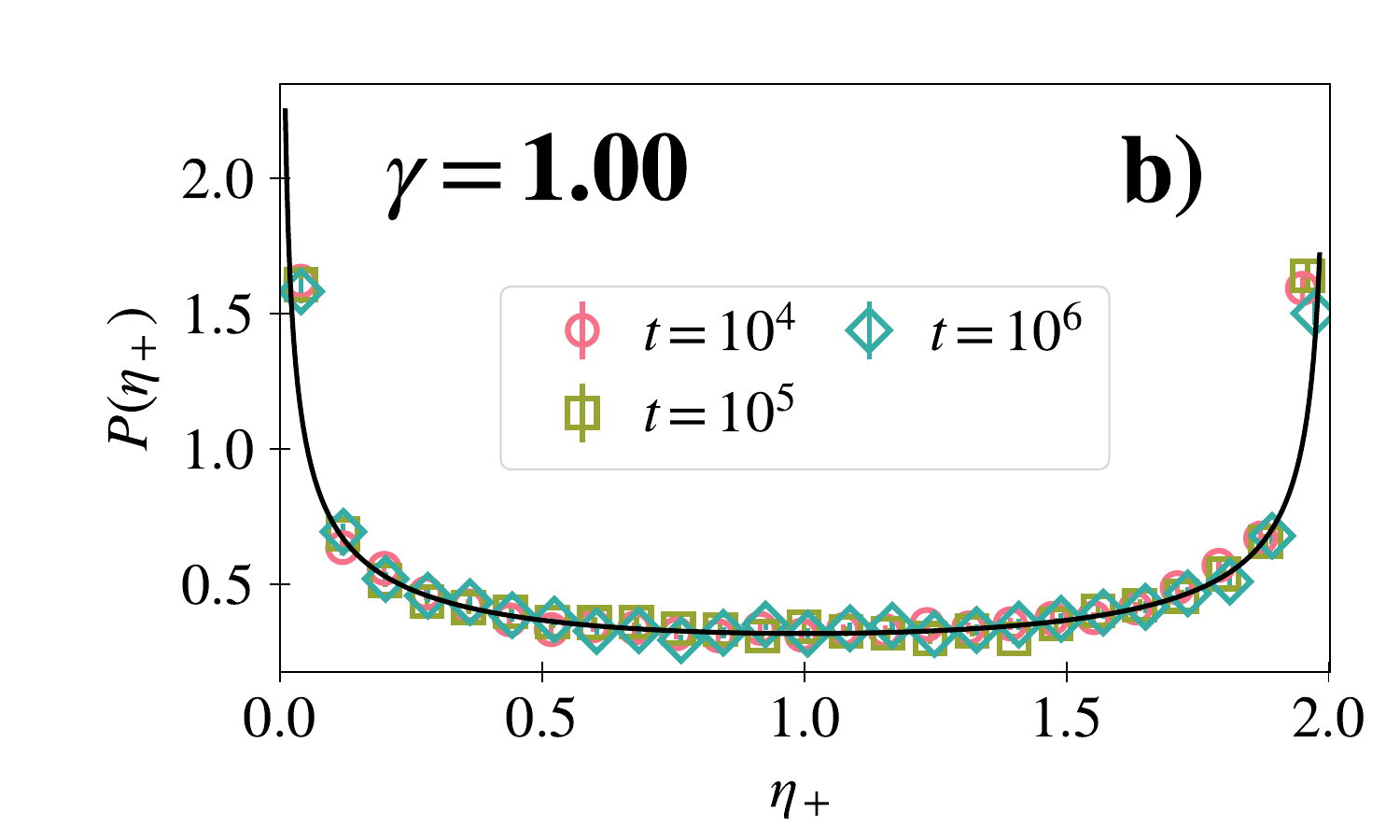}%
    \includegraphics[width=0.33\linewidth]{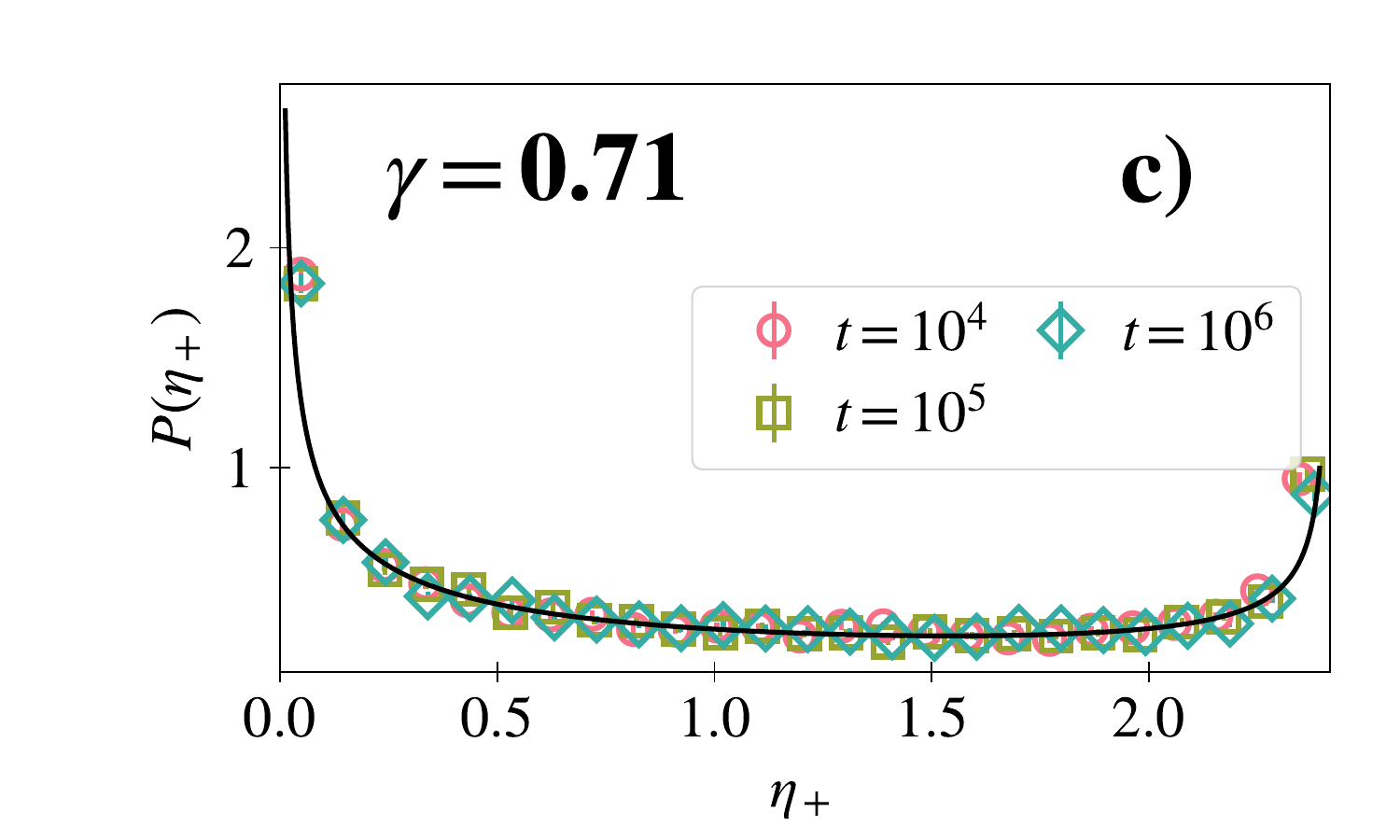}
    \caption{PDF of the time averaged half-occupation time for the piecewise heterogeneity. The different symbols are for trajectories with $t=10^4,10^5,$ and $10^6$. $D_{+}=0.5,1.0$, and $2.0$ in panels (a), (b), and (c), respectively. The solid line is computed wiht Eq. \eqref{Peta1} In all panels $D_{-}=1.0$, $x_0=0$, simulation timestep $dt=0.1$, and $N=10^4$ trajectories.}
    \label{fig:Timeaverage_pw_Tpl}
\end{figure}

Finally, we compute the PDF of the time average of the observable $\theta[x(t)]$, this is, the PDF of the time averaged occupation fraction
\begin{eqnarray}
  \eta_{+}=\lim_{t\to\infty}\frac{T^{+}}{\left\langle T^{+}(t)\right\rangle }
  \label{etamasd}
\end{eqnarray}
as defined in Eq. \eqref{eta}. It can be obtained from the limiting distribution $P(y|0)$ by simply accounting for the relation between the variables $y$ and $\eta_+$. We find from \eqref{ld1} the density
\begin{eqnarray}
    P(\eta_{+})&=&\frac{\gamma}{\gamma+1}P\left(y=\frac{\gamma}{\gamma+1}\eta_{+}\right)\nonumber\\
    &=&\frac{\gamma}{\pi\left(\gamma+\eta_{+}-\gamma\eta_{+}\right)\sqrt{\gamma\eta_{+}\left(1+\gamma-\gamma\eta_{+}\right)}}
    \label{Peta1}
\end{eqnarray}
where $0<\eta_+<(\gamma+1)/\gamma$. In Figure \ref{fig:Timeaverage_pw_Tpl} we compare Eq. \eqref{Peta1} with numeric simulations.

\subsection{Power law heterogeneity}
For the power law diffusion coefficient \eqref{dpl} we solve \eqref{ode1} in the regions $x_0<0$ and $x_0>0$ separately. The solution for the characteristic function is
\begin{eqnarray}
    \tilde{Q}(p,s|0)=\frac{1}{s+p}\left[1+\frac{p/s}{1+\left(\frac{s+p}{s}\right)^{\frac{1}{2-\alpha}}}\right],\quad\alpha<1
\label{Qsptmas}
\end{eqnarray}
for $x_0=0$ (see Appendix A for details of the derivation).

\begin{figure}[htbp]
    \centering
    \includegraphics[width=0.33\linewidth]{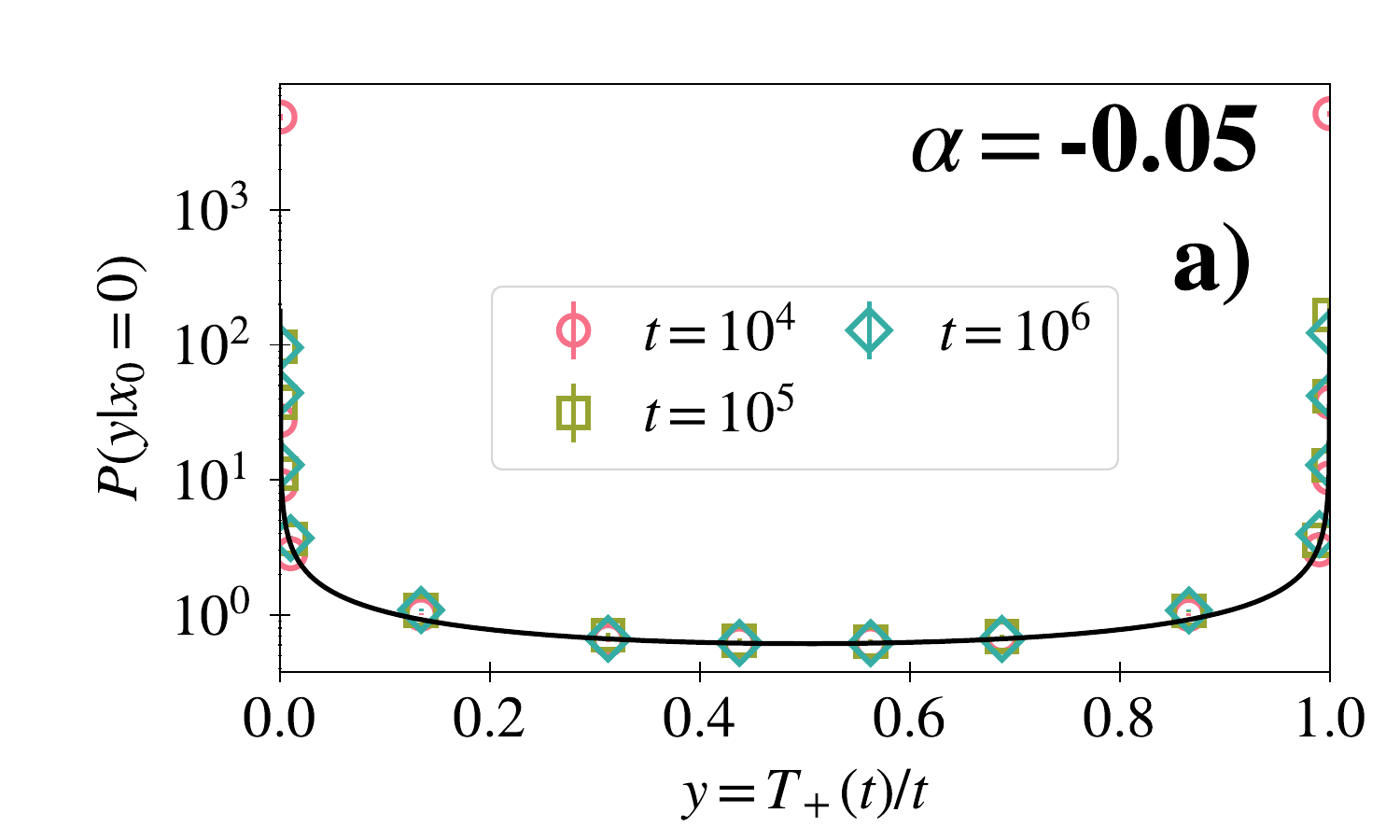}%
    \includegraphics[width=0.33\linewidth]{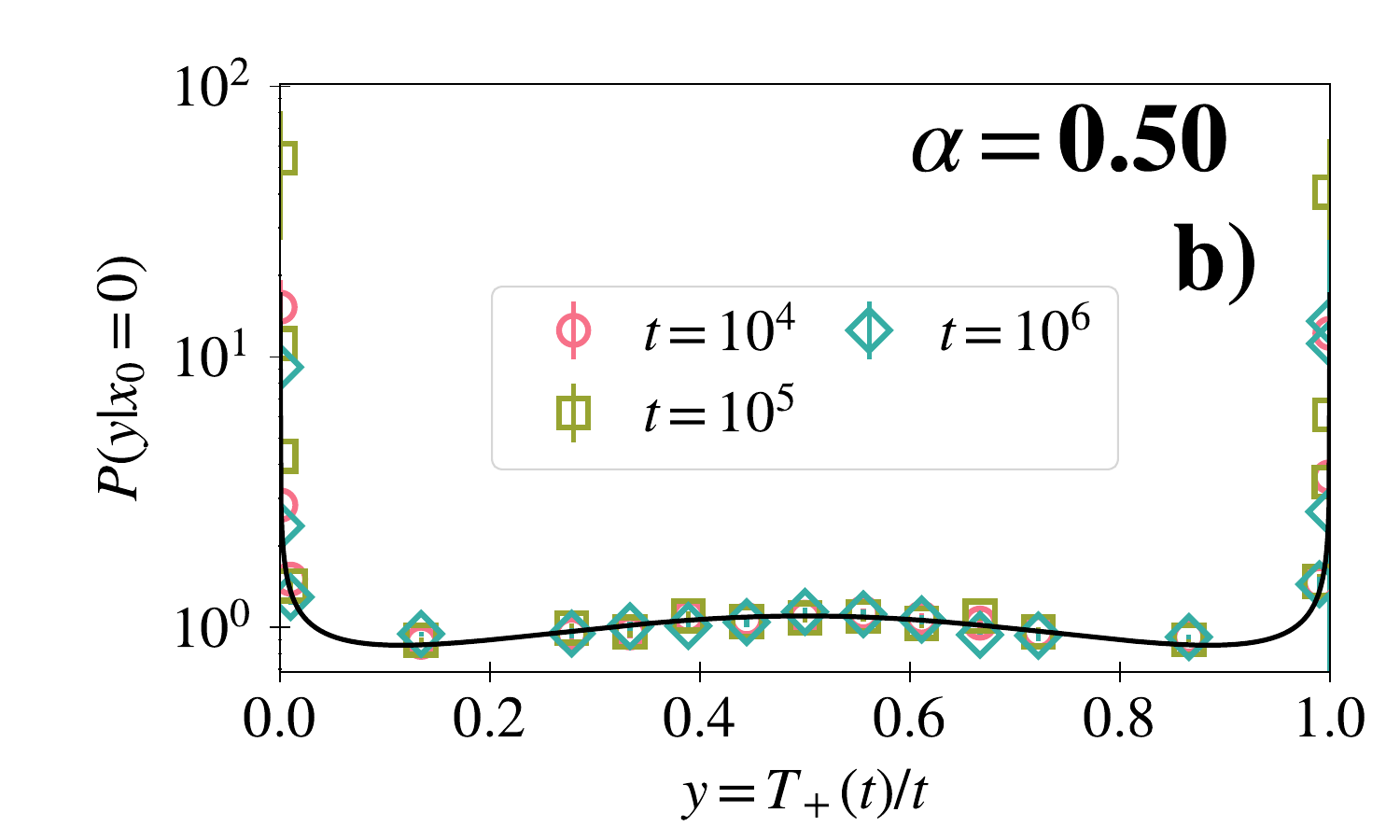}%
    \includegraphics[width=0.33\linewidth]{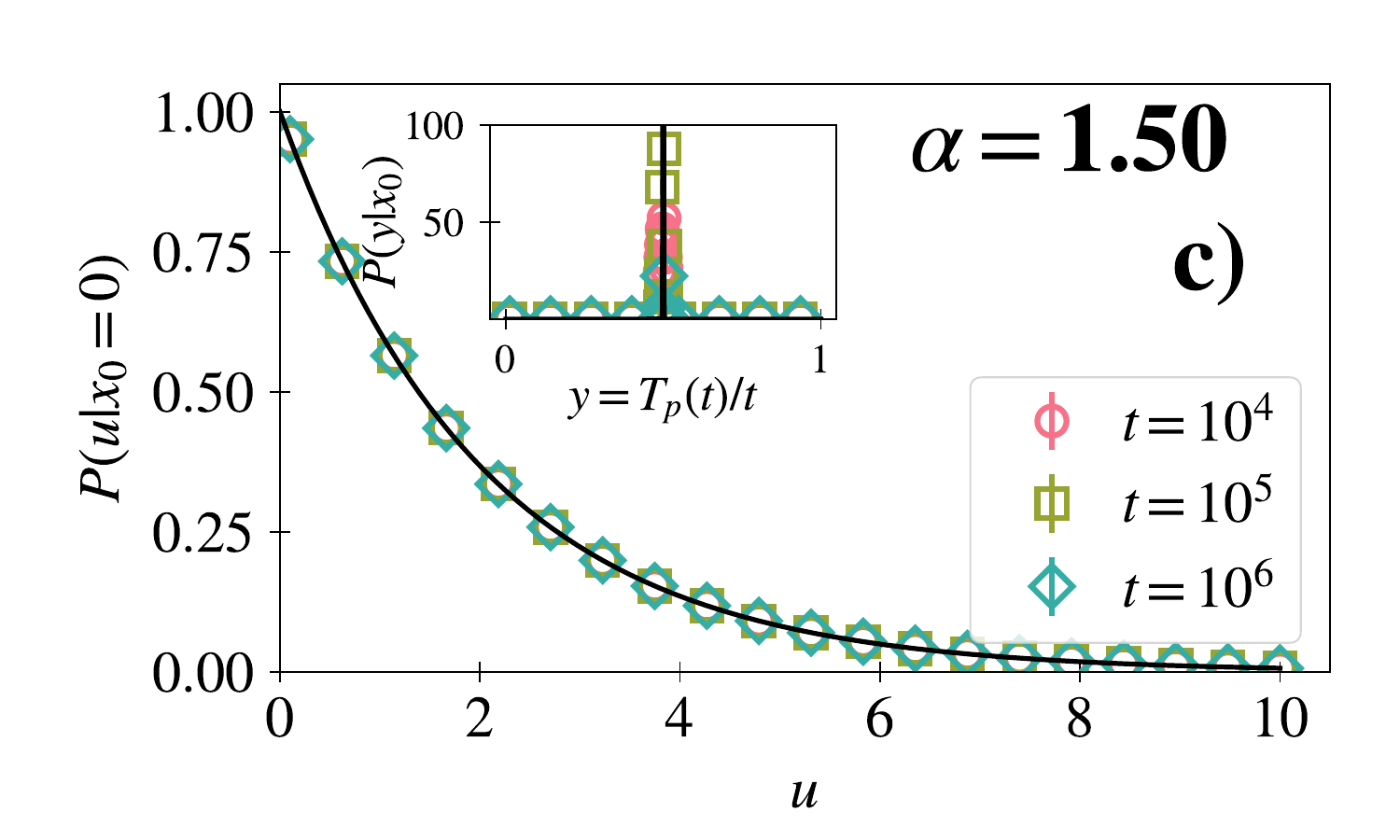}
    \caption{   
    Limiting distribution of the half-occupation time for the power-law heterogeneity. The different symbols are for trajectories with $t=10^4,10^5,$ and $10^6$. In panels (a), and (b) the black solid line corresponds to Eq. \eqref{lamperti}; In panel (c) corresponds to $e^{-0.5u}$, and to $\delta(t-0.5)$ in its inset. $\alpha=-0.05,0.5$, and $1.5$ in panels (a), (b), and (c), respectively. If $\alpha>0$ $\epsilon=10^{-10}$, and $\epsilon=10^{-3}$ if $\alpha<0$. In all panels $D_{0}=1.0$, $x_0=0$, simulation timestep $dt=0.1$, and $N=10^4$ trajectories.}
    \label{fig:Limitting_pl_Tpl}
\end{figure}

To invert  \eqref{Qsptmas} to the real space, one can follow the same method employed for the piecewise heterogeneity. Thus, from \eqref{ld} 
$$
g(\chi)=\frac{1}{1+\chi}\left[1+\frac{\chi}{1+(1+\chi)^{\frac{1}{2-\alpha}}}\right],
$$
so that 
\begin{eqnarray}
    P(T^+,t|0)= \frac{1}{t}G_{\frac{2}{2-\alpha}}\left( \frac{T^+}{t}\right)
    \label{ss}
\end{eqnarray}
where the scaling function is
\begin{eqnarray*}
    G_{\frac{2}{2-\alpha}}(y)=\frac{\sin\left(\frac{\pi}{2-\alpha}\right)}{\pi}\frac{\left[y(1-y)\right]^{\frac{-1+\alpha}{2-\alpha}}}{y^{\frac{2}{2-\alpha}}+(1-y)^{\frac{2}{2-\alpha}}+2\cos\left(\frac{\pi}{2-\alpha}\right)\left[y(1-y)\right]^{\frac{1}{2-\alpha}}}.
\end{eqnarray*}
for $\alpha <1$ with $y=T^+/t$. Thus, the PDF for $y$  reads
\begin{eqnarray}
P(y|0)=\frac{\sin\left(\frac{\pi}{2-\alpha}\right)}{\pi}\frac{\left[y(1-y)\right]^{\frac{-1+\alpha}{2-\alpha}}}{y^{\frac{2}{2-\alpha}}+(1-y)^{\frac{2}{2-\alpha}}+2\cos\left(\frac{\pi}{2-\alpha}\right)\left[y(1-y)\right]^{\frac{1}{2-\alpha}}}
    \label{lamperti}
\end{eqnarray}
for $0<y<1$. This result has been found in \cite{Si22} by other means, but it is included here for completeness. However, the ergodic properties discussed below have not been explored in \cite{Si22}. Eq. \eqref{ss} is the Lamperti PDF with index $2/(2-\alpha)$  which has been obtained in Refs \cite{CaBa10,La58} for a particle moving subdiffusively. However, Eq.  \eqref{lamperti} holds both when the underlying random walk is subdiffusive ($\alpha<0$) or superdiffusive ($0<\alpha<1$). Note also that in the limit $\alpha \to 0$ (homogeneous case) Eq. \eqref{lamperti} reduces to the arcsine law as expected. 
Moreover, it can be shown that Eq. \eqref{lamperti} posses a local maximum at $y=1/2$ when the heterogeneity is strong enough, this is, when the exponent $\alpha$ is high enough. The heterogeneity introduced through the spatial dependence given in Eq. \eqref{dpl} induces the particle to move in a region close to the origin, where the diffusion coefficient attains its minimum value. As $\alpha$ increases this region shrinks. This effect prevents the particle from performing excursions of long duration and forces the particle to revisit the origin more frequently. Thus, if $\alpha>\alpha_c$ a maximum of $P(y|0)$ appears at $y=1/2$ ($T^+=t/2$). When $\alpha<\alpha_c$ this peak disappears and $P(y|0)$ attains a U shape, indicating that the particle spends almost its entire time in one side ($x<0$ or $x>0$) only. To find $\alpha_c$ we impose the condition $(dP(y|0)/dy)_{y=1/2}=0$ for the PDF given in Eq. \eqref{lamperti}. Thus, $\alpha_c$ is the solution to the nonlinear equation
$$
\cos\left(\frac{\pi}{2-\alpha_{c}}\right)=\frac{4\alpha_{c}-2-\alpha_{c}^{2}}{(2-\alpha_{c})^{2}}
$$
whose numerical solution is $\alpha_c\simeq 0.318$. It is easy to show that $P(y|0)\sim y^{-\frac{1-\alpha}{2-\alpha}}$ and $$P(y|0)\sim (1-y)^{-\frac{1-\alpha}{2-\alpha}}$$ when $y$ is close to 0 and 1, respectively. Then, $P(y|0)$ diverges at $y=0$ and at $y=1$. In consequence, \eqref{lamperti} has two possible shapes, a U shape if $\alpha<0.318$ or a W shape if $\alpha>0.318$; both cases are shown in Figure \ref{fig:Limitting_pl_Tpl} (panels a and b). 
In panel c) we compare the limiting distribution obtained from simulations for $\alpha=1.5$ with \eqref{Py3} in the Laplace space, this is, with $P(u|0)=\mathcal{L}_{y\to u}[P(y|0)]=\mathcal{L}_{y\to u}[\delta(y-1)]=e^{-0.5u}$. In the inset we show the plot in the real space. In all panels we plot data for different $t$ to illustrate the convergence to the limiting distribution.   

Regarding the moments, the mean half occupation time and the mean squared half occupation time can be obtained from \eqref{Qsptmas} and \eqref{moments} as
\begin{eqnarray}
\left\langle T^{+}(t)\right\rangle =\frac{t}{2},\quad\left\langle T^{+}(t)^{2}\right\rangle =\frac{3-2\alpha}{4(2-\alpha)}t^{2}.
      \label{tm1}
\end{eqnarray}
The ergodicity breaking parameter can be straightforwardly found from \eqref{EB2} and \eqref{tm1}  as 
\begin{eqnarray}
\textrm{EB}_+=\frac{1-\alpha}{2-\alpha},\quad \alpha\leq 1.
\label{ebmas}
\end{eqnarray}

When $\alpha\to 1^-$ the expression for the characteristic function given by Eq. \eqref{Qsptmas} reads
\begin{eqnarray}
    \tilde{Q}(p,s|0)=\frac{2}{2s+p}
\end{eqnarray}
which after the double Laplace inversion reads
\begin{eqnarray}
    P(T^{+},t|0)=\delta\left(T^{+}-\frac{t}{2}\right).
    \label{ldtm}
\end{eqnarray}
In this case the PDF for $y$ reads
\begin{eqnarray}
    P(y|0)=\delta (y-1). 
    \label{Py3}
\end{eqnarray}

\begin{figure}[htbp]
    \centering
    \includegraphics[width=0.33\linewidth]{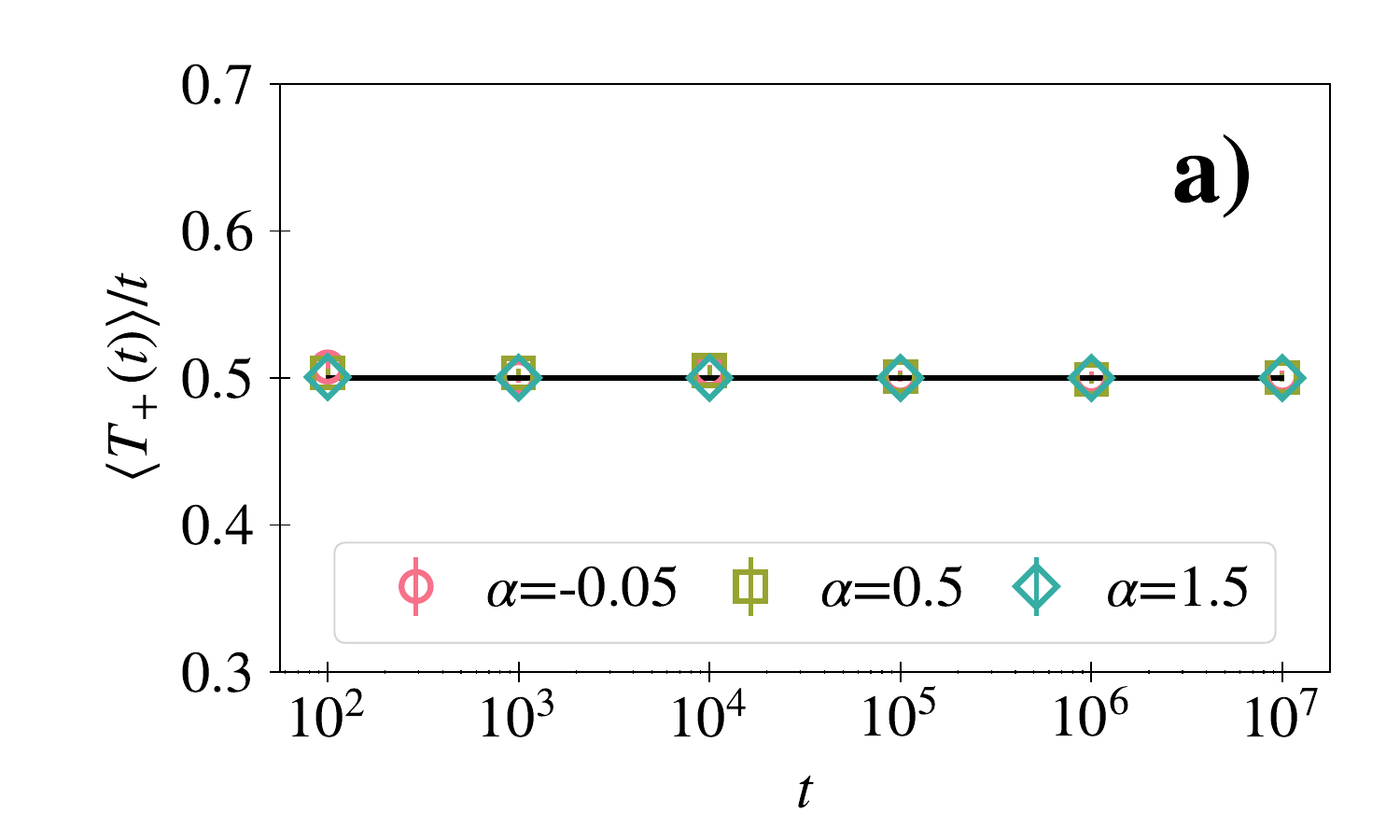}%
    \includegraphics[width=0.33\linewidth]{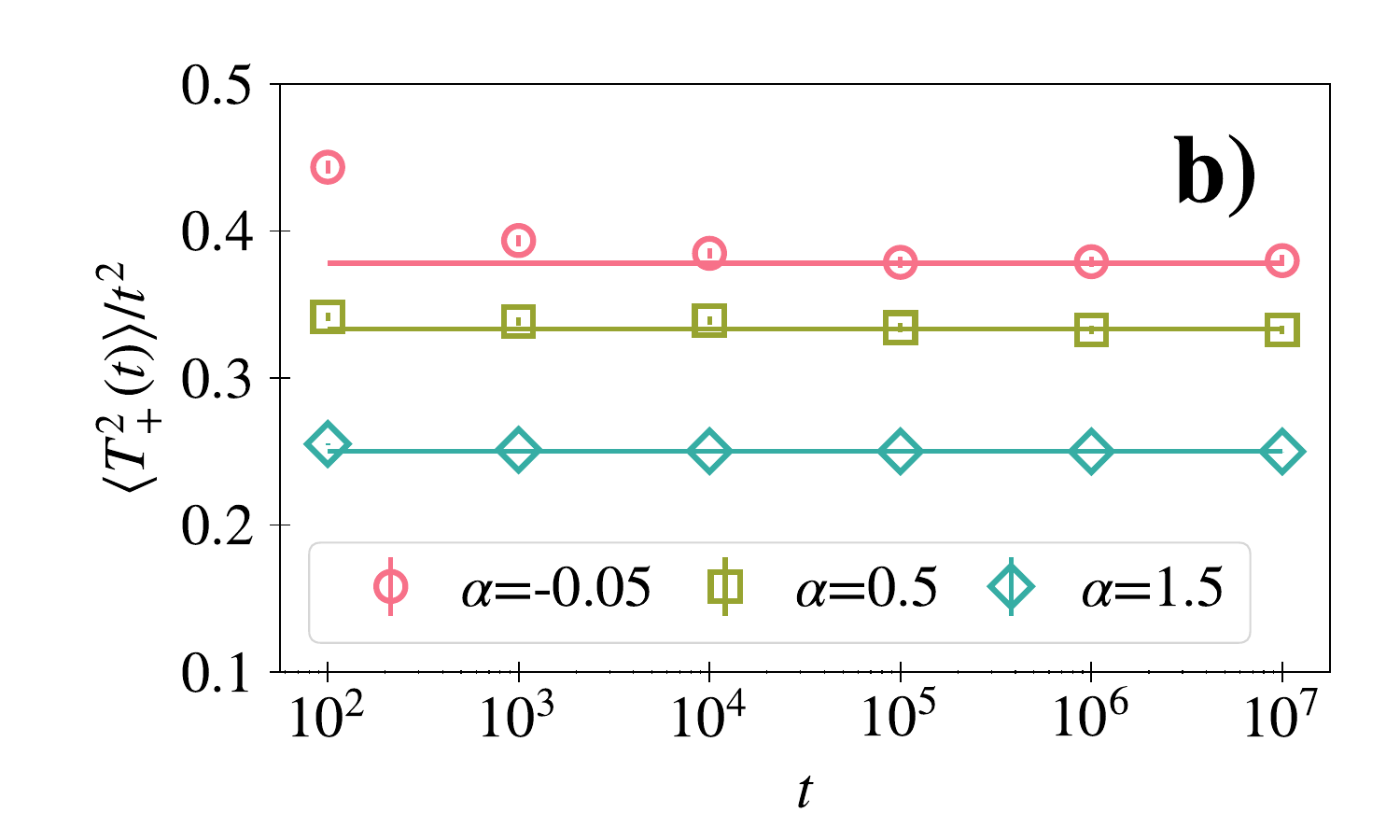}%
    \includegraphics[width=0.33\linewidth]{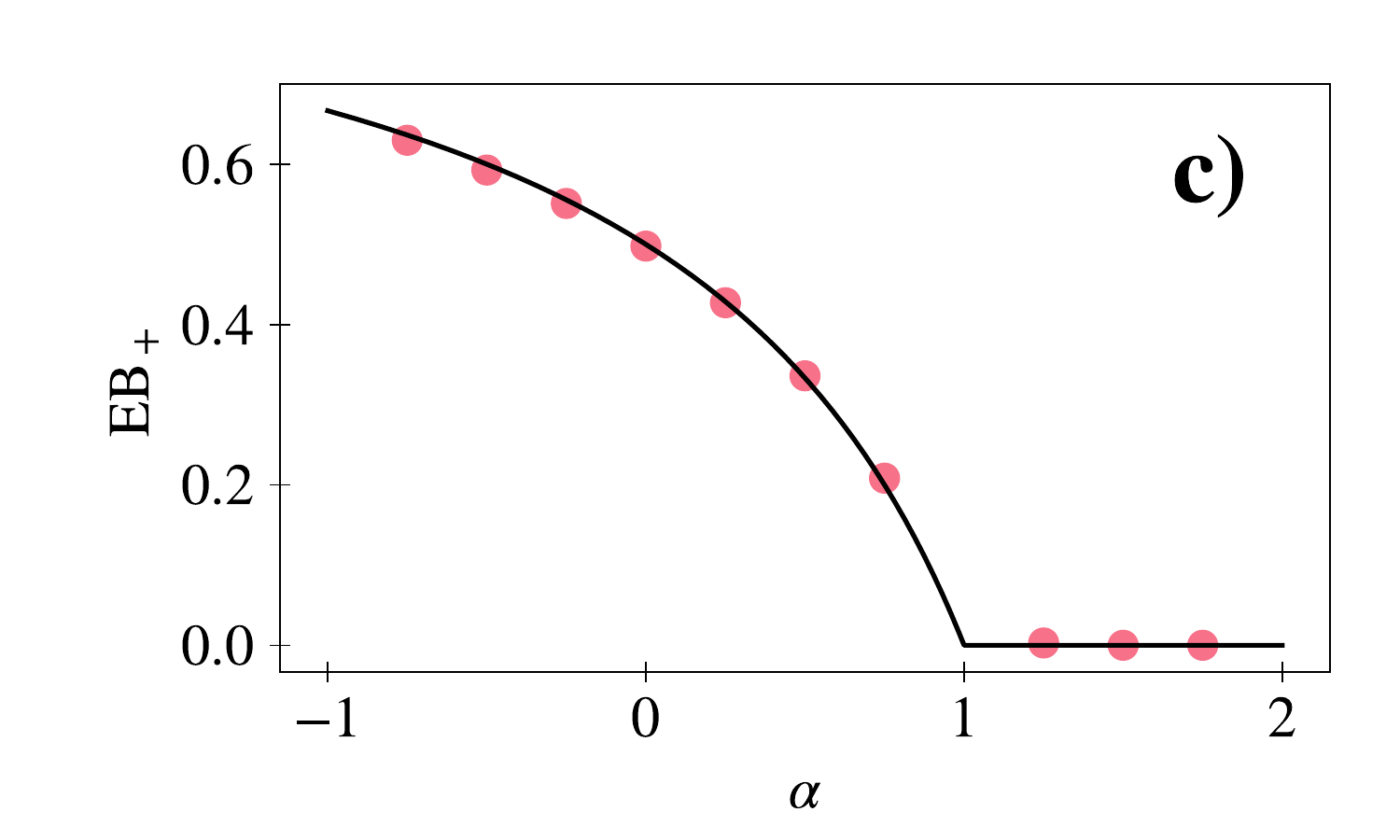}
    \caption{(a) $\langle T^{+}(t)\rangle$, (b) $\langle T^{+}(t)^2\rangle$, and (c) $\textrm{EB}_+$ of the half-occupation time for the power-law heterogeneity. In panels (a), and (b) the solid lines are computed with Eq. \eqref{tm1} and $\alpha=-0.05,0.5$, and $1.5$; In panel (c) the solid line is Eq. \eqref{ebmas}. If $\alpha>0$ $\epsilon=10^{-10}$, and $\epsilon=10^{-3}$ if $\alpha<0$. In all panels $D_{0}=1.0$, $x_0=0$, simulation timestep $dt=0.1$, and $N=10^4$ trajectories.}
    \label{fig:Pl-Tp}
\end{figure}

%This means that when $\alpha\to 1^-$ the half occupation time is not a random variable and always takes the value $t/2$ which coincides with the mean value. The effect of heterogeneity is stronger in this limit and tends to localize the particles near the origin, preventing large excursions. It can also be seen that in this case  EB$_+=0$, i.e., the process is ergodic, so that there is an ergodic transition at $\alpha=1$. 
This means that when $\alpha\to 1^-$, the half occupation time is always $t/2$, coinciding with its mean value. Likewise, it can also be seen that, in this case, EB$_+=0$, i.e., the process is ergodic. Therefore, there is an ergodic transition at $\alpha=1$. This transition can be understood by the effect of the heterogeneity, which, for $\alpha \geq 1$ is strong enough to localize the particles near the origin, preventing large excursions. 
Figure \ref{fig:Pl-Tp} compares \eqref{tm1} and \eqref{ebmas} with numerical simulations. It can be noticed the ergodic transition at $\alpha=1$ in panel \ref{fig:Pl-Tp}c.

\begin{figure}[htbp]
    \centering
    \includegraphics[width=0.33\linewidth]{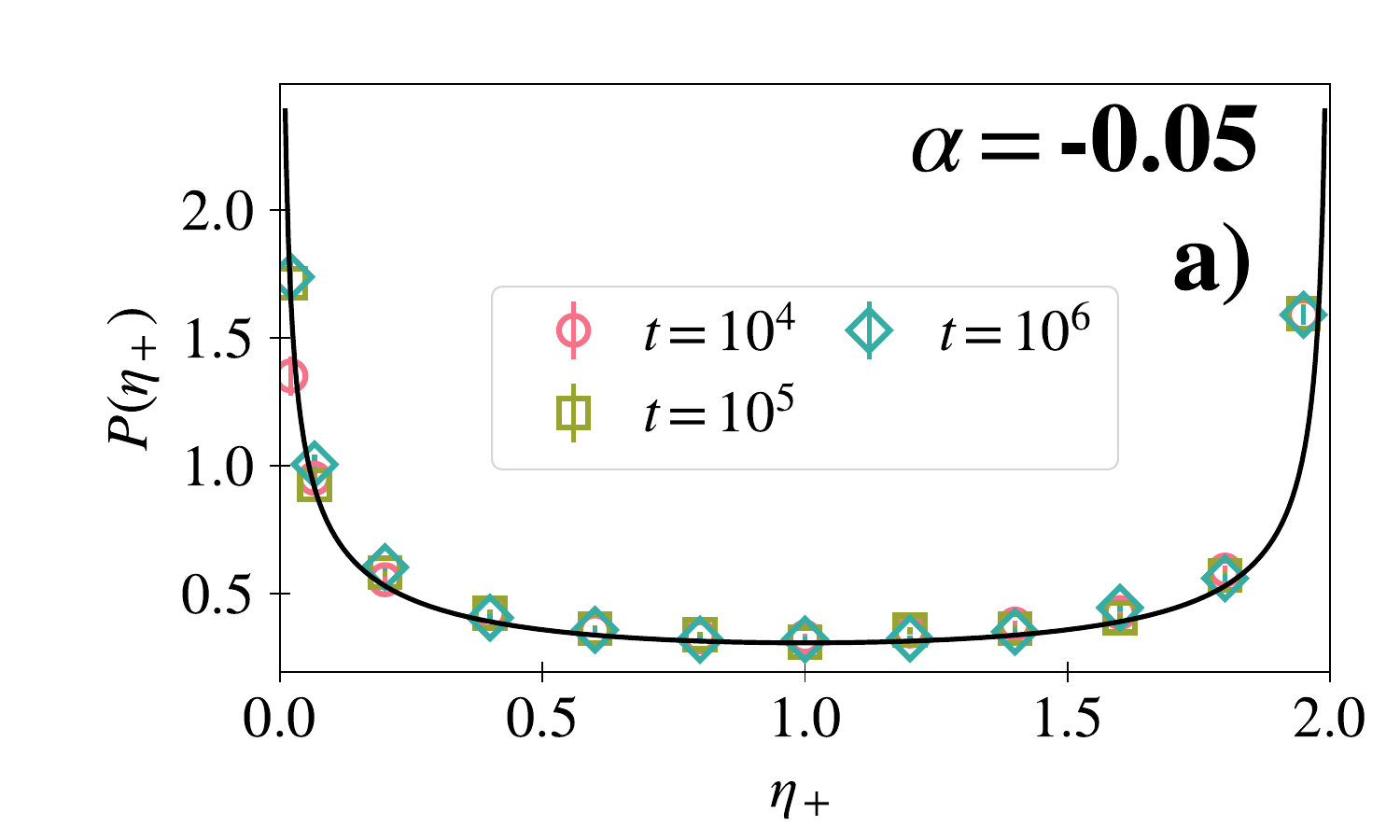}%
    \includegraphics[width=0.33\linewidth]{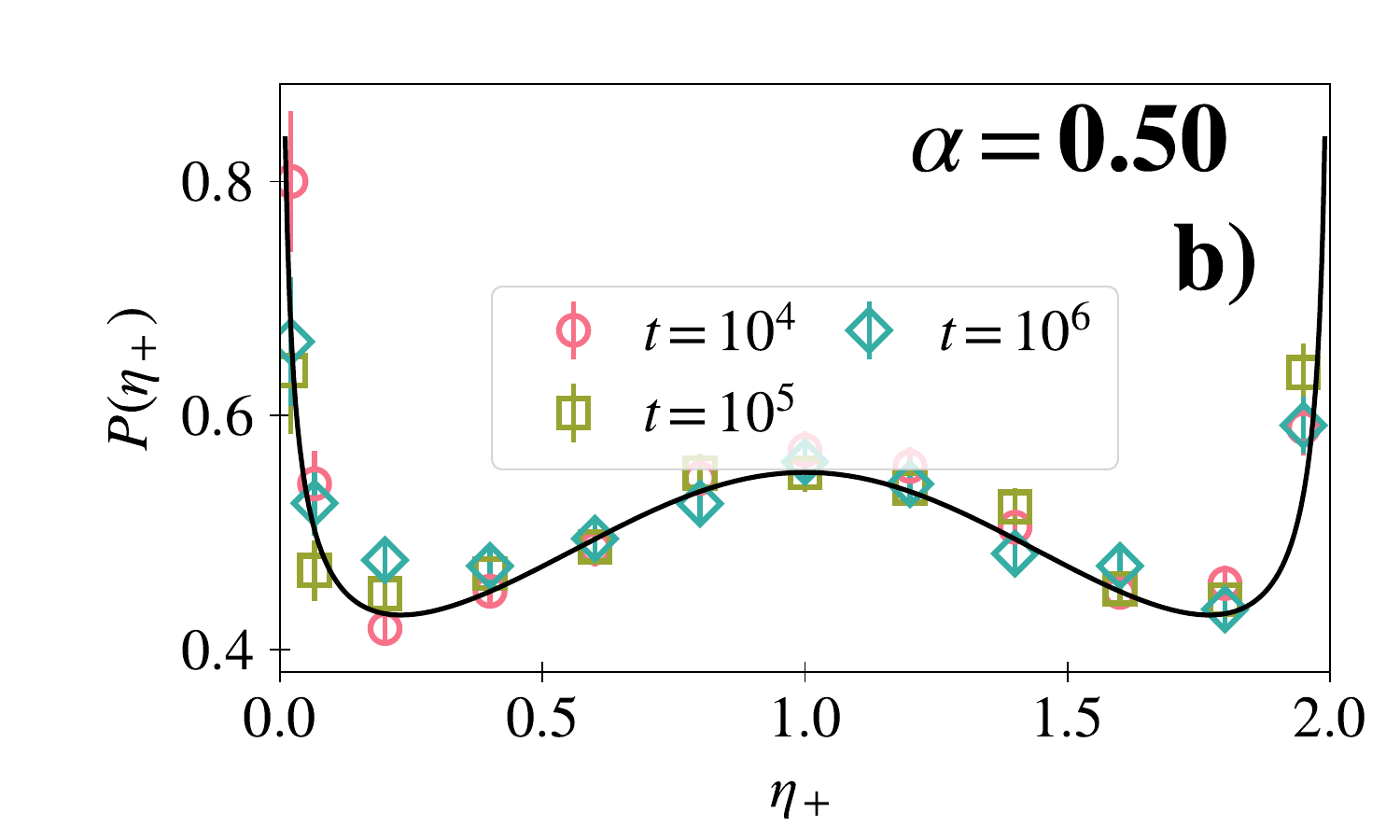}%
    \includegraphics[width=0.33\linewidth]{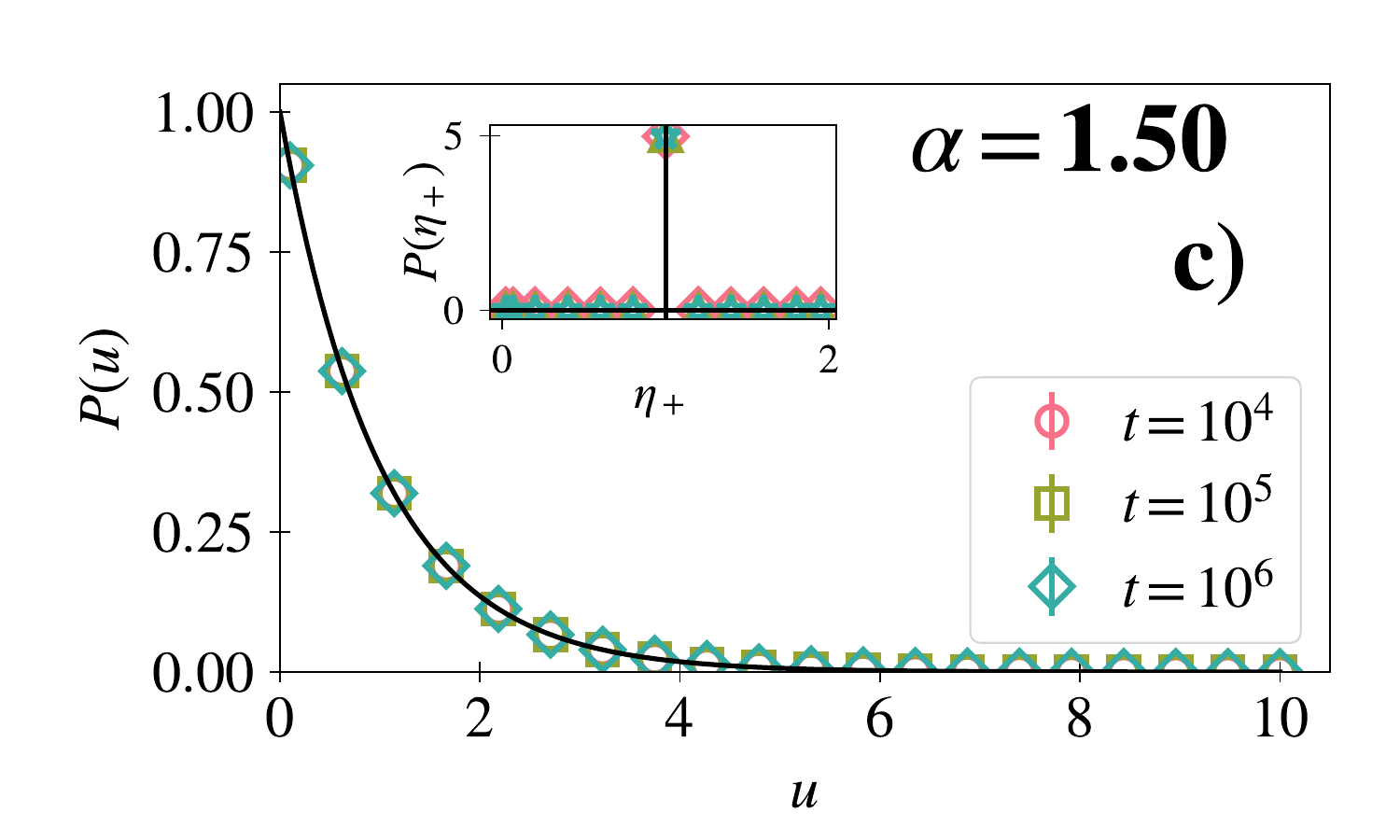}
    \caption{PDF of the time averaged half-occupation time for the power-law heterogeneity. The different symbols are for trajectories with $t=10^4,10^5,$ and $10^6$. The black solid line corresponds to Eq. \eqref{Peta2} in panels (a), and (b); to $e^{-u}$ in panel (c), and to $\delta(t-1)$ in its inset. $\alpha=-0.05,0.5$, and $1.5$ in panels (a), (b), and (c), respectively. If $\alpha>0$ we take $\epsilon=10^{-10}$, and $\epsilon=10^{-3}$ if $\alpha<0$. In all panels $D_{0}=1.0$, $x_0=0$, simulation timestep $dt=0.1$, and $N=10^4$ trajectories.}
    \label{fig:eta_pl_Tp}
\end{figure}

The PDF for the time averaged observable $\theta [x(t)]$ (this is $\eta_+$ defined in \eqref{etamasd}) can be obtained from  \eqref{lamperti} noting that $\eta_+=2y$ to get
\begin{eqnarray}
P(\eta_{+})=\frac{\sin\left(\frac{\pi}{2-\alpha}\right)}{\pi}\frac{2\left[\eta_{+}(2-\eta_{+})\right]^{\frac{-1+\alpha}{2-\alpha}}}{\eta_{+}^{\frac{2}{2-\alpha}}+(2-\eta_{+})^{\frac{2}{2-\alpha}}+2\cos\left(\frac{\pi}{2-\alpha}\right)\left[\eta_{+}(2-\eta_{+})\right]^{\frac{1}{2-\alpha}}}
\label{Peta2}
\end{eqnarray}
for $0<\eta_+<2$. In Figure \ref{fig:eta_pl_Tp} we compare Eq. \eqref{Peta2} to numerical simulations.

\section{Occupation time in an interval}
We consider now the occupation time $T_{a} (t|x_0)$ of the random particle in an interval $[-a,a]$ up to time $t$ if starting initially from $x_0$.  In this case, the function $U(x_0)=\theta (-a<x_0<a)$. The corresponding Feynman-Kac equation \eqref{FK} needs to be solved with suitable boundary conditions. For instance, if the starting point is at infinity i.e., $x_{0}\rightarrow \pm\infty$ the particle will never reach the interval so that $P(T_a,t|x_{0}\rightarrow \pm\infty)=\delta (T_a)$, i.e., $\tilde{Q}(p,s|x_{0}\rightarrow \pm\infty)=1/s$. The initial condition is such that $T_a(t=0|x_{0})=0$
so that $Q(p,t=0|x_{0})=1$. Thus, the Feynman-Kac equation for $\tilde{Q}(p,s|x_{0})$ tobe  solve in the Laplace space is
\begin{eqnarray}
    s\tilde{Q}(p,s|x_{0})-1=D(x_{0})\frac{d\tilde{Q}(p,s|x_{0})}{dx_{0}^{2}}-p\theta(-a<x_{0}<a)\tilde{Q}(p,s|x_{0})
    \label{ode2}
\end{eqnarray}
under the above boundary conditions. It must be explicitly found in the regions I ($x_0<-a$), II ($-a<x_0<0$), III ($0<x_0<a$) and IV ($x_0>a$).

\subsection{Piecewise heterogeneity}
The solution of Eq. \eqref{ode2} for the piecewise diffusion coefficient \eqref{pwd} is given by (setting $x_0=0$)
 \begin{eqnarray}
     \tilde{Q}(p,s|0)=\frac{1}{s+p}\left[1+\frac{p}{sF(s,p)}\right]
     \label{Qtaph}
 \end{eqnarray}
where
$$
F(s,p)=\frac{M_{-}(s,p)N_{+}(s,p)+\gamma M_{+}(s,p)N_{-}(s,p)}{N_{+}(s,p)+\gamma N_{-}(s,p)}
$$
and
\begin{eqnarray*}
   M_{\pm}(s,p)&=&\cosh\left(a\sqrt{\frac{s+p}{D_{\pm}}}\right)+\sqrt{\frac{s+p}{s}}\sinh\left(a\sqrt{\frac{s+p}{D_{\pm}}}\right) \\
   N_{\pm}(s,p)&=&\sinh\left(a\sqrt{\frac{s+p}{D_{\pm}}}\right)+\sqrt{\frac{s+p}{s}}\cosh\left(a\sqrt{\frac{s+p}{D_{\pm}}}\right).
\end{eqnarray*}

Since the double Laplace inversion of $\tilde{Q}(p,s|0)$ cannot be found explicitly we derive an approximate expression which holds when $t$ and $T_a$ are large. To do this, we consider $s$ and $p$ small and comparable in Eq. \eqref{Qtaph} and find
\begin{eqnarray}
   \tilde{Q}(p,s|0)\simeq \frac{1}{s+p}\left[1+\frac{p}{s+\mathcal{C}\sqrt{s}(s+p)}\right] 
   \label{Qbulk1}
\end{eqnarray}
where
$$
\mathcal{C}=\frac{a}{\sqrt{D_{+}}+\sqrt{D_{-}}}\left(\sqrt{\frac{D_{+}}{D_{-}}}+\sqrt{\frac{D_{-}}{D_{+}}}\right).
$$
Performing the double Laplace inversion of \eqref{Qbulk1}, we find that the PDF of $T_a$ obeys the half-Gaussian density

\begin{eqnarray}
P(T_{a},t|0)\sim\frac{1}{t^{1/2}}H\left(\frac{\sqrt{D_{*}}T_{a}}{2a\sqrt{t}}\right)
    \label{halfg}
\end{eqnarray}

\begin{figure}[htbp]
    \centering
    \includegraphics[width=0.33\linewidth]{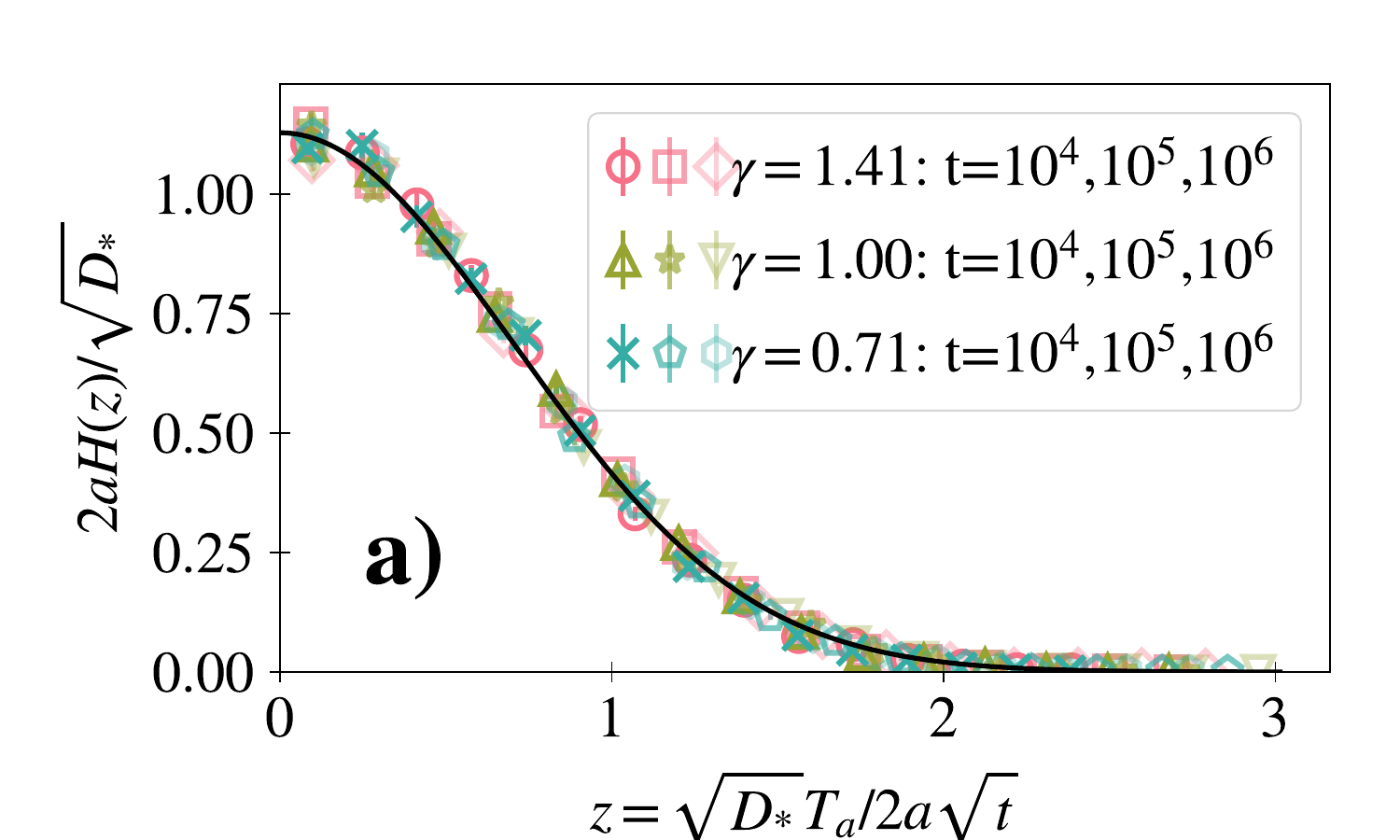}%
    \includegraphics[width=0.33\linewidth]{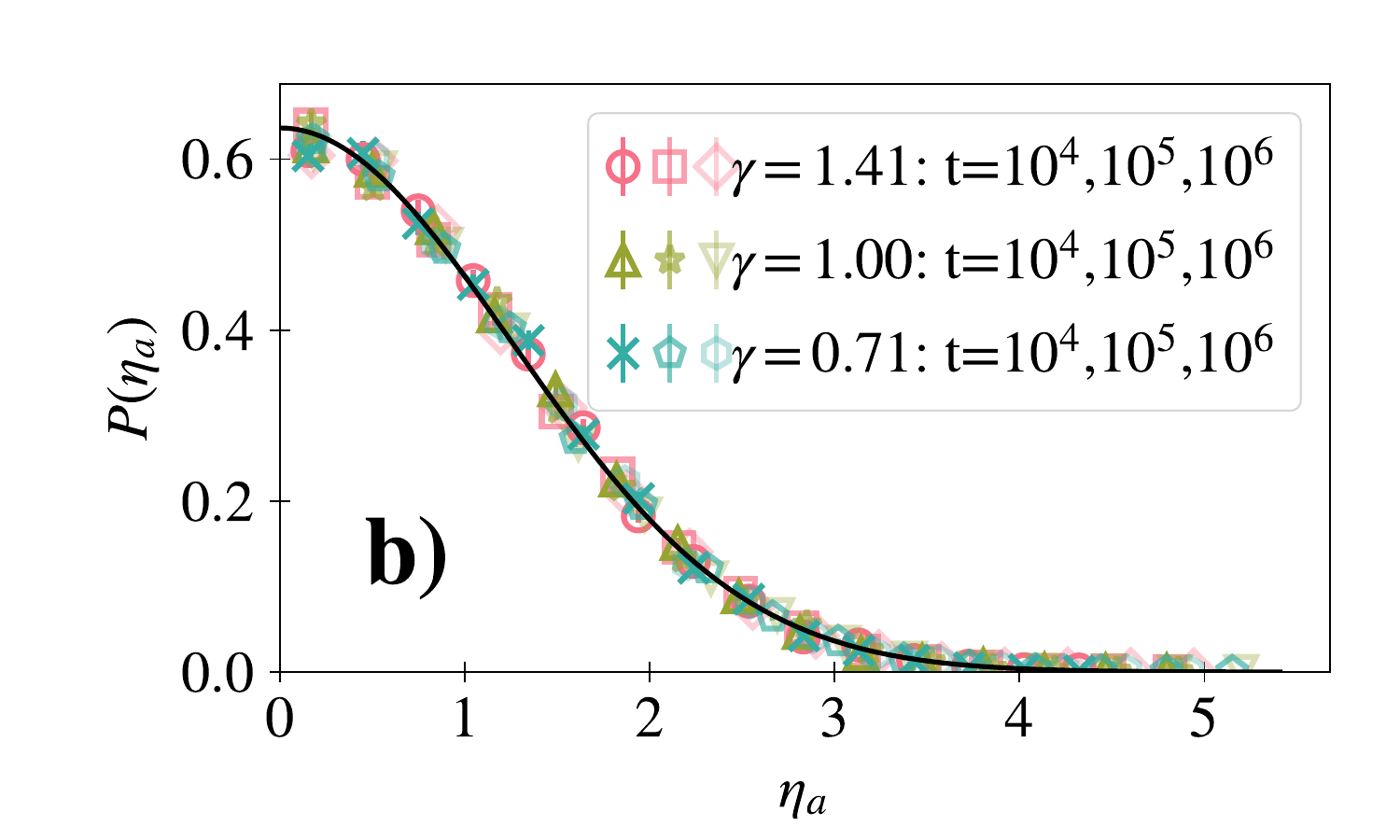}
    \caption{(a) Limiting distribution of the occupation time in the interval $[-3,3]$ for the piecewise heterogeneity. The black solid line corresponds to Eq. \eqref{halfg}. (b) PDF of the time averaged occupation time in the interval [-3,3] for the piecewise heterogeneity. The black solid line corresponds to Eq. \eqref{ml12}. In all panels, the symbols of the same color are for $D_{+}=0.5,1.0$, and $2.0$ salmon, green, and blue, respectively. The different symbol shapes are for trajectories with $t=10^4, 10^5,$ and $10^6$. $D_{-}=1.0$, $x_0=0$, simulation timestep $dt=0.1$, and $N=10^4$ trajectories.}
    \label{fig:pw_Ta}
\end{figure}

where the scaling function is
$$
H(z)=\sqrt{\frac{D_{*}}{a^{2}\pi}}e^{-z^{2}},\quad D_{*}=\left(\frac{D_{+}\sqrt{D_{-}}+D_{-}\sqrt{D_{+}}}{D_{+}+D_{-}}\right)^{2}.
$$

In panel a) of Figure \ref{fig:pw_Ta} we check \eqref{halfg} against numerical solutions for different values of $\gamma$. We also add data computed for different large values of $t$ to show that when $t$ is sufficiently large the numerical results fall on the theoretical curve.

The mean and mean square occupation time can be computed from \eqref{Qtaph} and \eqref{moments} as
\begin{eqnarray}
    \left\langle T_{a}(t)\right\rangle \simeq \frac{2a\sqrt{t}}{\sqrt{\pi D_{*}}},\quad
     \left\langle T_{a}(t)^{2}\right\rangle \simeq \frac{2a^{2}}{D_{*}}t.
    \label{ta12}
\end{eqnarray}

\begin{figure}[htbp]
    \centering
    \includegraphics[width=0.33\linewidth]{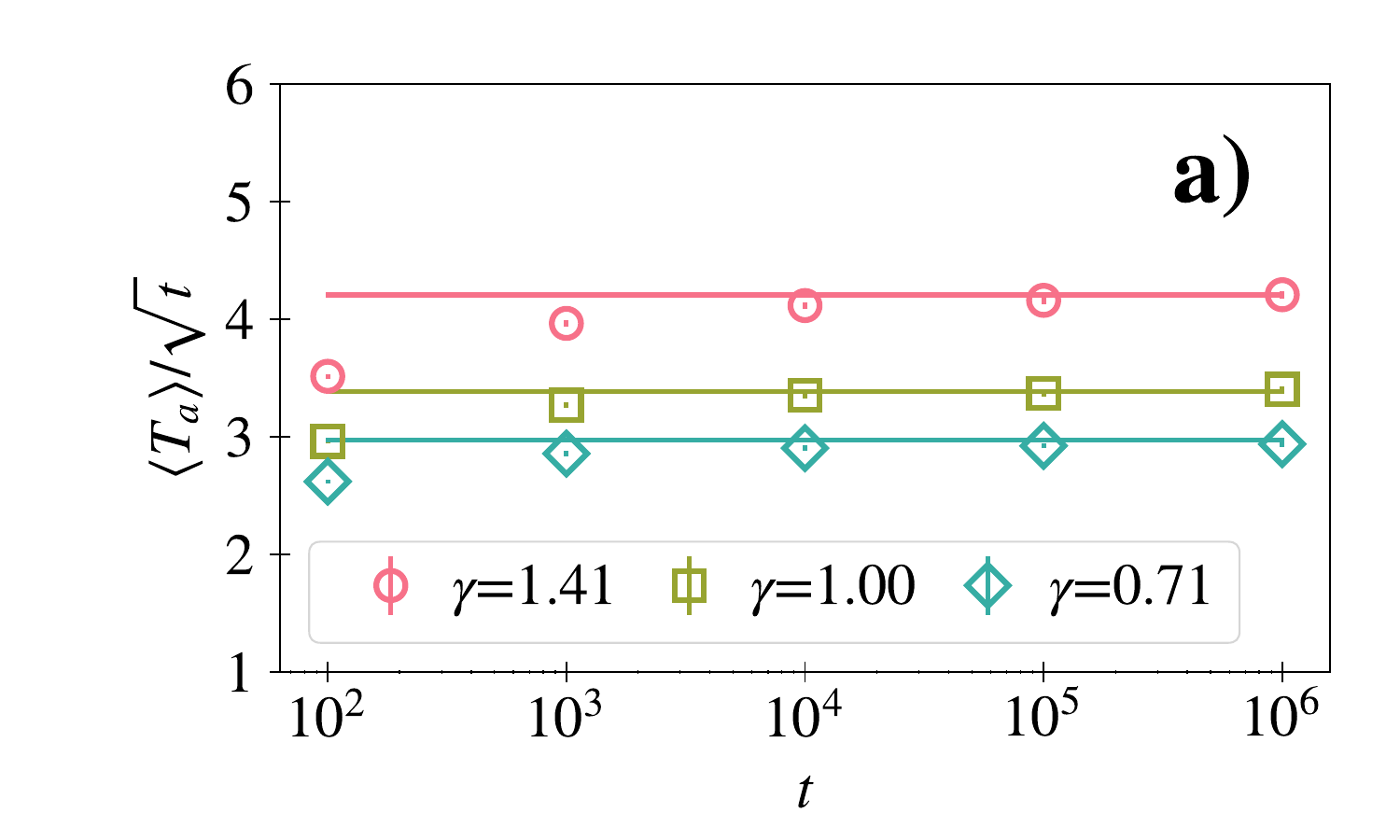}%
    \includegraphics[width=0.33\linewidth]{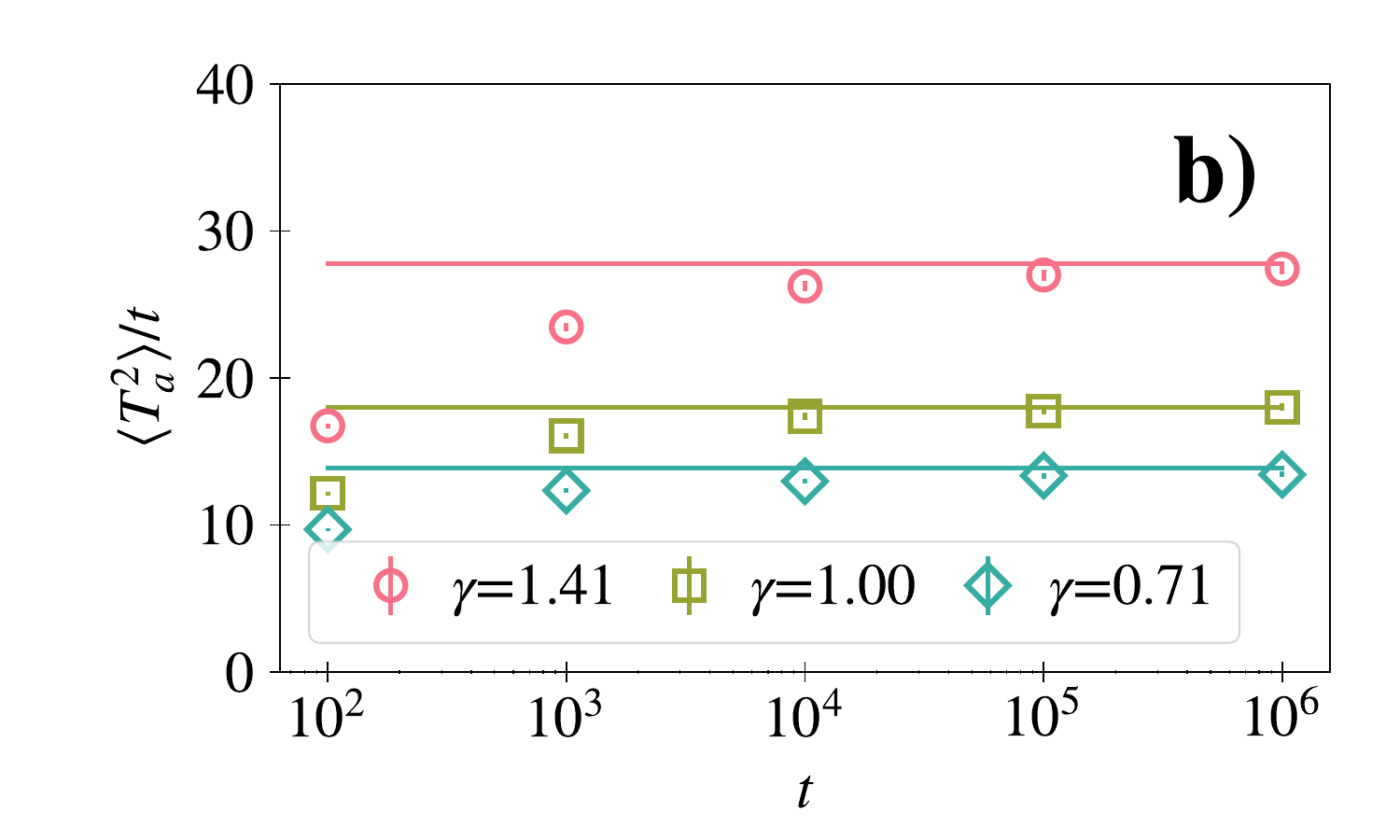}%
    \includegraphics[width=0.33\linewidth]{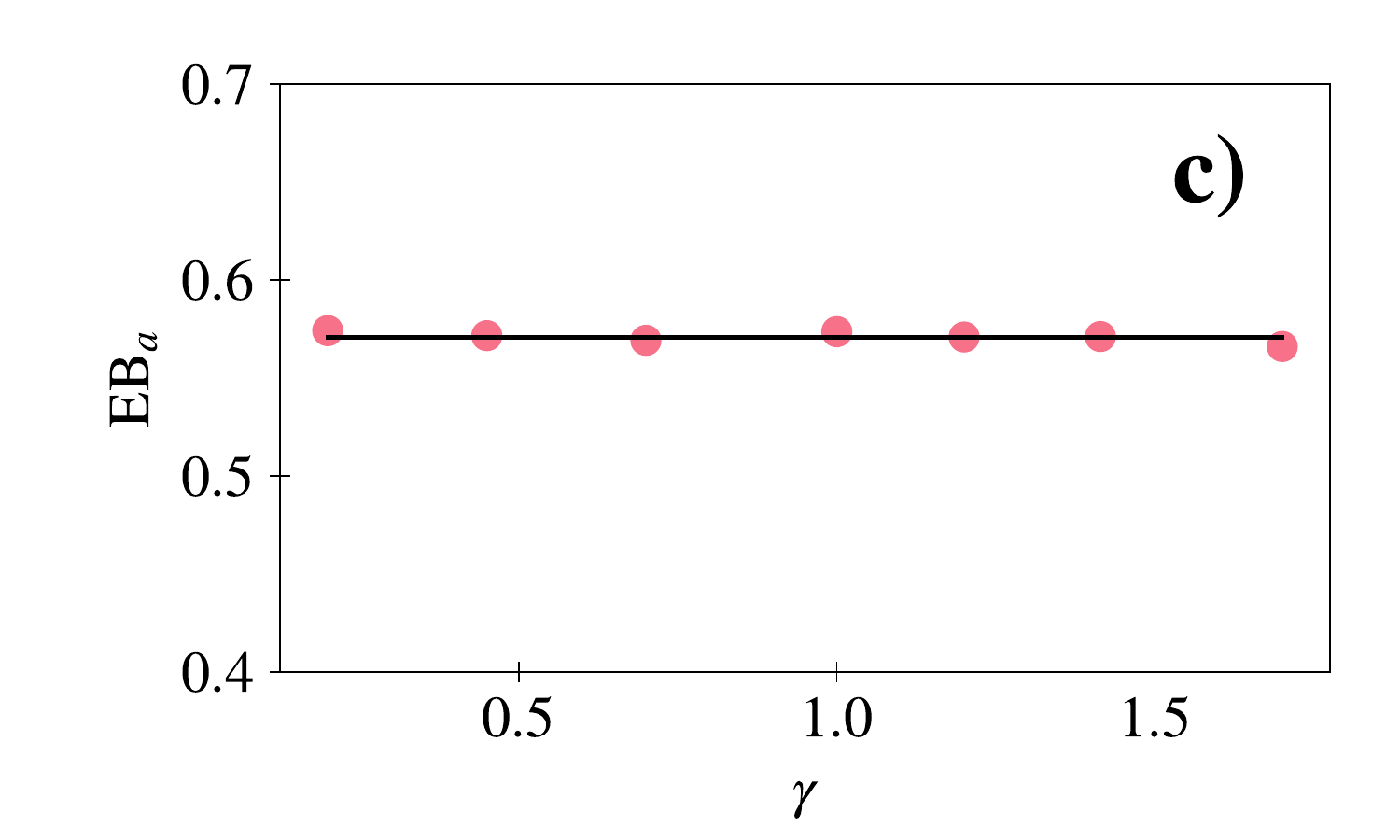}
    \caption{
    (a) $\langle T_{a} (t)\rangle$, (b) $\langle T_{a}(t)^2\rangle$, and (c) $\textrm{EB}_a$ of the half-occupation time for the piecewise heterogeneity. In panels (a), and (b) the solid lines are computed with Eq. \eqref{ta12}, $D_{+}=0.5,1.0$, and $2.0$; In panel (c) the solid line is Eq. \eqref{EBad}. In all panels $D_{-}=1.0$, $x_0=0$, simulation timestep $dt=0.1$, and $N=10^4$ trajectories. In panels (a), and (b) the interval is $[-3,3]$ and in panel (c) the interval is $[-0.5,0.5]$.}
    \label{fig:Pw-Ta}
\end{figure}

It is interesting to note that in the long time limit the PDF $P(T_a,t|0)$ can be corresponds to the PDF for the Brownian motion moving in a homogeneous media \cite{BaFlMe23} with an effective diffusion coefficient $D_*$. 
The ergodicity breaking parameter can be found from \eqref{EB2} and \eqref{ta12}  to get
\begin{eqnarray}
    \textrm{EB}_a=\frac{\left\langle T_{a}(t)^{2}\right\rangle }{\left\langle T_{a}(t)\right\rangle ^{2}}-1=\frac{\pi}{2}-1
    \label{EBad}
\end{eqnarray}
which is the same as the EB parameter for the Brownian motion in homogeneous media. This means that in the long time limit the ergodic properties in terms of the occupation time in an interval for a Brownian particle are not altered by the piecewise heterogeneity. In Figure \ref{fig:Pw-Ta} we compare \eqref{ta12} and \eqref{EBad} with numerical simulations. 

Now we find the PDF of the time average of the observable $\theta[-a<x(t)<a]$ in the long time limit, this is, the PDF of 
\begin{eqnarray}
    \eta_{a}=\lim_{t\to\infty}\frac{T_a}{\left\langle T_a(t)\right\rangle },
    \label{etaa}
\end{eqnarray}
as defined in Eq. \eqref{eta}. For the piecewise diffusion coefficient
$$
\eta_{a}=\frac{T_{a}}{2a}\sqrt{\frac{\pi D_{*}}{t}}
$$
and its PDF follows from \eqref{peta} and \eqref{halfg}. In this case we readily find
 \begin{eqnarray}
     P(\eta_{a})=\mathcal{M}_{1/2}(\eta_a)=\frac{2}{\pi}e^{-\frac{\eta_{a}^{2}}{\pi}},
     \label{ml12}
 \end{eqnarray}
 which is the same result as for BM in a homogeneous media. In panel b) of Figure \ref{fig:pw_Ta} we compare \eqref{ml12} with numerical simulations. We note that the PDF $P(\eta_a)$ computed for different values of the asymmetry parameter $\gamma$ collapse on the same curve, which confirms that $P(\eta_a)$ does not depend on $\gamma$. In addition, it can be observed how the distribution converges to the theoretical result \eqref{ml12} in the long time limit.

\subsection{Power law heterogeneity}
In this case the expression for the characteristic function in the long time limit is given by (see details of the derivation in Appendix C)
\begin{eqnarray}
\tilde{Q}(p,s|0)\simeq\frac{\lambda_{\alpha}+s^{-1+\frac{1}{2-\alpha}}}{s^{\frac{1}{2-\alpha}}+\lambda_{\alpha}(s+p)}
\label{Qpsta2}
\end{eqnarray}
which holds for $\alpha<1$ and where we have defined
$$
\lambda_{\alpha}=\frac{a^{1-\alpha}}{\left[(2-\alpha)\sqrt{D_{0}}\right]^{\frac{2-2\alpha}{2-\alpha}}}\frac{\Gamma\left(\frac{1}{2-\alpha}\right)}{\Gamma\left(\frac{3-2\alpha}{2-\alpha}\right)}.
$$

Before proceeding to invert in Laplace this result we note that for $\alpha<1$ we can further simplify this expression by noting that in the limit $s\to 0$ we have $s\ll s^{\frac{1}{2-\alpha}}$ and thus
\begin{eqnarray}
\tilde{Q}(p,s|0)\simeq\frac{s^{-1+\frac{1}{2-\alpha}}}{s^{\frac{1}{2-\alpha}}+\lambda_{\alpha}p}.
\label{Qpsta22}
\end{eqnarray}
Performing the Laplace inversion of \eqref{Qpsta22} with respect to $p$ one readily finds
\begin{eqnarray}
    Q(T_{a},s|0)\simeq\frac{s^{-1+\frac{1}{2-\alpha}}}{\lambda_{\alpha}}\exp\left(-s^{\frac{1}{2-\alpha}}\frac{T_{a}}{\lambda_{\alpha}}\right)
    \label{Qpsta222}
\end{eqnarray}
We next compute the Laplace inversion of \eqref{Qpsta222} with respect to $s$ using 
$$
\mathcal{L}_{s}^{-1}\left(s^{\beta-1}e^{-us^{\beta}}\right)=\frac{t}{\beta u^{1+1/\beta}}l_{\beta}\left(\frac{t}{u^{1/\beta}}\right)
$$
where $l_{\beta}(\cdot)$ is the one-sided L\'evy density with order $\beta$.
We thus find the scaling form
\begin{eqnarray}
 P(T_{a},t|0)\sim\frac{1}{t^{\frac{1}{2-\alpha}}}N_{\alpha}\left(\frac{T_{a}}{\lambda_{\alpha}t^{\frac{1}{2-\alpha}}}\right)
   \label{Pat}
\end{eqnarray}
for $\alpha<1$, where the scaling function $N_{\alpha}(z)$ reads
$$
N_{\alpha}(z)=\frac{2-\alpha}{\lambda_{\alpha}z^{3-\alpha}}l_{\frac{1}{2-\alpha}}\left(\frac{1}{z^{2-\alpha}}\right).
$$
In the limit $\alpha \to 1^-$ we see that \eqref{Qpsta22} turns into
\begin{eqnarray}
   \tilde{Q}(p,s|0)\simeq\frac{1}{s+p} 
\end{eqnarray}
which after Laplace inversion yields
\begin{eqnarray}
 P(T_{a},t|0)\simeq  \delta(T_{a}-t),\quad   \alpha \to 1^-.
 \label{Ptapl}
\end{eqnarray}

\begin{figure}[htbp]
    \centering
    \includegraphics[width=0.33\linewidth]{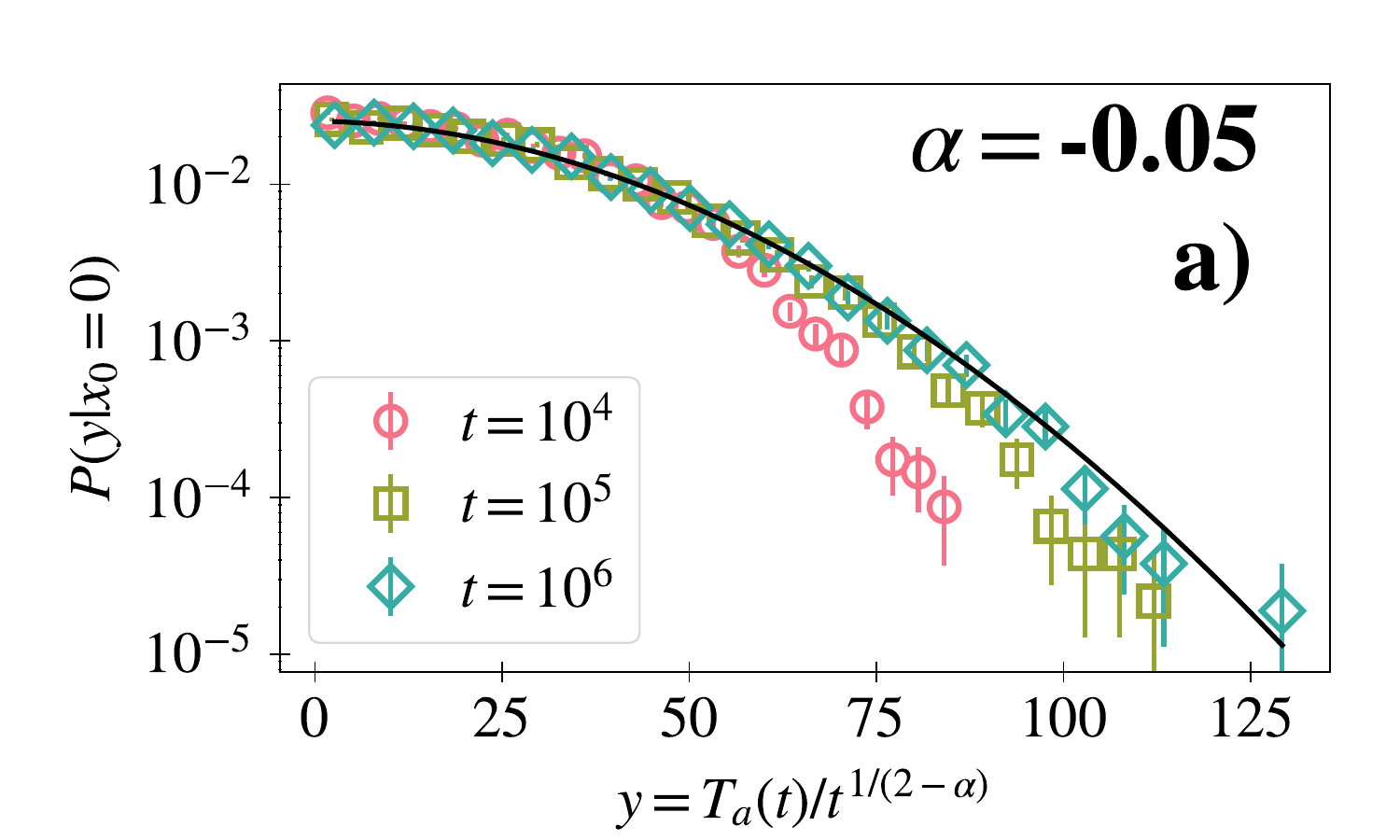}%
    \includegraphics[width=0.33\linewidth]{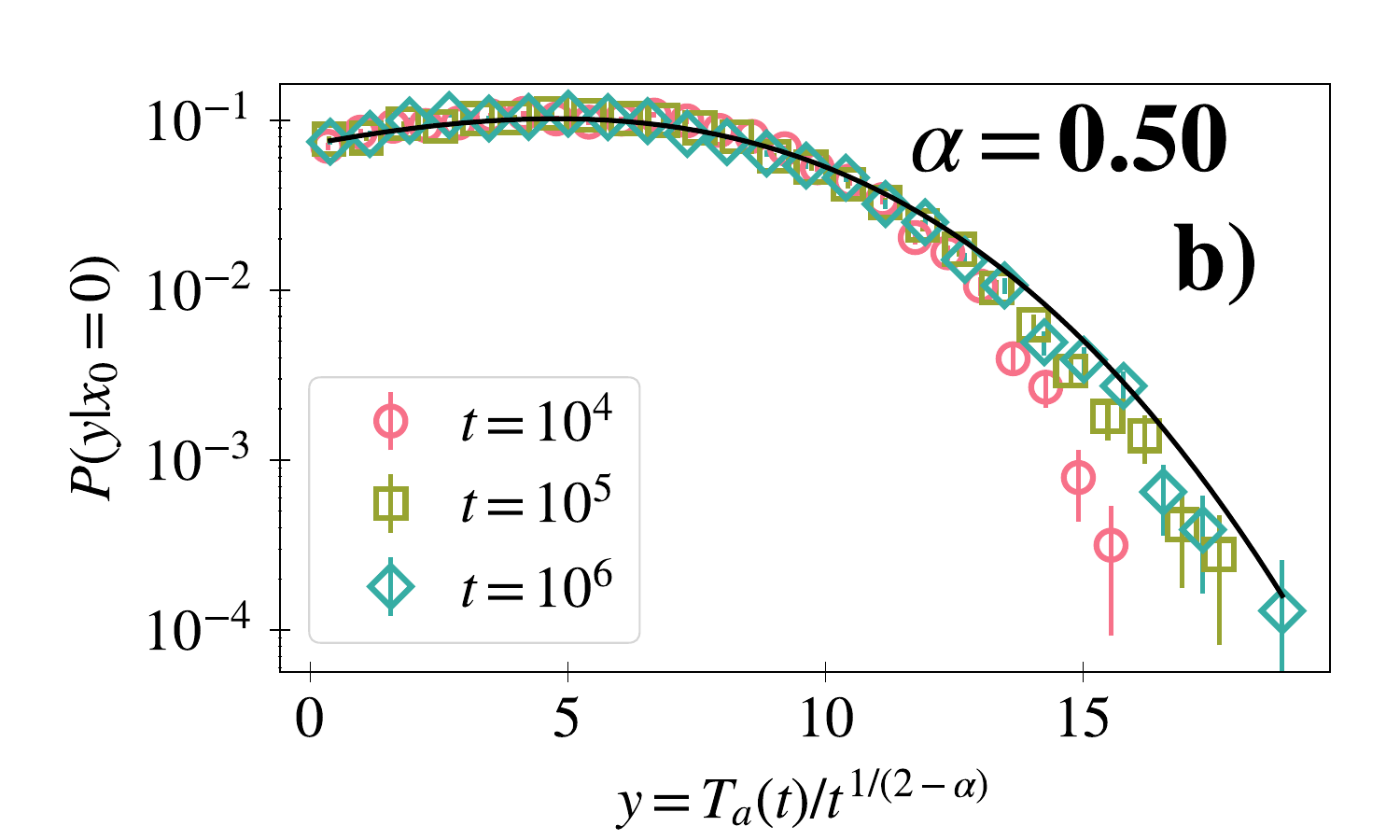}%
    \includegraphics[width=0.33\linewidth]{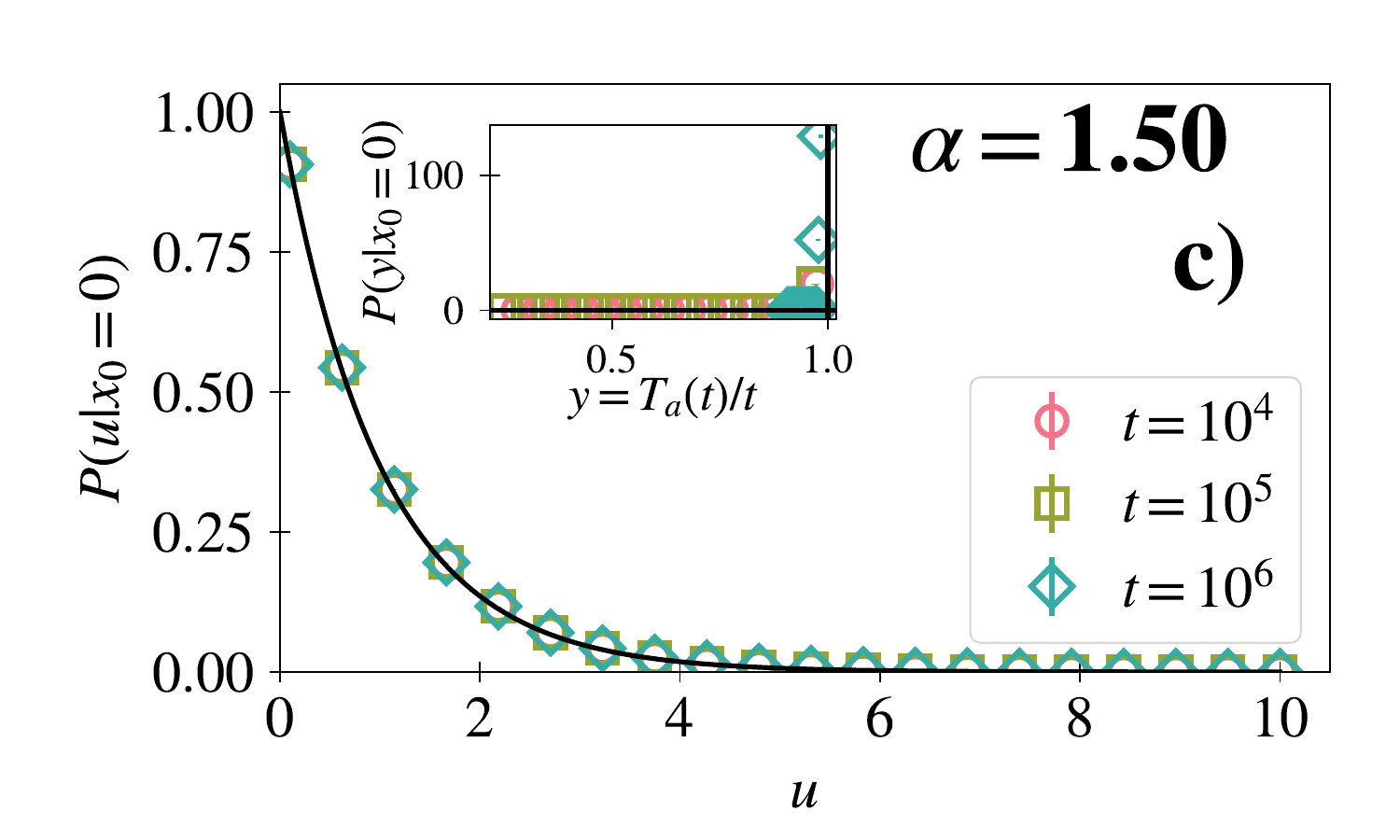}
    \caption{Limiting distribution of the occupation time in the interval $[-20,20]$ for the power-law heterogeneity. The different symbols are for trajectories with $t=10^4,10^5,$ and $10^6$. The black solid line corresponds to Eq. \eqref{Pat} in panels (a), and (b); to $e^{-u}$ in panel (c), and to Eq. \eqref{Ptapl} in its inset. $\alpha=-0.05,0.5$, and $1.5$ in panels (a), (b), and (c), respectively. If $\alpha>0$ $\epsilon=10^{-10}$, and $\epsilon=10^{-3}$ if $\alpha<0$. In all panels $D_{0}=1.0$, $x_0=0$, simulation timestep $dt=0.1$, and $N=10^4$ trajectories.}
    \label{fig:Limitting_pl_Ta}
\end{figure}

It is worth observing that at $\alpha =1$ there is a transition in the behavior of the PDF $ P(T_{a},t|0)$. For $1\leq \alpha<2$ the heterogeneity is so strong that the particle cannot escape the interval $[-a,a]$ for any $a$, which is reflected in the fact that $T_a$ is equal to $t$.  

The comparison of the limiting distributions given in Eqs. \eqref{Pat} and \eqref{Ptapl} with numerical simulations is shown in Figure \ref{fig:Limitting_pl_Ta}. We note that the PDFs converge to the theoretical results in the long time limit in panels a) and b). In panel c) we show the limiting distribution for a value of $\alpha>1$ in the Laplace space. In this case the observable is ergodic, i.e., the limiting distribution is given by \eqref{Ptapl}. The inset shows the limiting distribution in the real space.

On the other hand, using \eqref{Qpsta22} we can compute de two first moments. 
The mean and mean square occupation time are
\begin{eqnarray}
 \left\langle T_{a}(t)\right\rangle \simeq\frac{\lambda_{\alpha}}{\Gamma\left(\frac{3-\alpha}{2-\alpha}\right)}t^{\frac{1}{2-\alpha}},\quad    \left\langle T_{a}(t)^{2}\right\rangle \simeq\frac{2\lambda_{\alpha}^{2}}{\Gamma\left(\frac{4-\alpha}{2-\alpha}\right)}t^{\frac{2}{2-\alpha}} . 
 \label{ta1pl}
\end{eqnarray}
Likewise, the ergodicity breaking parameter can be found from \eqref{EB2} and \eqref{ta1pl} to get
\begin{eqnarray}
\textrm{EB}_{a}=\frac{\left\langle T_{a}(t)^{2}\right\rangle }{\left\langle T_{a}(t)\right\rangle ^{2}}-1=\left\{ \begin{array}{cc}
\frac{2\Gamma\left(\frac{3-\alpha}{2-\alpha}\right)^{2}}{\Gamma\left(\frac{4-\alpha}{2-\alpha}\right)}-1, & \alpha<1\\
0, & 1<\alpha<2
\end{array}\right.  
\label{ebpat}
\end{eqnarray}
Note that at $\alpha=1$ there is an ergodic transition between non-ergodic phase $\alpha <1$ and ergodic phases $\alpha >1$, like for the half occupation time. The results given in \eqref{Pat} and \eqref{ebpat} have been previously obtained in Ref. \cite{LeBa19} by assuming that the PDF of $T_a$ can be given by the PDF of the number of returns to the interval $[-a,a]$. This assumption requires that the returns constitute a renewal process \cite{Ba06}, i.e., the times between successive returns are uncorrelated, which is not true for any general process.
However, we have obtained the same results without making any further assumption, but by directly solving the corresponding Feynman-Kac equation.

\begin{figure}[htbp]
    \centering
    \includegraphics[width=0.33\linewidth]{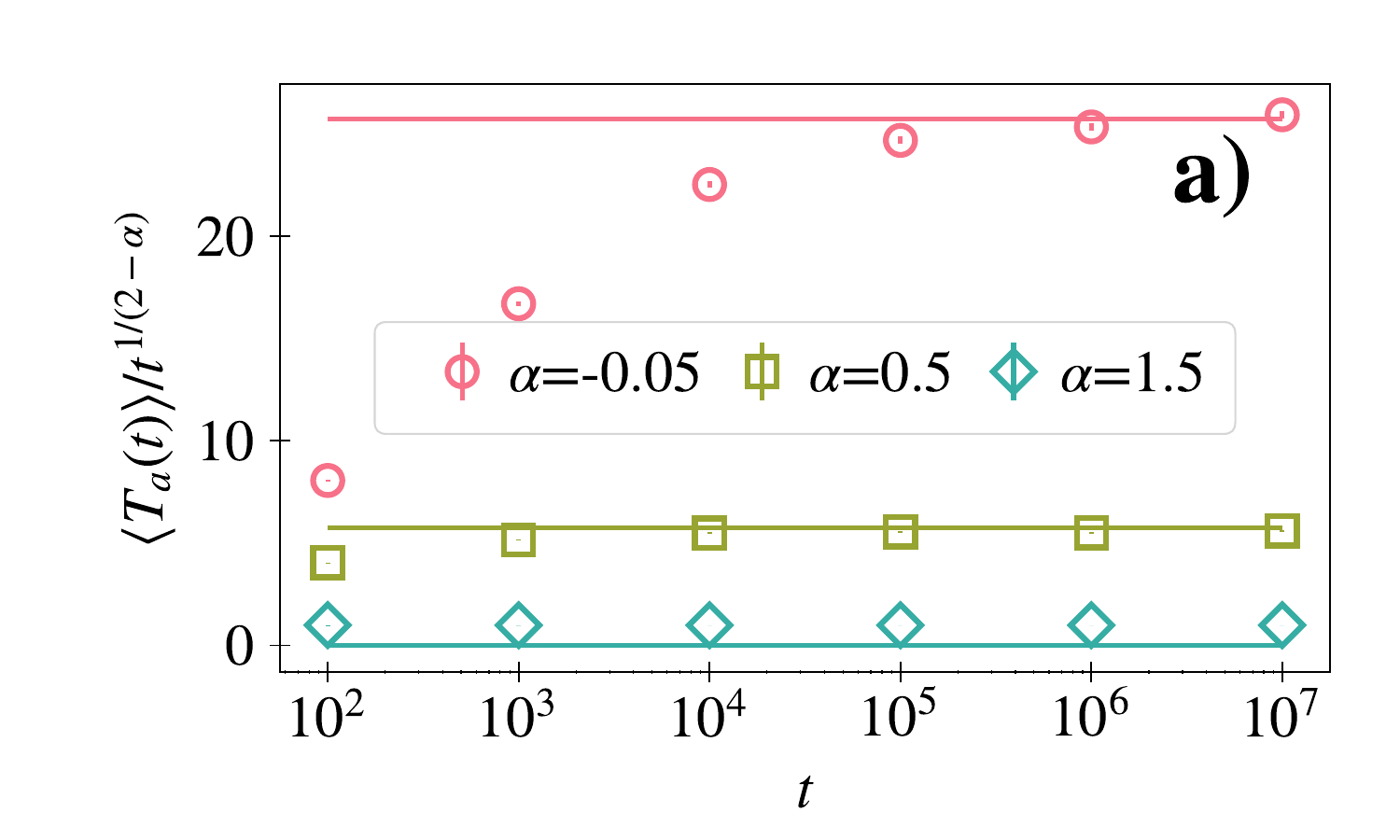}%
    \includegraphics[width=0.33\linewidth]{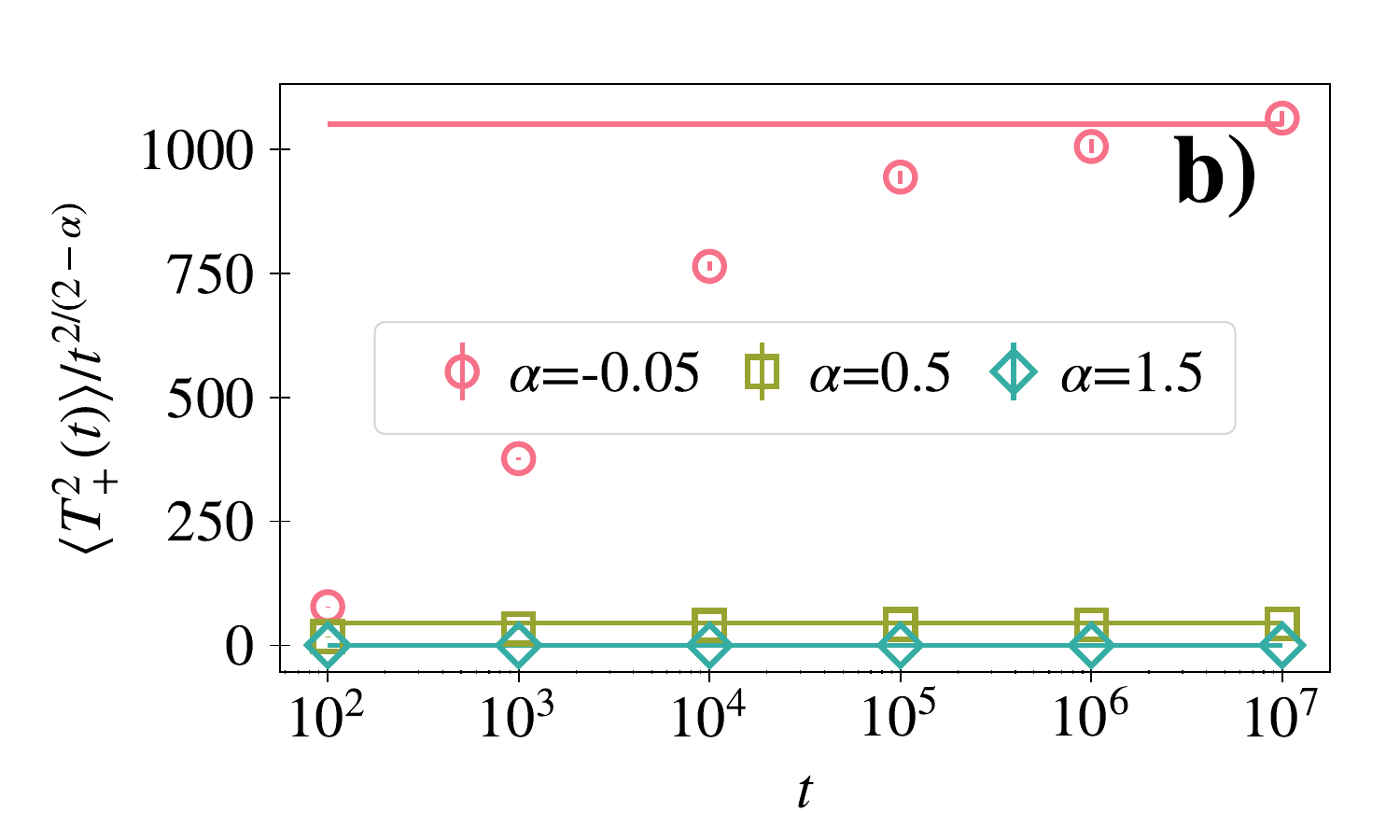}%
    \includegraphics[width=0.33\linewidth]{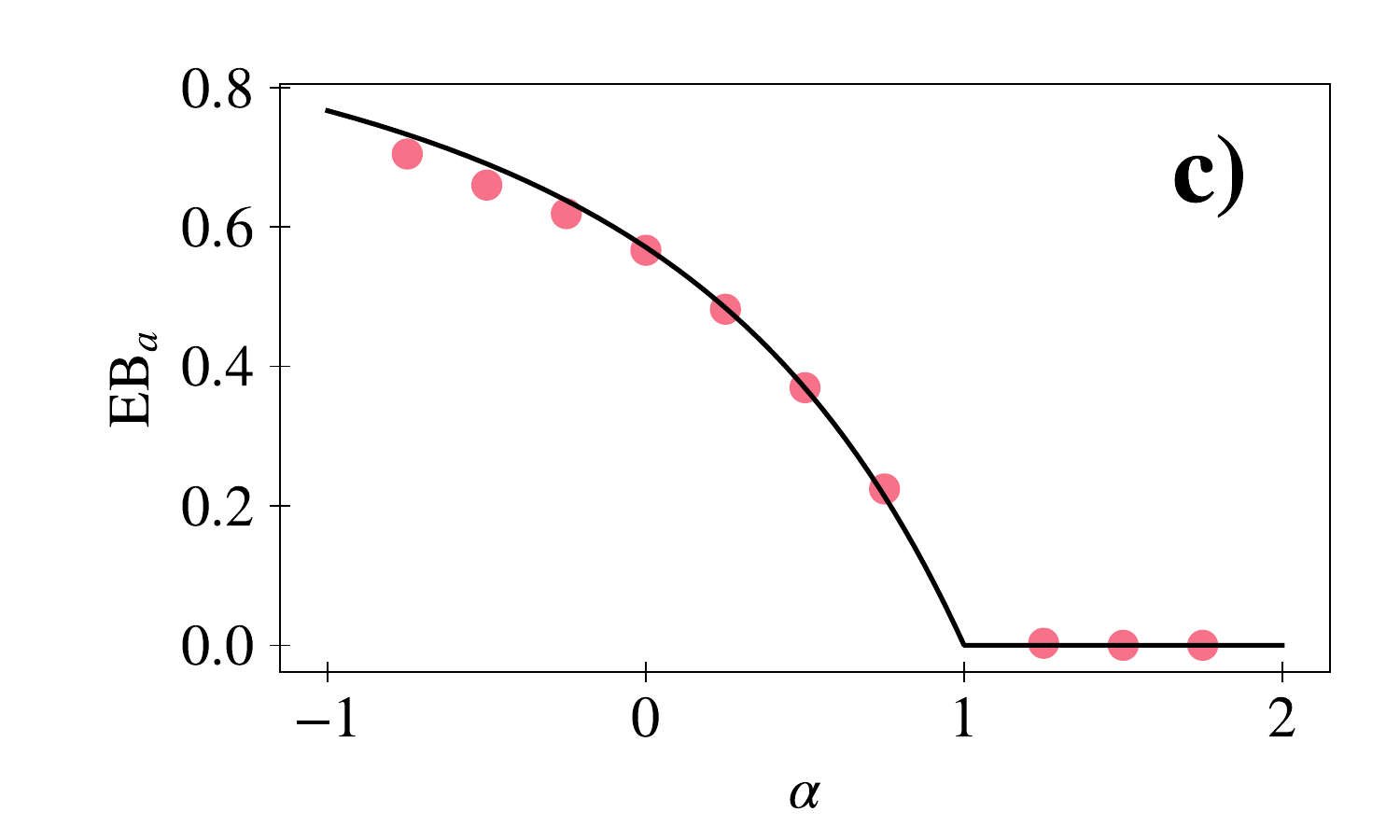}
    \caption{(a) $\langle T_{a} (t)\rangle$, (b) $\langle T_{a}(t)^2\rangle$, and (c) $\textrm{EB}_a$ of the occupation time in the interval [-20,20] for the power-law heterogeneity. In panels (a), and (b) the solid lines are computed with Eq. \eqref{ta1pl} and $\alpha=-0.05,0.5$, and $1.5$; In panel (c) the solid line is Eq. \eqref{ebpat}. We take $\alpha>0$ if $\epsilon=10^{-10}$, and $\epsilon=10^{-3}$ if $\alpha<0$. In all panels $D_{0}=1.0$, $x_0=0$, simulation timestep $dt=0.1$, and $N=10^4$ trajectories.}
    \label{fig:Pl-Ta}
\end{figure}

In Figure \ref{fig:Pl-Ta} we compare the theoretical results for the two first moments given in Eqs. \eqref{ta1pl} and \eqref{ebpat} with numerical simulations. It is worth mentioning that for $\alpha<0$ the convergence to the analytic result is slower the smaller $\alpha$ is. Note also that EB$_a$ and EB$_+$ vs $\alpha$ in panels \ref{fig:Pl-Tp}c and \ref{fig:Pl-Ta}c look qualitatively very similar and be regarded as phase diagrams of the ergodic properties.

\begin{figure}[htbp]
    \centering
    \includegraphics[width=0.33\linewidth]{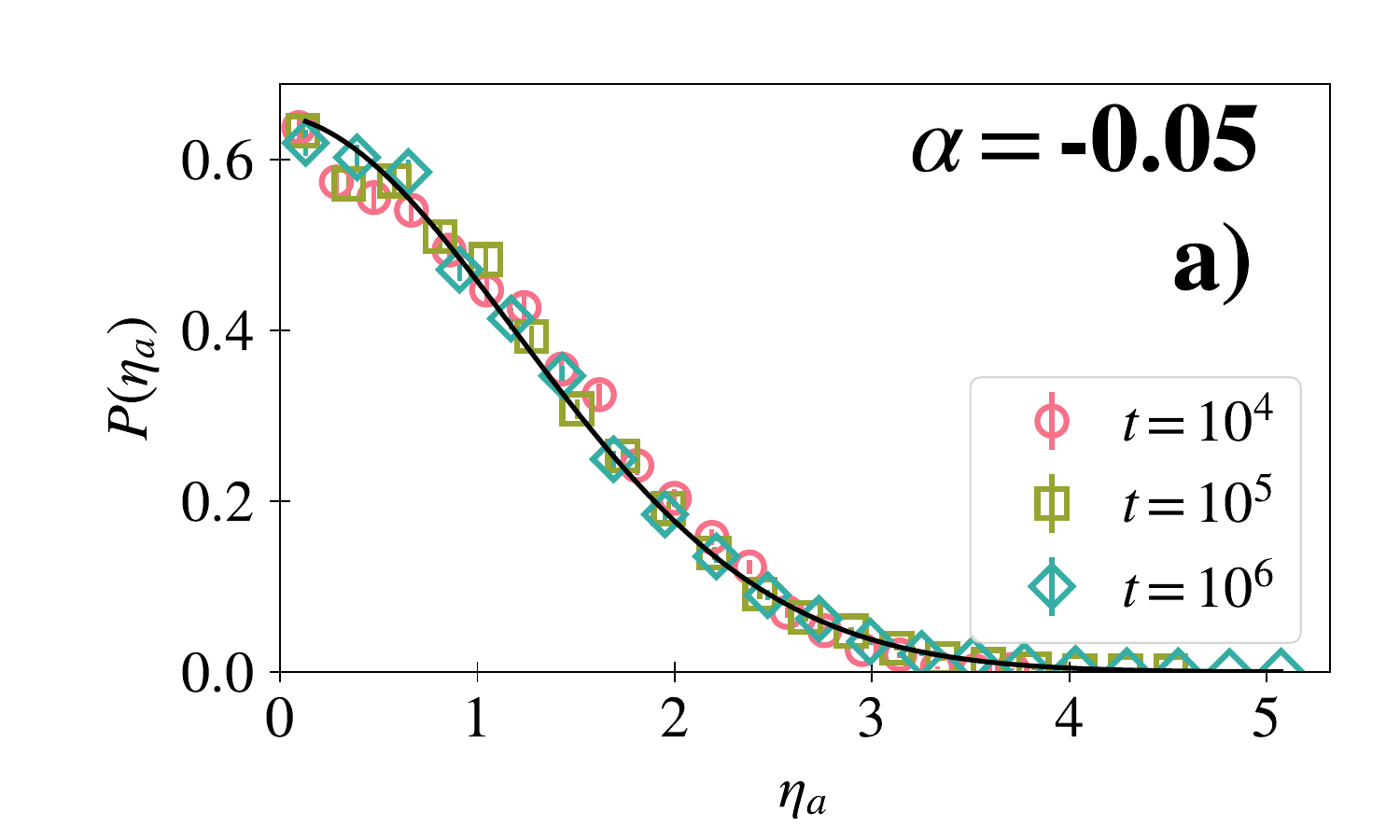}%
    \includegraphics[width=0.33\linewidth]{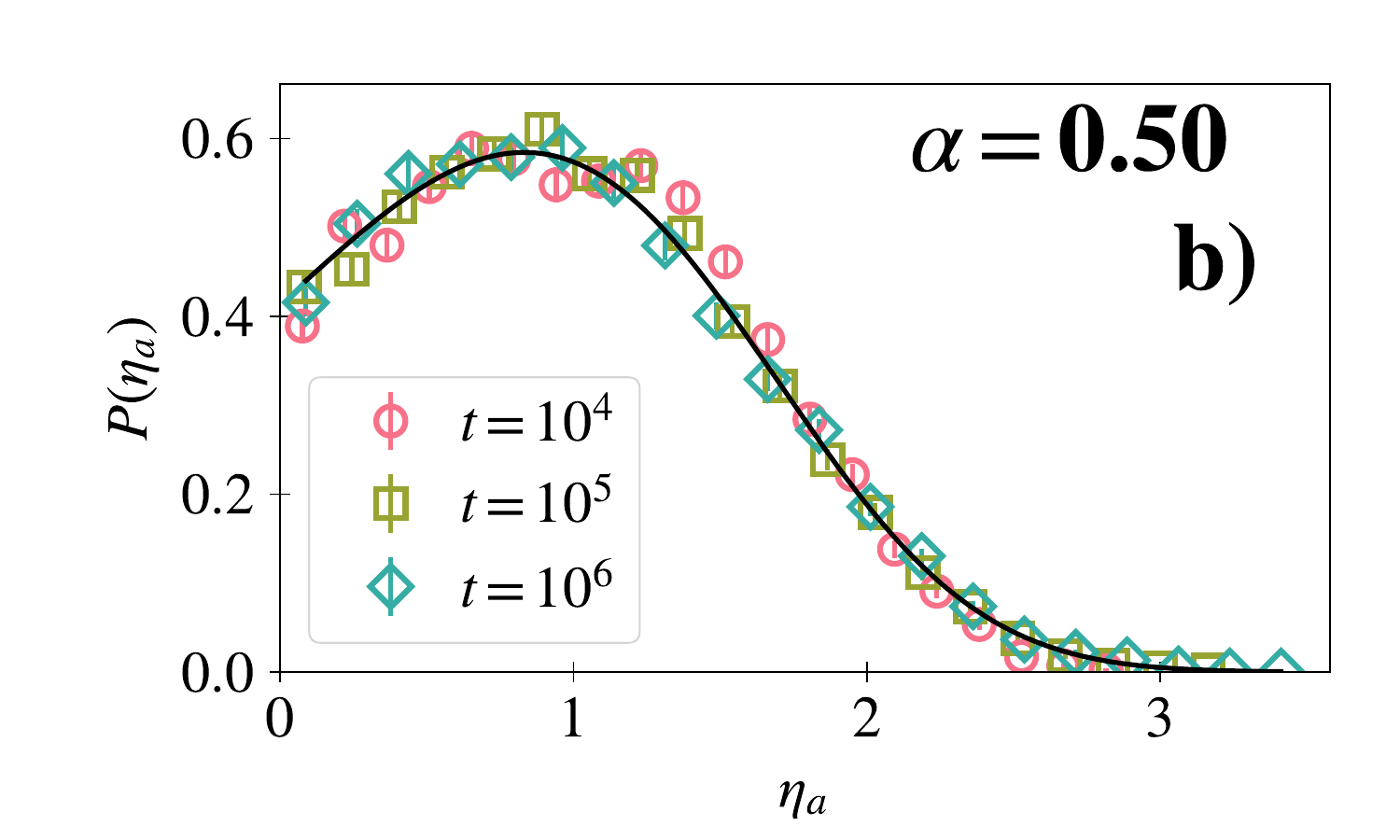}%
    \includegraphics[width=0.33\linewidth]{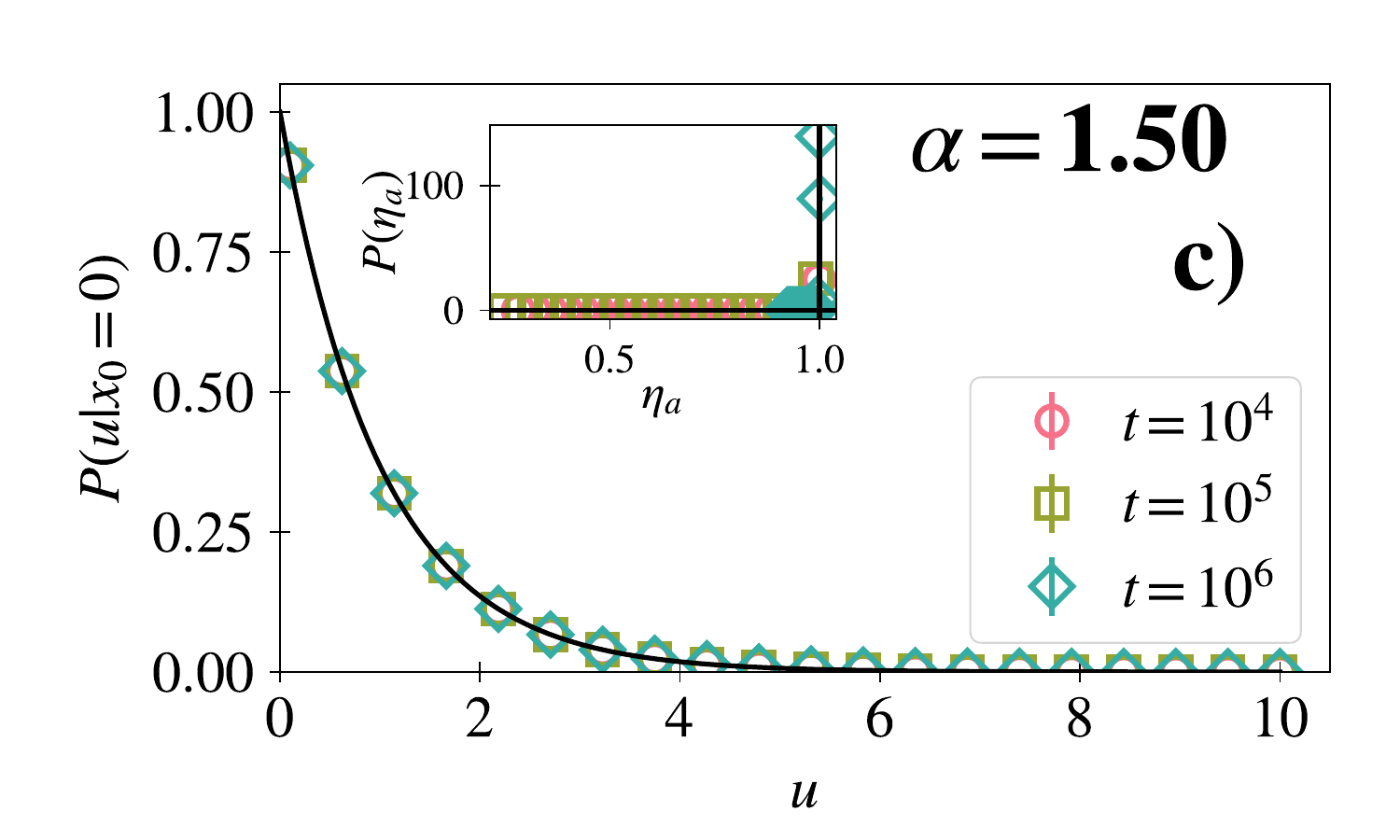}
    \caption{
    PDF of the time averaged occupation time in the interval $[-20,20]$ for the power-law heterogeneity. The different symbols are for trajectories with $t=10^4,10^5,$ and $10^6$. The black solid line corresponds to Eq. \eqref{ML} in panels (a), and (b); to $e^{-u}$ in panel (c), and to $\delta(\eta_a-1)$ in its inset. $\alpha=-0.05,0.5$, and $1.5$ in panels (a), (b), and (c), respectively. If $\alpha>0$ $\epsilon=10^{-10}$, and $\epsilon=10^{-3}$ if $\alpha<0$. In all panels $D_{0}=1.0$, $x_0=0$, simulation timestep $dt=0.1$, and $N=10^4$ trajectories.}
    \label{fig:eta_pl_Ta}
\end{figure}

Next, we compute the PDF for the time average of the observable $\theta[-a<x(t)<a]$ in the long time limit. From \eqref{etaa} and \eqref{ta12}
 $$
 \eta_{a}=\frac{T_{a}}{\lambda_{\alpha}t^{\frac{1}{2-\alpha}}}\Gamma\left(\frac{3-\alpha}{2-\alpha}\right)
 $$
and from \eqref{Pat}, we find that in the long time limit $P(\eta _a)$ is given by the Mittag-Leffler (ML) distribution of order $\beta$:
\begin{eqnarray}
    P(\eta _a)=\mathcal{M}_{\beta}(\eta_{a})=\frac{\Gamma\left(1+\beta\right)^{\frac{1}{\beta}}}{\beta\eta_{a}^{1+\frac{1}{\beta}}}l_{\beta}\left[\frac{\Gamma\left(1+\beta\right)^{\frac{1}{\beta}}}{\eta_{a}^{\frac{1}{\beta}}}\right],\quad \beta =\frac{1}{2-\alpha}
    \label{ML}
\end{eqnarray}
for $\alpha<1$. The PDF given in \eqref{ML} depends on neither $D_0$ nor $a$ but only on $\alpha$. This result is checked with numerical simulations as shown in Figure  \ref{fig:eta_pl_Ta}. For $\alpha>1$ the PDF of time averaged occupation time in the interval is $P(\eta_a)=\delta(\eta_a-1)$, according to \eqref{Ptapl}. In panels a) and b) we show two cases for $\alpha<1$ while in panel c) we show a case for $\alpha>1$ in the Laplace space. In the inset of panel c) we show the same PDF but in the real space.

\section{Conclusions}
In this work, we analyzed the ergodic properties of Brownian motion in heterogeneous media through the statistics of the occupation time. By solving the Feynman–Kac equation associated with the characteristic function of the occupation times, we obtained analytical results for two relevant forms of spatially varying diffusion coefficient: a piecewise-constant model and a power-law dependence. From these solutions, we derived the limiting distributions of the occupation times, their first two moments, and the corresponding ergodicity-breaking (EB) parameters, finding excellent agreement with numerical simulations.

For the piecewise diffusion coefficient, we showed that the probability density of the half occupation time follows an asymmetric arcsine law, with a strictly positive EB parameter. The distribution of the occupation time within an interval is half-Gaussian—analogous to the homogeneous case but with an effective diffusion coefficient—and the EB parameter matches that of standard Brownian motion, confirming non-ergodic behavior in terms of both occupation times.

For the power-law diffusion coefficient, our results reproduce the occupation-time distributions previously obtained in Refs. \cite{PS22,LeBa19}, but reveal a transition between non-ergodic and ergodic phases. As the exponent $\alpha$ increases, the system evolves from a non-ergodic regime for $\alpha<1$ to an ergodic for $\alpha>1$.

Overall, this study illustrates how the Feynman–Kac formalism provides a powerful framework to characterize ergodic properties of diffusion processes in heterogeneous media. While we focused on two specific spatial dependencies of the diffusion coefficient, the approach can be extended to more general forms \cite{ChMe20}.

\section*{Acknowledgements}
The authors acknowledge the financial support of the Ministerio de Ciencia e Innovaci\'on (Spanish government) under
Grant No. PID2021-122893NB-C22.

\appendix
\section{Derivation of Eq. \eqref{Qsptmas}}
In the region I, $x_0\in (-\infty,0]$, the equation for the characteristic function is
$$
D_{0}|x_{0}|^{\alpha}\frac{d\tilde{Q}_{I}(p,s|x_{0})}{dx_{0}^{2}}-s\tilde{Q}_{I}(p,s|x_{0})=-1
$$
while in the region II, $x_0\in [0,+\infty)$, one has
$$
D_{0}|x_{0}|^{\alpha}\frac{d\tilde{Q}_{II}(p,s|x_{0})}{dx_{0}^{2}}-(s+p)\tilde{Q}_{II}(p,s|x_{0})=-1.
$$
The above equations can be identified with a Lommel-type equation (see Eq. 8.491.11 in Ref. \cite{Gr15} )
$$\frac{d^{2}u(z)}{dz^{2}}-\omega^{2}\rho^{2}z^{2\omega-2}u(z)=0
$$
which admits the solution 
$$
u(z)=\sqrt{z}\left[c_{1}K_{\frac{1}{2\omega}}(\rho z^{\omega})+c_{2}I_{\frac{1}{2\omega}}(\rho z^{\omega})\right],
$$
where $K(\cdot)$ and $I(\cdot)$ are the modified Bessel functions.
Identifying
$$
\omega=\frac{2-\alpha}{2},\quad\rho=\frac{2}{2-\alpha}\sqrt{\frac{s}{D_{0}}}
$$
the solutions in regions I and II under the boundary conditions $\tilde{Q}(p,s|x_{0}\to\pm\infty)<\infty$ are
\begin{eqnarray}
    \tilde{Q}_{I}(p,s|x_{0})&=&\frac{1}{s}+A_{1}|x_{0}|^{1/2}K_{\frac{1}{2-\alpha}}\left(\frac{2}{2-\alpha}|x_{0}|^{1-\frac{\alpha}{2}}\sqrt{\frac{s}{D_{0}}}\right),\;x_{0}<0\nonumber\\
    \tilde{Q}_{II}(p,s|x_{0})&=&\frac{1}{s+p}+A_{2}x_{0}^{1/2}K_{\frac{1}{2-\alpha}}\left(\frac{2}{2-\alpha}x_{0}^{1-\frac{\alpha}{2}}\sqrt{\frac{s+p}{D_{0}}}\right),\;x_{0}>0.
    \label{spl}
\end{eqnarray}
To compute the constants $A_1$ and $A_2$ we proceed to match both solutions and their derivatives at $x_0=0$:
\begin{eqnarray}
    \tilde{Q}_{I}(p,s|x_{0}\to0)&=&\tilde{Q}_{II}(p,s|x_{0}\to0)\nonumber\\
    \left[\frac{\partial\tilde{Q}_{I}(p,s|x_{0})}{\partial x_{0}}\right]_{x_{0}=0}&=&\left[\frac{\partial\tilde{Q}_{II}(p,s|x_{0})}{\partial x_{0}}\right]_{x_{0}=0}
    \label{bcpl}
\end{eqnarray}
Introducing \eqref{spl} into \eqref{bcpl} and making use of the approximation of the modified Bessel function for small argument 
\begin{eqnarray}
 \lim_{x_{0}\to0}\left[x_{0}^{\mu}K_{\nu}\left(cx_{0}^{\sigma}\right)\right]=2^{\nu-1}\frac{\Gamma(\nu)}{c^{\nu}}\lim_{x_{0}\to0}\left(x_{0}^{\mu-\sigma\nu}\right)=\left\{ \begin{array}{cc}
0, & \mu>\sigma\nu\\
2^{\nu-1}\frac{\Gamma(\nu)}{c^{\nu}}, & \mu=\sigma\nu\\
\infty, & \mu<\sigma\nu
\end{array}\right.  
\label{limit}
\end{eqnarray}
one gets a system of to equations for $A_1$ and $A_2$. 
We note that continuity of the derivative is well-defined only for $\alpha <1$. For $\alpha\geq1$ the derivatives of $\tilde{Q}_I$ and $\tilde{Q}_{II}$ diverge in the limit $x_0\to 0$ as we show in turn. To compute the derivatives we need to make use of the properties
\begin{eqnarray}
\frac{dK_{\nu}(z)}{dz}=-K_{\nu-1}(z)-\frac{\nu}{z}K_{\nu}(z)   
\label{prop}
\end{eqnarray}
and $K_{-\nu}(z)=K_{\nu}(z)$. The derivatives of $\tilde{Q}$ given in Eq. \eqref{spl} are of the form
$$
\left[\frac{\partial\tilde{Q}_{I,II}(p,s|x_{0})}{\partial x_{0}}\right]_{x_{0}=0}=A_{1,2}\frac{2-\alpha}{2}c\lim_{x_{0}\to0}\left[x_{0}^{\frac{1-\alpha}{2}}K_{\frac{1-\alpha}{2-\alpha}}\left(cx_{0}^{1-\frac{\alpha}{2}}\right)\right]
$$
where $c$ is the coefficient of $x_{0}^{1-\frac{\alpha}{2}}$ in the arguments of the Bessel functions of Eq. \eqref{spl}. When $\alpha <1$ the parameter $\nu$ in \eqref{limit} is $\nu=(1-\alpha)/(2-\alpha)>0$ and $\mu=(1-\alpha)/2$, $\sigma=1-\alpha/2$. In this case, $\mu=\sigma\nu$ so that
$$
\left[\frac{\partial\tilde{Q}_{I,II}(p,s|x_{0})}{\partial x_{0}}\right]_{x_{0}=0}=A_{1,2}\frac{2-\alpha}{2^{\frac{3-\alpha}{2-\alpha}}}c^{\frac{1}{2-\alpha}}\Gamma\left(\frac{1-\alpha}{2-\alpha}\right),\;\alpha<1
$$
When $\alpha\geq 1$ we have $\nu <0$ but since $K_\nu(\cdot)=K_{-\nu}(\cdot)$ the derivative in the limit $x_0\to0$ reads
$$
\left[\frac{\partial\tilde{Q}_{I,II}(p,s|x_{0})}{\partial x_{0}}\right]_{x_{0}=0}=A_{1,2}\frac{2-\alpha}{2}c\lim_{x_{0}\to0}\left[x_{0}^{\frac{1-\alpha}{2}}K_{\frac{\alpha-1}{2-\alpha}}\left(cx_{0}^{1-\frac{\alpha}{2}}\right)\right]=\infty,\;\alpha>1.
$$
Finally, plugging $A_1$ and $B_1$ in Eq. \eqref{spl} one finds \eqref{Qsptmas}.

\section{Derivation of Eq. \eqref{Qtaph}}
The solution of \eqref{ode2} for the piecewise diffusion coefficient can be written
as
\begin{eqnarray}
    \tilde{Q}_{I}(p,s|x_{0})&=&\frac{1}{s}+C_{1}e^{x_{0}\sqrt{\frac{s}{D_{-}}}},\nonumber\\
    \tilde{Q}_{II}(p,s|x_{0})&=&\frac{1}{s+p}+C_{2}e^{-x_{0}\sqrt{\frac{s+p}{D_{-}}}}+C_{3}e^{x_{0}\sqrt{\frac{s+p}{D_{-}}}},\nonumber\\
\tilde{Q}_{III}(p,s|x_{0})&=&\frac{1}{s+p}+C_{4}e^{-x_{0}\sqrt{\frac{s+p}{D_{+}}}}+C_{5}e^{x_{0}\sqrt{\frac{s+p}{D_{+}}}},\nonumber\\
\tilde{Q}_{IV}(p,s|x_{0})&=&\frac{1}{s}+C_{6}e^{-x_{0}\sqrt{\frac{s}{D_{+}}}},\;x_{0}>a.
\end{eqnarray}
The constants $C_i$ with $i=1,...,6$ can be found from the matching conditions
\begin{eqnarray}
\tilde{Q}_{I}(p,s|x_{0}=-a)&=&\tilde{Q}_{II}(p,s|x_{0}=-a)\nonumber\\
\tilde{Q}_{II}(p,s|x_{0}=0)&=&\tilde{Q}_{III}(p,s|x_{0}=0)\nonumber\\
\tilde{Q}_{III}(p,s|x_{0}=a)&=&\tilde{Q}_{IV}(p,s|x_{0}=a)\nonumber\\
\left[\frac{\partial\tilde{Q}_{I}(p,s|x_{0})}{\partial x_{0}}\right]_{x_{0}=-a}&=&\left[\frac{\partial\tilde{Q}_{II}(p,s|x_{0})}{\partial x_{0}}\right]_{x_{0}=-a}\nonumber\\
\left[\frac{\partial\tilde{Q}_{II}(p,s|x_{0})}{\partial x_{0}}\right]_{x_{0}=0}&=&\left[\frac{\partial\tilde{Q}_{III}(p,s|x_{0})}{\partial x_{0}}\right]_{x_{0}=0}\nonumber\\
\left[\frac{\partial\tilde{Q}_{III}(p,s|x_{0})}{\partial x_{0}}\right]_{x_{0}=a}&=&\left[\frac{\partial\tilde{Q}_{IV}(p,s|x_{0})}{\partial x_{0}}\right]_{x_{0}=a}
    \label{bcpl2}
\end{eqnarray}
 Solving the system of algebraic equations for $C_i$ we finally obtain the solution for the characteristic equation given by Eq. \eqref{Qtaph}.

\section{Derivation of Eq. \eqref{Qpsta2}}
Following the same procedure as in Appendix A, the solution of Eq. \eqref{ode2} for the power law diffusion coefficient \eqref{dpl} is given by 
\begin{eqnarray}
    \tilde{Q}_{I}(p,s|x_{0})&=&\frac{1}{s}+A_{1}|x_{0}|^{1/2}K_{\frac{1}{2-\alpha}}\left(\frac{2}{2-\alpha}|x_{0}|^{1-\frac{\alpha}{2}}\sqrt{\frac{s}{D_{0}}}\right),\nonumber\\
\tilde{Q}_{II}(p,s|x_{0})&=&\frac{1}{s+p}+A_{2}|x_{0}|^{1/2}K_{\frac{1}{2-\alpha}}\left(\frac{2}{2-\alpha}|x_{0}|^{1-\frac{\alpha}{2}}\sqrt{\frac{s+p}{D_{0}}}\right)\nonumber\\
&+&A_{3}|x_{0}|^{1/2}I_{\frac{1}{2-\alpha}}\left(\frac{2}{2-\alpha}|x_{0}|^{1-\frac{\alpha}{2}}\sqrt{\frac{s+p}{D_{0}}}\right),\nonumber\\
\tilde{Q}_{III}(p,s|x_{0})&=&\frac{1}{s+p}+A_{4}x_{0}^{1/2}K_{\frac{1}{2-\alpha}}\left(\frac{2}{2-\alpha}x_{0}{}^{1-\frac{\alpha}{2}}\sqrt{\frac{s+p}{D_{0}}}\right)\nonumber\\
&+&A_{5}x_{0}{}^{1/2}I_{\frac{1}{2-\alpha}}\left(\frac{2}{2-\alpha}x_{0}{}^{1-\frac{\alpha}{2}}\sqrt{\frac{s+p}{D_{0}}}\right),\nonumber\\
\tilde{Q}_{IV}(p,s|x_{0})&=&\frac{1}{s}+A_{6}x_{0}^{1/2}|K_{\frac{1}{2-\alpha}}\left(\frac{2}{2-\alpha}x_{0}{}^{1-\frac{\alpha}{2}}\sqrt{\frac{s}{D_{0}}}\right).
\end{eqnarray}
The constant $A_i$ with $i=1,...,6$ can be found using the matching conditions \eqref{bcpl2}. At the boundary $x_0=0$ we employ the property \eqref{limit}. The system of equations for the constants is
\begin{eqnarray}
    A_{2}\varphi(s+p,a)+A_{3}\phi(s+p,a)-A_{1}\varphi(s,a)=\frac{p}{s(s+p)}\nonumber\\
A_{4}\varphi(s+p,a)+A_{5}\phi(s+p,a)-A_{6}\varphi(s,a)=\frac{p}{s(s+p)}\nonumber\\
A_{2}=A_{4}\nonumber\\
A_{1}\left[\frac{d\varphi(s,x_{0})}{dx_{0}}\right]_{x_{0}=a}=A_{2}\left[\frac{d\varphi(s+p,x_{0})}{dx_{0}}\right]_{x_{0}=a}+A_{3}\left[\frac{d\phi(s+p,x_{0})}{dx_{0}}\right]_{x_{0}=a}\nonumber\\
(A_{2}+A_{4})\left[\frac{d\varphi(s+p,x_{0})}{dx_{0}}\right]_{x_{0}=0}+(A_{3}+A_{5})\left[\frac{d\phi(s+p,x_{0})}{dx_{0}}\right]_{x_{0}=0}=0\nonumber\\
A_{4}\left[\frac{d\varphi(s+p,x_{0})}{dx_{0}}\right]_{x_{0}=a}+A_{5}\left[\frac{d\phi(s+p,x_{0})}{dx_{0}}\right]_{x_{0}=a}=A_{6}\left[\frac{d\varphi(s,x_{0})}{dx_{0}}\right]_{x_{0}=a}
\label{As}
\end{eqnarray}
where we have defined the functions
\begin{eqnarray}
    \varphi(s,x_{0})=\sqrt{x_{0}}K_{\frac{1}{2-\alpha}}\left(\frac{2}{2-\alpha}x_{0}{}^{1-\frac{\alpha}{2}}\sqrt{\frac{s}{D_{0}}}\right)\nonumber\\
\phi(s,x_{0})=\sqrt{x_{0}}I_{\frac{1}{2-\alpha}}\left(\frac{2}{2-\alpha}x_{0}{}^{1-\frac{\alpha}{2}}\sqrt{\frac{s}{D_{0}}}\right)
\end{eqnarray}
The derivatives in Eqs. \eqref{As} evaluated at $x_0=0$ are divergent unless $\alpha<1$. Computing the derivatives using the properties of the Bessel functions using \eqref{prop} and evaluating them at the boundaries $x_0=0$ and $x_0=a$ we solve the system of equations for the constants and find
 \begin{eqnarray}
    \tilde{Q}(p,s|0)=\frac{1}{s+p}\left[1+\frac{p}{s}\frac{\mathcal{H}_{1}(s,p)}{\mathcal{H}_{2}(s,p)}\right]
     \label{Qtaph2}
 \end{eqnarray}
after setting $x_0=0$. The functions $\mathcal{H}_{1}(s,p)$ and $\mathcal{H}_{2}(s,p)$ are
\begin{eqnarray}
    \mathcal{H}_{1}(s,p)=\frac{\Gamma\left(\frac{1}{2-\alpha}\right)}{2\left(\frac{a^{1-\frac{\alpha}{2}}}{2-\alpha}\sqrt{\frac{s+p}{D_{0}}}\right)^{\frac{1}{2-\alpha}}}\quad\textrm{and}\quad \mathcal{H}_{2}(s,p)=\psi_{1}-\psi_{2}\psi_{3}
    \label{Hs}
\end{eqnarray}
where
\begin{eqnarray}
    \psi_{1}&=&K_{\frac{1}{2-\alpha}}\left(\frac{2}{2-\alpha}a{}^{1-\frac{\alpha}{2}}\sqrt{\frac{s+p}{D_{0}}}\right)+\sigma_{\alpha}I_{\frac{1}{2-\alpha}}\left(\frac{2}{2-\alpha}a{}^{1-\frac{\alpha}{2}}\sqrt{\frac{s+p}{D_{0}}}\right)\nonumber\\
\psi_{2}&=&\frac{2}{2-\alpha}a{}^{1-\frac{\alpha}{2}}\sqrt{\frac{s+p}{D_{0}}}\left[K_{\frac{1-\alpha}{2-\alpha}}\left(\frac{2}{2-\alpha}a{}^{1-\frac{\alpha}{2}}\sqrt{\frac{s+p}{D_{0}}}\right)-\sigma_{\alpha}I_{\frac{3-\alpha}{2-\alpha}}\left(\frac{2}{2-\alpha}a{}^{1-\frac{\alpha}{2}}\sqrt{\frac{s+p}{D_{0}}}\right)\right]\nonumber\\
&-&\frac{2\sigma_{\alpha}}{2-\alpha}I_{\frac{1}{2-\alpha}}\left(\frac{2}{2-\alpha}a{}^{1-\frac{\alpha}{2}}\sqrt{\frac{s+p}{D_{0}}}\right)\nonumber\\
\psi_{3}&=&\frac{K_{\frac{1}{2-\alpha}}\left(\frac{2}{2-\alpha}a{}^{1-\frac{\alpha}{2}}\sqrt{\frac{s}{D_{0}}}\right)}{\frac{2}{2-\alpha}a{}^{1-\frac{\alpha}{2}}\sqrt{\frac{s}{D_{0}}}K_{\frac{1-\alpha}{2-\alpha}}\left(\frac{2}{2-\alpha}a{}^{1-\frac{\alpha}{2}}\sqrt{\frac{s}{D_{0}}}\right)}
\label{psis}
\end{eqnarray}

and 
$$
\sigma_{\alpha}=\frac{\pi/2}{\sin\left(\pi\frac{1-\alpha}{2-\alpha}\right)}.
$$
Before proceeding to perform the double Laplace inversion we consider the long time limit for which we consider $s$ and $p$ very small and comparable. In this limit, the arguments of the Bessel functions in Eq. \eqref{psis} are small and to approximate $\psi_1$ and $\psi_3$ we can make use of  \eqref{limit} and 
$$
I_\nu (z)\simeq \frac{1}{\Gamma\left(1+\nu\right)}\left(\frac{z}{2}\right)^{\nu}\quad \textrm{as}\quad z\to 0.
$$
However, to approximate $\psi_2$ we need to go to the next order in the expansion of $K_\nu (\cdot)$ for small argument so that we have to consider
$$
K_{\nu}(z)\simeq\frac{\Gamma\left(\nu\right)\Gamma\left(1-\nu\right)}{2}\left[\frac{1}{\Gamma\left(1-\nu\right)}\left(\frac{z}{2}\right)^{-\nu}-\frac{1}{\Gamma\left(1+\nu\right)}\left(\frac{z}{2}\right)^{\nu}\right]
$$
for $\nu<1$. Under these considerations we find
\begin{eqnarray*}
    \psi_{1}&\simeq&\frac{1}{2}\Gamma\left(\frac{1}{2-\alpha}\right)\left(\frac{a^{1-\frac{\alpha}{2}}}{2-\alpha}\sqrt{\frac{s+p}{D_{0}}}\right)^{-\frac{1}{2-\alpha}}\nonumber\\
\psi_{2}&\simeq&-\frac{2-\alpha}{1-\alpha}\Gamma\left(\frac{1}{2-\alpha}\right)\left(\frac{a^{1-\frac{\alpha}{2}}}{2-\alpha}\sqrt{\frac{s+p}{D_{0}}}\right)^{\frac{3-2\alpha}{2-\alpha}}\nonumber\\
\psi_{3}&\simeq& \frac{\Gamma\left(\frac{1}{2-\alpha}\right)}{2\Gamma\left(\frac{1-\alpha}{2-\alpha}\right)}\left(\frac{a^{1-\frac{\alpha}{2}}}{2-\alpha}\sqrt{\frac{s}{D_{0}}}\right)^{-\frac{2}{2-\alpha}}
\end{eqnarray*}
and finally, from \eqref{Qtaph2} we find the expression for the characteristic function \eqref{Qpsta2} in the long time limit for both $T_a$ and $t$.

\bibliography{main}

@article{GoLu01,
author = {Godreche, C. and Luck, Jean-Marc},
year = {2001},
month = {11},
pages = {489–524},
title = {Statistics of the Occupation Time of Renewal Processes},
volume = {104},
journal = {Journal of Statistical Physics},
doi = {10.1023/A:1010364003250}
}

@article{Ma05,
  TITLE = {{Brownian Functionals in Physics and Computer Science}},
  AUTHOR = {Majumdar, Satya N.},
  URL = {https://hal.archives-ouvertes.fr/hal-00165789},
  JOURNAL = {{Current Science}},
  PUBLISHER = {{Indian Academy of Sciences}},
  VOLUME = {89},
  PAGES = {2076},
  YEAR = {2005},
  HAL_ID = {hal-00165789},
  HAL_VERSION = {v1},
}

@article{Le40,
author = {L\'evy, Paul},
journal = {Compositio Mathematica},
pages = {283-339},
publisher = {Johnson Reprint Corporation},
title = {Sur certains processus stochastiques homog\`enes},
url = {http://eudml.org/doc/88744},
volume = {7},
year = {1940},
}

@article{La58,
	author = {John Lamperti},
	doi = {10.1090/s0002-9947-1958-0094863-x},
	journal = {Transactions of the American Mathematical Society},
	number = {2},
	pages = {380--387},
	publisher = {American Mathematical Society ({AMS})},
	title = {An occupation time theorem for a class of stochastic processes},
	url = {https://doi.org/10.1090%2Fs0002-9947-1958-0094863-x},
	volume = {88},
	year = 1958}

@article{CaBa10,
	author = {Carmi, Shai and Turgeman, Lior and Barkai, Eli},
	date = {2010/12/01},
	doi = {10.1007/s10955-010-0086-6},
	id = {Carmi2010},
	isbn = {1572-9613},
	journal = {Journal of Statistical Physics},
	number = {6},
	pages = {1071--1092},
	title = {On Distributions of Functionals of Anomalous Diffusion Paths},
	url = {https://doi.org/10.1007/s10955-010-0086-6},
	volume = {141},
	year = {2010}}

@article{Ba06,
	author = {Barkai, E. },
	date = {2006/05/01},
	doi = {10.1007/s10955-006-9109-8},
	id = {Barkai2006},
	isbn = {1572-9613},
	journal = {Journal of Statistical Physics},
	number = {4},
	pages = {883--907},
	title = {Residence Time Statistics for Normal and Fractional Diffusion in a Force Field},
	url = {https://doi.org/10.1007/s10955-006-9109-8},
	volume = {123},
	year = {2006}}

@article{Me14,
	author = {Metzler, Ralf and Jeon, Jae-Hyung and Cherstvy, Andrey G. and Barkai, Eli},
	doi = {10.1039/C4CP03465A},
	issue = {44},
	journal = {Phys. Chem. Chem. Phys.},
	pages = {24128-24164},
	publisher = {The Royal Society of Chemistry},
	title = {Anomalous diffusion models and their properties: non-stationarity{,} non-ergodicity{,} and ageing at the centenary of single particle tracking},
	url = {http://dx.doi.org/10.1039/C4CP03465A},
	volume = {16},
	year = {2014}}

@article{BaFlMe23,
  title = {Ergodic properties of Brownian motion under stochastic resetting},
  author = {Barkai, E. and Flaquer-Galm\'es, R. and M\'endez, V.},
  journal = {Phys. Rev. E},
  volume = {108},
  issue = {6},
  pages = {064102},
  numpages = {20},
  year = {2023},
  month = {Dec},
  publisher = {American Physical Society},
  doi = {10.1103/PhysRevE.108.064102},
  url = {https://link.aps.org/doi/10.1103/PhysRevE.108.064102}
}

@article{De18,
  title = {Feynman-Kac equation revisited},
  author = {Wang, Xudong and Chen, Yao and Deng, Weihua},
  journal = {Phys. Rev. E},
  volume = {98},
  issue = {5},
  pages = {052114},
  numpages = {13},
  year = {2018},
  month = {Nov},
  publisher = {American Physical Society},
  doi = {10.1103/PhysRevE.98.052114},
  url = {https://link.aps.org/doi/10.1103/PhysRevE.98.052114}
}

@article{Sr07,
  title = {Non-Markovian L\'evy diffusion in nonhomogeneous media},
  author = {Srokowski, T.},
  journal = {Phys. Rev. E},
  volume = {75},
  issue = {5},
  pages = {051105},
  numpages = {8},
  year = {2007},
  month = {May},
  publisher = {American Physical Society},
  doi = {10.1103/PhysRevE.75.051105},
  url = {https://link.aps.org/doi/10.1103/PhysRevE.75.051105}
}

@article{LaBa01,
	author = {{Lan{\c c}on, P.} and {Batrouni, G.} and {Lobry, L.} and {Ostrowsky, N.}},
	doi = {10.1209/epl/i2001-00103-6},
	journal = {Europhys. Lett.},
	number = 1,
	pages = {28-34},
	title = {Drift without flux: Brownian walker with a space-dependent diffusion coefficient},
	url = {https://doi.org/10.1209/epl/i2001-00103-6},
	volume = 54,
	year = 2001}

@article{PeJi16,
	author = {Pe{\v s}ek, Ji{\v r}{\'\i} and Baerts, Pieter and Smeets, Bart and Maes, Christian and Ramon, Herman},
	doi = {10.1039/C5SM03106K},
	issue = {14},
	journal = {Soft Matter},
	pages = {3360-3387},
	publisher = {The Royal Society of Chemistry},
	title = {Mathematical model suitable for efficient simulation of thin semi-flexible polymers in complex environments},
	url = {http://dx.doi.org/10.1039/C5SM03106K},
	volume = {12},
	year = {2016}}

@article{ReSh16,
  title = {Isothermal Langevin dynamics in systems with power-law spatially dependent friction},
  author = {Regev, Shaked and Gr\o{}nbech-Jensen, Niels and Farago, Oded},
  journal = {Phys. Rev. E},
  volume = {94},
  issue = {1},
  pages = {012116},
  numpages = {6},
  year = {2016},
  month = {Jul},
  publisher = {American Physical Society},
  doi = {10.1103/PhysRevE.94.012116},
  url = {https://link.aps.org/doi/10.1103/PhysRevE.94.012116}
}

@book{Ga04,
  address = {Berlin},
  author = {Gardiner, C. W.},
  edition = {Third},
  isbn = {3-540-20882-8},
  mrclass = {00A69 (60-01 60Hxx 60Jxx 82C31)},
  mrnumber = {2053476 (2004m:00008)},
  pages = {xviii+415},
  publisher = {Springer-Verlag},
  series = {Springer Series in Synergetics},
  title = {Handbook of stochastic methods for physics, chemistry and the
              natural sciences},
  volume = 13,
  year = 2004
}

@book{BaBo01, 
place={Cambridge}, title={Lévy Statistics and Laser Cooling: How Rare Events Bring Atoms to Rest}, publisher={Cambridge University Press}, author={Bardou, François and Bouchaud, Jean-Philippe and Aspect, Alain and Cohen-Tannoudji, Claude}, year={2001}}

@article{FaGr16,
	author = {Farago, Oded and Gr{\o}nbech-Jensen, Niels},
	doi = {10.1063/1.4942114},
	eprint = {https://pubs.aip.org/aip/jcp/article-pdf/doi/10.1063/1.4942114/15508460/084102\_1\_online.pdf},
	issn = {0021-9606},
	journal = {The Journal of Chemical Physics},
	month = {02},
	number = {8},
	pages = {084102},
	title = {On the connection between dissipative particle dynamics and the It{\^o}-Stratonovich dilemma},
	url = {https://doi.org/10.1063/1.4942114},
	volume = {144},
	year = {2016}}

@article{Br00,
  title = {Random walks in logarithmic and power-law potentials, nonuniversal persistence, and vortex dynamics in the two-dimensional $\mathrm{XY}$ model},
  author = {Bray, A. J.},
  journal = {Phys. Rev. E},
  volume = {62},
  issue = {1},
  pages = {103--112},
  numpages = {0},
  year = {2000},
  month = {Jul},
  publisher = {American Physical Society},
  doi = {10.1103/PhysRevE.62.103},
  url = {https://link.aps.org/doi/10.1103/PhysRevE.62.103}
}

@article{PiHe16,
	author = {Pieprzyk, S. and Heyes, D. M. and Bra{\'n}ka, A. C.},
	doi = {10.1063/1.4964935},
	eprint = {https://pubs.aip.org/aip/bmf/article-pdf/doi/10.1063/1.4964935/14593775/054118\_1\_online.pdf},
	issn = {1932-1058},
	journal = {Biomicrofluidics},
	month = {10},
	number = {5},
	pages = {054118},
	title = {Spatially dependent diffusion coefficient as a model for pH sensitive microgel particles in microchannels},
	url = {https://doi.org/10.1063/1.4964935},
	volume = {10},
	year = {2016}}

@article{BeMa17,
	author = {Berezhkovskii, Alexander M. and Makarov, Dmitrii E.},
	doi = {10.1063/1.5006456},
	eprint = {https://pubs.aip.org/aip/jcp/article-pdf/doi/10.1063/1.5006456/13771145/201102\_1\_online.pdf},
	issn = {0021-9606},
	journal = {The Journal of Chemical Physics},
	month = {11},
	number = {20},
	pages = {201102},
	title = {Communication: Coordinate-dependent diffusivity from single molecule trajectories},
	url = {https://doi.org/10.1063/1.5006456},
	volume = {147},
	year = {2017}}

@article{Ku11,
	author = {K{\"u}hn, Thomas AND Ihalainen, Teemu O. AND Hyv{\"a}luoma, Jari AND Dross, Nicolas AND Willman, Sami F. AND Langowski, J{\"o}rg AND Vihinen-Ranta, Maija AND Timonen, Jussi},
	doi = {10.1371/journal.pone.0022962},
	journal = {PLOS ONE},
	month = {08},
	number = {8},
	pages = {1-12},
	publisher = {Public Library of Science},
	title = {Protein Diffusion in Mammalian Cell Cytoplasm},
	url = {https://doi.org/10.1371/journal.pone.0022962},
	volume = {6},
	year = {2011}}

@article{KaAl09,
	author = {B Kaulakys and M Alaburda},
	doi = {10.1088/1742-5468/2009/02/P02051},
	journal = {Journal of Statistical Mechanics: Theory and Experiment},
	month = {feb},
	number = {02},
	pages = {P02051},
	title = {Modeling scaled processes and $1/f^\beta$ noise using nonlinear stochastic differential equations},
	url = {https://dx.doi.org/10.1088/1742-5468/2009/02/P02051},
	volume = {2009},
	year = {2009}}

@article{Che13,
	author = {Andrey G Cherstvy and Aleksei V Chechkin and Ralf Metzler},
	doi = {10.1088/1367-2630/15/8/083039},
	journal = {New Journal of Physics},
	month = {aug},
	number = {8},
	pages = {083039},
	publisher = {IOP Publishing},
	title = {Anomalous diffusion and ergodicity breaking in heterogeneous diffusion processes},
	url = {https://dx.doi.org/10.1088/1367-2630/15/8/083039},
	volume = {15},
	year = {2013}}

@article{LeBa19,
  title = {Infinite ergodic theory for heterogeneous diffusion processes},
  author = {Leibovich, N. and Barkai, E.},
  journal = {Phys. Rev. E},
  volume = {99},
  issue = {4},
  pages = {042138},
  numpages = {15},
  year = {2019},
  month = {Apr},
  publisher = {American Physical Society},
  doi = {10.1103/PhysRevE.99.042138},
  url = {https://link.aps.org/doi/10.1103/PhysRevE.99.042138}
}

@article{Radice23,
  title = {First-passage functionals of Brownian motion in logarithmic potentials and heterogeneous diffusion},
  author = {Radice, Mattia},
  journal = {Phys. Rev. E},
  volume = {108},
  issue = {4},
  pages = {044151},
  numpages = {18},
  year = {2023},
  month = {Oct},
  publisher = {American Physical Society},
  doi = {10.1103/PhysRevE.108.044151},
  url = {https://link.aps.org/doi/10.1103/PhysRevE.108.044151}
}

@article{Si22,
  title = {Extreme value statistics and arcsine laws for heterogeneous diffusion processes},
  author = {Singh, Prashant},
  journal = {Phys. Rev. E},
  volume = {105},
  issue = {2},
  pages = {024113},
  numpages = {12},
  year = {2022},
  month = {Feb},
  publisher = {American Physical Society},
  doi = {10.1103/PhysRevE.105.024113},
  url = {https://link.aps.org/doi/10.1103/PhysRevE.105.024113}
}

@article{WaDeCh19,
	author = {Wang, Xudong and Deng, Weihua and Chen, Yao},
	doi = {10.1063/1.5090594},
    issn = {0021-9606},
	journal = {The Journal of Chemical Physics},
	month = {04},
	number = {16},
	pages = {164121},
	title = {Ergodic properties of heterogeneous diffusion processes in a potential well},
	url = {https://doi.org/10.1063/1.5090594},
	volume = {150},
	year = {2019}}

@article{BrLa17,
  title = {Temporal disorder as a mechanism for spatially heterogeneous diffusion},
  author = {Bressloff, Paul C. and Lawley, Sean D.},
  journal = {Phys. Rev. E},
  volume = {95},
  issue = {6},
  pages = {060101},
  numpages = {5},
  year = {2017},
  month = {Jun},
  publisher = {American Physical Society},
  doi = {10.1103/PhysRevE.95.060101},
  url = {https://link.aps.org/doi/10.1103/PhysRevE.95.060101}
}

@article{PS22,
  title = {Extreme value statistics and arcsine laws for heterogeneous diffusion processes},
  author = {Singh, Prashant},
  journal = {Phys. Rev. E},
  volume = {105},
  issue = {2},
  pages = {024113},
  numpages = {12},
  year = {2022},
  month = {Feb},
  publisher = {American Physical Society},
  doi = {10.1103/PhysRevE.105.024113},
  url = {https://link.aps.org/doi/10.1103/PhysRevE.105.024113}
}

@article{Pa24,
  title = {Langevin Equation in Heterogeneous Landscapes: How to Choose the Interpretation},
  author = {Pacheco-Pozo, Adrian and Balcerek, Micha\l{} and Wy\l{}omanska, Agnieszka and Burnecki, Krzysztof and Sokolov, Igor M. and Krapf, Diego},
  journal = {Phys. Rev. Lett.},
  volume = {133},
  issue = {6},
  pages = {067102},
  numpages = {7},
  year = {2024},
  month = {Aug},
  publisher = {American Physical Society},
  doi = {10.1103/PhysRevLett.133.067102},
  url = {https://link.aps.org/doi/10.1103/PhysRevLett.133.067102}
}

@article{ChMe20,
	author = {Cherstvy, Andrey G. and Metzler, Ralf},
	doi = {10.1039/C3CP53056F},
	issue = {46},
	journal = {Phys. Chem. Chem. Phys.},
	pages = {20220-20235},
	publisher = {The Royal Society of Chemistry},
	title = {Population splitting{,} trapping{,} and non-ergodicity in heterogeneous diffusion processes},
	url = {http://dx.doi.org/10.1039/C3CP53056F},
	volume = {15},
	year = {2013}}

@article{vM05,
	Author = {van Milligen, B. P. and Bons, P. D. and Carreras, B. A. and Sanchez, R.},
	Date = {SEP 2005},
	Doi = {10.1088/0143-0807/26/5/023},
	Isi = {WOS:000232321900033},
	Issn = {0143-0807},
	Journal = {EUROPEAN JOURNAL OF PHYSICS},
	Month = {Sep},
	Number = {5},
	Pages = {913--925},
	Publication-Type = {J},
	RI = {Sanchez, Raul/C-2328-2008; Bons, Paul/F-2942-2011},
	Times-Cited = {22},
	Title = {On the applicability of Fick's law to diffusion in inhomogeneous systems},
	Volume = {26},
	Year = {2005},
	Z8 = {0},
	Z9 = {22},
	ZB = {4},
	ZS = {0}}

@article{We79,
title = {Stochastic processes with non-additive fluctuations: I. Itô and Stratonovich calculus and the effects of correlations},
journal = {Physica A: Statistical Mechanics and its Applications},
volume = {97},
number = {2},
pages = {211-233},
year = {1979},
issn = {0378-4371},
doi = {https://doi.org/10.1016/0378-4371(79)90103-1},
url = {https://www.sciencedirect.com/science/article/pii/0378437179901031},
author = {B.J. West and A.R. Bulsara and K. Lindenberg and V. Seshadri and K.E. Shuler}
}

@book{Gr15,
      author        = "Gradshteyn, Izrail Solomonovich and Ryzhik, I M and
                       Zwillinger, Daniel and Moll, Victor",
      title         = "{Table of integrals, series, and products}",
      publisher     = "Academic Press",
      address       = "Amsterdam",
      year          = "2015",
      url           = "https://cds.cern.ch/record/1702455",
      doi           = "0123849330",
}

@article{Ya12,
	author = {Yang, Mingcheng and Ripoll, Marisol},
	doi = {10.1063/1.4723685},
	issn = {0021-9606},
	journal = {The Journal of Chemical Physics},
	month = {05},
	number = {20},
	pages = {204508},
	title = {Drift velocity in non-isothermal inhomogeneous systems},
	url = {https://doi.org/10.1063/1.4723685},
	volume = {136},
	year = {2012}}

@article{Ma21,
	author = {Maris, J. J. Erik and Fu, Donglong and Meirer, Florian and Weckhuysen, Bert M.},
	date = {2021/04/01},
	doi = {10.1007/s10450-020-00292-7},
	id = {Maris2021},
	isbn = {1572-8757},
	journal = {Adsorption},
	number = {3},
	pages = {423--452},
	title = {Single-molecule observation of diffusion and catalysis in nanoporous solids},
	url = {https://doi.org/10.1007/s10450-020-00292-7},
	volume = {27},
	year = {2021}}

@article{Mi08,
  title = {Layering and Position-Dependent Diffusive Dynamics of Confined Fluids},
  author = {Mittal, Jeetain and Truskett, Thomas M. and Errington, Jeffrey R. and Hummer, Gerhard},
  journal = {Phys. Rev. Lett.},
  volume = {100},
  issue = {14},
  pages = {145901},
  numpages = {4},
  year = {2008},
  month = {Apr},
  publisher = {American Physical Society},
  doi = {10.1103/PhysRevLett.100.145901},
  url = {https://link.aps.org/doi/10.1103/PhysRevLett.100.145901}
}

@article{Ma15,
	author = {Manzo, Carlo and Garcia-Parajo, Maria F},
	doi = {10.1088/0034-4885/78/12/124601},
	journal = {Reports on Progress in Physics},
	month = {oct},
	number = {12},
	pages = {124601},
	publisher = {IOP Publishing},
	title = {A review of progress in single particle tracking: from methods to biophysical insights},
	url = {https://doi.org/10.1088/0034-4885/78/12/124601},
	volume = {78},
	year = {2015}}

@article{We17,
	author = {Weron, Aleksander and Burnecki, Krzysztof and Akin, Elizabeth J. and Sol{\'e}, Laura and Balcerek, Micha{\l} and Tamkun, Michael M. and Krapf, Diego},
	date = {2017/07/14},
	doi = {10.1038/s41598-017-05911-y},
	id = {Weron2017},
	isbn = {2045-2322},
	journal = {Scientific Reports},
	number = {1},
	pages = {5404},
	title = {Ergodicity breaking on the neuronal surface emerges from random switching between diffusive states},
	url = {https://doi.org/10.1038/s41598-017-05911-y},
	volume = {7},
	year = {2017}}

@article{No17,
	author = {Norregaard, Kamilla and Metzler, Ralf and Ritter, Christine M. and Berg-S{\o}rensen, Kirstine and Oddershede, Lene B.},
	date = {2017/03/08},
	doi = {10.1021/acs.chemrev.6b00638},
	isbn = {0009-2665},
	journal = {Chemical Reviews},
	journal1 = {Chemical Reviews},
	journal2 = {Chem. Rev.},
	month = {03},
	number = {5},
	pages = {4342--4375},
	publisher = {American Chemical Society},
	title = {Manipulation and Motion of Organelles and Single Molecules in Living Cells},
	type = {doi: 10.1021/acs.chemrev.6b00638},
	url = {https://doi.org/10.1021/acs.chemrev.6b00638},
	volume = {117},
	year = {2017},
	year1 = {2017}}

@article{Che17,
  title = {Brownian yet Non-Gaussian Diffusion: From Superstatistics to Subordination of Diffusing Diffusivities},
  author = {Chechkin, Aleksei V. and Seno, Flavio and Metzler, Ralf and Sokolov, Igor M.},
  journal = {Phys. Rev. X},
  volume = {7},
  issue = {2},
  pages = {021002},
  numpages = {20},
  year = {2017},
  month = {Apr},
  publisher = {American Physical Society},
  doi = {10.1103/PhysRevX.7.021002},
  url = {https://link.aps.org/doi/10.1103/PhysRevX.7.021002}
}

@ARTICLE{Kr19,
       author = {{Krapf}, Diego and {Metzler}, Ralf},
        title = "{Strange interfacial molecular dynamics}",
      journal = {Physics Today},
         year = 2019,
        month = sep,
       volume = {72},
       number = {9},
        pages = {48-54},
          doi = {10.1063/PT.3.4294},
       adsurl = {https://ui.adsabs.harvard.edu/abs/2019PhT....72i..48K}
}

@article{En11,
	address = {Department of Cell and Molecular Biology, Uppsala University, Uppsala, Sweden.},
	author = {English, Brian P and Hauryliuk, Vasili and Sanamrad, Arash and Tankov, Stoyan and Dekker, Nynke H and Elf, Johan},
	cois = {The authors declare no conflict of interest.},
	crdt = {2011/07/07 06:00},
	date = {2011 Aug 2},
	dcom = {20111027},
	dep = {20110705},
	doi = {10.1073/pnas.1102255108},
	edat = {2011/07/07 06:00},
	issn = {1091-6490 (Electronic); 0027-8424 (Print); 0027-8424 (Linking)},
	jid = {7505876},
	journal = {Proc Natl Acad Sci U S A},
	jt = {Proceedings of the National Academy of Sciences of the United States of America},
	lid = {10.1073/pnas.1102255108 {$[$}doi{$]$}},
	lr = {20240320},
	mh = {Amino Acids/pharmacology; Biocatalysis; Blotting, Western; Cell Division/drug effects; Computer Simulation; Cytosol/metabolism; Escherichia coli/cytology/genetics/*metabolism; Escherichia coli Proteins/genetics/*metabolism; Flow Cytometry; Guanosine Tetraphosphate/metabolism; Kinetics; Ligases/genetics/*metabolism; Luminescent Proteins/genetics/metabolism; Methionine/pharmacology; Microscopy, Confocal; Ribosomes/*metabolism; Time Factors; Tryptophan/pharmacology},
	mhda = {2011/10/28 06:00},
	month = {Aug},
	number = {31},
	own = {NLM},
	pages = {E365-73},
	pl = {United States},
	pmc = {PMC3150888},
	pmcr = {2011/07/05},
	pmid = {21730169},
	pst = {ppublish},
	sb = {IM},
	status = {MEDLINE},
	title = {Single-molecule investigations of the stringent response machinery in living bacterial cells},
	volume = {108},
	year = {2011}}

@article{Ri26,
	author = {Lewis Fry Richardson},
	journal = {Proceedings of The Royal Society A: Mathematical, Physical and Engineering Sciences},
	pages = {709-737},
	title = {Atmospheric Diffusion Shown on a Distance-Neighbour Graph},
	url = {https://api.semanticscholar.org/CorpusID:124997027},
	volume = {110},
	year = {1926}}

@article{De12,
	author = {Marco Dentz and Philippe Gouze and Anna Russian and Jalal Dweik and Frederick Delay},
	doi = {https://doi.org/10.1016/j.advwatres.2012.07.015},
	issn = {0309-1708},
	journal = {Advances in Water Resources},
	pages = {13-22},
	title = {Diffusion and trapping in heterogeneous media: An inhomogeneous continuous time random walk approach},
	url = {https://www.sciencedirect.com/science/article/pii/S0309170812002035},
	volume = {49},
	year = {2012}}

@article{Lo09,
  title = {Quantifying Hopping and Jumping in Facilitated Diffusion of DNA-Binding Proteins},
  author = {Loverdo, C. and B\'enichou, O. and Voituriez, R. and Biebricher, A. and Bonnet, I. and Desbiolles, P.},
  journal = {Phys. Rev. Lett.},
  volume = {102},
  issue = {18},
  pages = {188101},
  numpages = {4},
  year = {2009},
  month = {May},
  publisher = {American Physical Society},
  doi = {10.1103/PhysRevLett.102.188101},
  url = {https://link.aps.org/doi/10.1103/PhysRevLett.102.188101}
}

@article{Che14,
	author = {Cherstvy, Andrey G. and Chechkin, Aleksei V. and Metzler, Ralf},
	doi = {10.1039/C3SM52846D},
	issue = {10},
	journal = {Soft Matter},
	pages = {1591-1601},
	publisher = {The Royal Society of Chemistry},
	title = {Particle invasion{,} survival{,} and non-ergodicity in 2D diffusion processes with space-dependent diffusivity},
	url = {http://dx.doi.org/10.1039/C3SM52846D},
	volume = {10},
	year = {2014}}

\end{document}